\documentclass[acmlarge]{acmart}

\AtBeginDocument{
  \providecommand\BibTeX{{
    \normalfont B\kern-0.5em{\scshape i\kern-0.25em b}\kern-0.8em\TeX}}}

\setcopyright{acmcopyright}
\copyrightyear{2020}
\acmYear{2020}
\acmDOI{10.1145/1122445.1122456}

\acmJournal{TOSEM}
\acmVolume{1}
\acmNumber{1}
\acmArticle{1}
\acmMonth{1}

\usepackage{graphicx}
\usepackage{subfigure}
\graphicspath{figures/}
\DeclareGraphicsExtensions{.pdf,.jpeg,.png}

\usepackage{epstopdf}
\usepackage{fancybox}
\usepackage{amsfonts}
\usepackage{amsmath}
\usepackage{caption}
\usepackage{fancyvrb}
\usepackage{color}
\usepackage{algorithmic}
\usepackage{framed}
\usepackage{array}
\usepackage{multirow}
\usepackage{booktabs}
\usepackage{balance}
\usepackage{flushend}
\usepackage{rotating}

\usepackage{bm}

\usepackage{tikz}
\usetikzlibrary {arrows}
\usetikzlibrary {snakes}

\PassOptionsToPackage{hyphens}{url}

\newcommand{\PreserveBackslash}[1]{\let\temp=\\#1\let\\=\temp}
\newcolumntype{C}[1]{>{\PreserveBackslash\centering}p{#1}}
\newcolumntype{R}[1]{>{\PreserveBackslash\raggedleft}p{#1}}
\newcolumntype{L}[1]{>{\PreserveBackslash\raggedright}p{#1}}

\definecolor{gray50}{gray}{.5}
\definecolor{gray40}{gray}{.6}
\definecolor{gray30}{gray}{.7}
\definecolor{gray20}{gray}{.8}
\definecolor{gray10}{gray}{.9}
\definecolor{gray05}{gray}{.95}

\newcommand{\tabincell}[2]{\begin{tabular}{@{}#1@{}}#2\end{tabular}} 

\newlength\Linewidth
\def\findlength{\setlength\Linewidth\linewidth
\addtolength\Linewidth{-4\fboxrule}
\addtolength\Linewidth{-3\fboxsep}
}
\newenvironment{examplebox}{\par\begingroup
  \setlength{\fboxsep}{5pt}\findlength
  \setbox0=\vbox\bgroup\noindent
  \hsize=0.95\linewidth
  \begin{minipage}{0.95\linewidth}\normalsize}
    {\end{minipage}\egroup
    \textcolor{gray20}{\fboxsep1.5pt\fbox
     {\fboxsep5pt\colorbox{gray05}{\normalcolor\box0}}}
    \endgroup\par\noindent
    \normalcolor\ignorespacesafterend}

\newcommand{\RQ}[2]{%
 \vspace{5pt}
 \begin{center}	
  \begin{examplebox}
  \textbf{RQ#1.}~#2
  \end{examplebox}	 
 \end{center}
}

\begin{document}

\title{On the Replicability and Reproducibility of Deep Learning in Software Engineering}

\author{Chao Liu}
\email{liuchaoo@zju.edu.cn}
\affiliation{
  \institution{College of Computer Science and Technology, Zhejiang University, China, and PengCheng Laboratory, China}}

\author{Cuiyun Gao}
\email{gaocuiyun@hit.edu.cn}
\affiliation{
  \institution{Harbin Institute of Technology (Shenzhen), China}}

\author{Xin Xia}
\email{Xia@monash.edu}

\authornote{Corresponding Author: Xin Xia.}
\affiliation{
  \institution{Monash University, Australia}}

\author{David Lo}
\email{davidlo@smu.edu.sg}
\affiliation{
 \institution{Singapore Management University, Singapore}}

\author{John Grundy}
\email{John.Grundy@monash.edu}
\affiliation{
  \institution{Monash University, Australia}}

\author{Xiaohu Yang}
\email{yangxh@zju.edu.cn}
\affiliation{
  \institution{College of Computer Science and Technology, Zhejiang University, China}}

\renewcommand{\shortauthors}{Liu et al.}

\begin{abstract}

Deep learning (DL) techniques have gained significant popularity among software engineering (SE) researchers in recent years. This is because they can often solve many SE challenges without enormous manual feature engineering effort and complex domain knowledge. Although many DL studies have reported substantial advantages over other state-of-the-art models on effectiveness, they often ignore two factors: \emph{(1) replicability} -- whether the reported experimental result can be approximately reproduced in high probability with the same DL model and the same data; and \emph{(2) reproducibility} -- whether one reported experimental findings can be reproduced by new experiments with the same experimental protocol and DL model, but different sampled real-world data. Unlike traditional machine learning (ML) models, DL studies commonly overlook these two factors and declare them as minor threats or leave them for future work. This is mainly due to high model complexity with many manually set parameters and the time-consuming optimization process. In this study, we conducted a literature review on 93 DL studies recently published in twenty SE journals or conferences. Our statistics show the urgency of investigating these two factors in SE, where only 10.8\% of the studies discussed any research questions affecting replicability and/or reproducibility. More than 74.2\% of the studies do not even share source code and data to support the replicability of their complex models. Moreover, we re-ran four representative DL models in SE. Experimental results show the importance of replicability and reproducibility, where the reported performance of a DL model could not be replicated for an unstable optimization process. Reproducibility could be substantially compromised if the model training is not convergent, or if performance is sensitive to the size of vocabulary and testing data. It is therefore urgent for the SE community to provide a long-lasting link to a replication package, enhance DL-based solution stability and convergence, and avoid performance sensitivity on different sampled data.

\end{abstract}



\maketitle

\section{Introduction}\label{intro}

Deep learning (DL) has become a key branch of Machine Learning (ML) \cite{lecun2015deep,schmidhuber2015deep,hanprogrammers}, and now is a core component in systems for many aspects of modern society, such as autonomous cars \cite{tian2018deeptest}, medical image diagnosis \cite{litjens2017survey}, financial market prediction \cite{fischer2018deep}, etc. The popularity of DL technologies mainly derive from their success in computer vision \cite{forsyth2002computer,szeliski2010computer}, natural language processing \cite{collobert2011natural,manning2014stanford}, machine translation \cite{deselaers2009deep,vaswani2018tensor2tensor}, etc. Generally, DL models aim to facilitate  representation learning by leveraging big data and a specially designed neural network \cite{lecun2015deep}, such as the Recurrent Neural Network (RNN) \cite{mikolov2010recurrent,mikolov2011extensions} and the Convolutional Neural Network (CNN) \cite{lawrence1997face,kalchbrenner2014convolutional}.

In recent years, DL models have been increasingly used in the Software Engineering (SE) domain for building things such as code search engines \cite{gu2018deep,wan2019multi,shuaiimproving}, code summarization tools \cite{leclair2019neural,wan2018improving}, vulnerability identification tools \cite{han2017learning,godefroid2017learn}, and so forth. Most studies have shown that a DL algorithm can achieve higher \emph{effectiveness} over the state-of-the-art approaches that employ alternative solutions \cite{wan2019multi,wan2018improving,liu2018recommending,ge2018deep,zhou2019deep}. The main advantage of these neural network based models over other machine learning models is easier multi-level representation learning from raw data (e.g., code or text) for a specific task (e.g., code classification), substantially mitigating the hard work involved in manual feature engineering \cite{lecun2015deep,arpteg2018software,liu2018recommender,zhou2019deep}.

Apart from effectiveness, \emph{replicability} and \emph{reproducibility} are also widely accepted as important considerations in scientific research \cite{juristo2010replication,neto2019strategy,mahmood2018reproducibility,branco2017replicability,boylan2015reproducibility}. Commonly, replicability refers to whether a reported experimental result can be exactly reproduced with the same model and the same data. This is usually achieved by sharing a replication package including source code and data \cite{juristo2010replication,branco2017replicability,louridas2012note}. This is a key way to help researchers build confidence in the scientific merit of a reported result \cite{boylan2015reproducibility,branco2017replicability,li2019random}. However, most DL models are likely fail when exact reproduction is attempted due to the strong randomness in model initialization and optimization, unlike many other machine learning (ML) models \cite{li2019random}. Thus, for DL studies, we need to consider replicability under a broader definition: whether a reported experimental result can be \emph{approximately} reproduced to a high probability with the same model and the same data. 

Nonetheless, high replicability is just a requirement for the repeated experiment and it does not guarantee that the reported result represents the reality under new experiments \cite{juristo2010replication,boylan2015reproducibility,anda2008variability}. Therefore, reproducibility is required, which refers to whether one reported experimental finding can be achieved by another new experiment with the same experimental protocol, the same model, but different sampled real-world data \cite{juristo2010replication,mahmood2018reproducibility}. 

Despite the scientific merit of replicability and reproducibility, many DL-based studies in SE often ignore these two issues. It is uncertain the degree to which their reported experimental results can be replicated or reproduced with various manually set model parameters, such as the number of training iterations, the size of vocabulary that converts words into numbers, etc. Although some studies realized the importance of these factors, they regard them as mere threats and left solving them to future work \cite{tufano2019learning,leclair2019neural,gu2018deep,bhatia2018neuro}. 

To understand the prevalence and importance of replicability and reproducibility issues for DL studies in SE, we conducted this study. We wanted to investigate the following seven key research questions (RQs) in two parts. 

\vspace{5pt}\noindent\textbf{Part I: Literature Review on DL Replicability and Reproducibility.} 
The first three RQs explore the prevalence of replicability and reproducibility issues via a detailed literature review. 

\begin{itemize}
    \item \textbf{RQ1. How are DL models used and evaluated in SE studies?} We performed a meta-analysis on 93 DL models published in the last five years in twenty SE journals or conferences. Our statistics show that 77.4\% of them are published in the last two years, indicating DL's rising popularity in SE. 76.3\% of these DL studies address the challenges in code/text representation learning via RNN or CNN based models. \vspace{2pt}

    \item \textbf{RQ2. How often do SE studies provide replication packages to support the replicability of DL models?} We observed that \emph{only 25.8\%} of the reviewed DL studies provide publicly accessible replication packages. We thus highly recommend  that DL all studies must share source code and data to strengthen their replicability, considering the complexity of the DL technique. \vspace{2pt}
    
    \item \textbf{RQ3. How often do SE studies investigate RQs affecting DL replicability/reproducibility?} We found that \emph{only 10.8\%} of the DL studies discuss at least one RQ related to DL replicability/reproducibility. Therefore, the replicability and reproducibility of most DL studies in SE left unknown and future DL studies in SE must pay more attention to these two factors.
\end{itemize}

\vspace{5pt}\noindent\textbf{Part II: Experiments on DL Replicability and Reproducibility.} 
To analyze the importance of replicability and reproducibility, the following four RQs re-run four typical DL models and investigate four representative elements affecting replicability/reproducibility. These four DL models include an information retrieval model DeepCS \cite{gu2018deep}, an RNN based generation model RRGen \cite{gao2019automating}, a generation model based on reinforcement learning plus neural network (RLNN)  \cite{liu2019automatic}, and a classification model ASTNN \cite{zhang2019novel}. They were selected because they cover four different representative SE tasks that now often use ML, and they were published within the last two years in top SE venues with accessible replication packages.

\begin{itemize}
    \item \textbf{RQ4. How does model stability affect replicability?} Model stability is usually influenced by randomly initialized network weights and randomly selected training data in model optimization. After running each DL model multiple times, our experimental results show that \emph{the performance of two of the DL models is substantially overestimated by 12.2\% and 12.5\%}, in terms of the average compared to their reported results. Therefore, overestimation caused by instability will strongly influence model replicability, as the reported results would be reproduced with a low probability. \vspace{2pt}
    
    \item \textbf{RQ5. How does model convergence affect reproducibility?} Model convergence is determined by many aspects, such as quality of training data, optimization stop conditions, etc. When training each DL model with more iterations, we observe that \emph{two of the DL models show strongly turbulent performance} when compared with the ones in RQ4. Here, performance may be improved by over 8\% or reduced by less than -8\%. Thus, the high level of performance turbulence derived from low convergence would strongly affect its reproducibility, due to the low probability of reproducing a good reported result. \vspace{2pt}
    
    \item \textbf{RQ6. How does the out-of-vocabulary issue affect reproducibility?} The out-of-vocabulary (OOV) issue widely occurs when applying a DL model for natural language processing. This is because the vocabulary built for model training cannot cover new words in testing data. To test how the OOV issue affects reproducibility, we train a model with different vocabulary sizes. We find that \emph{the performance of one DL model is highly sensitive to the vocabulary size}, and whose performance increased by 3.4\% to 104.2\% for larger sizes. The results imply a low reproducibility of this model as its reported performance can not be reproduced under the new testing data with many out-of-vocabulary words. We also observed that the pointer generator \cite{see2017get} in RLNN \cite{liu2019automatic} can largely avoid the OOV issue by copying related existing word from the training data. The abstract syntax tree (AST) that captures the code structural information in ASTNN \cite{zhang2019novel} can substantially mitigate the influence of OOV. \vspace{2pt}
    
    \item \textbf{RQ7. How does testing data size affect reproducibility?} DL models are commonly verified on a subset of real-world data, e.g., GitHub\footnote{https://github.com/} and Stack Overflow\footnote{https://stackoverflow.com/}. It is often implicitly assumed that experimental findings could be reproduced for larger-scale testing data. To investigate the relationship between testing data size and DL reproducibility, we test each pre-trained DL model under different sizes of testing data. We notice that \emph{two of the DL models show considerable sensitivity to the testing data size}, where performance may be improved by 7\% or decreased by -9\%. In this case, results for a model whose performance is highly sensitive to  testing data size can not be satisfactorily reproduced.
\end{itemize}

In summary, ignoring the importance of DL replicability and reproducibility can substantially threaten the validity of a DL model used in SE. Hence, in order to strengthen DL replicability/reproducibility, it is essential and urgent for the SE community to consider the following experimental guidelines:

\begin{itemize}
    \item \emph{To ensure DL replicability}, it is recommended for all SE DL-based studies to provide a long-lasting link to a replication package with an automated evaluation approach to enhance the DL stability and convergence.
    
    \item \emph{To strengthen DL reproducibility}, SE studies using DL should mitigate issues brought by newly sampled data from other reproduced experiments, such as the OOV issue and sensitivity to the testing data scale.
    
    \item \emph{To promote DL replicability/reproducibility studies}, SE studies using DL should reduce model training time so that the burden of computation resources will not be an obstacle for replicability/reproducibility studies
\end{itemize}

The main contributions of this study are:

\begin{itemize}
    \item Conducting a detailed literature review on many recent DL studies in SE pubished in top SE venues to analyze the prevalence of DL replicability and reproducibility issues.
    
    \item Performing large-scale in-depth experiments on four representative SE studies using DL-based models to investigate the importance of DL replicability and reproducibility.
    
    \item Providing some guidelines for the SE community to mitigate the replicability and reproducibility issues in many DL-based SE studies.
\end{itemize}

The remainder of this paper is organized as follows. Section \ref{background} provides a background and surveys key related work, followed by the detailed investigated RQs in Section \ref{questions}. Section \ref{models} and Section \ref{experiment} present four studied DL models and their experimental setup. Section \ref{results} and Section \ref{discussion} show the results of our replication studies and discuss implications of these. Section \ref{conclusion} summarizes this study.

\section{Background and Related Works}\label{background}
\subsection{DL technology in SE}
The popularity of DL models in SE is mainly due to the advantages of representation learning from raw data \cite{wan2019multi,liu2019automatic,li2019random}. For example, in many recent SE studies, a large number of challenges derive from the semantic comprehension of code in programming languages \cite{zhang2019novel,liu2019two,yan2017automated,liu2018cross,yan2016self}, text in natural languages \cite{gao2019automating,yan2017learning}, or their mutual transformation \cite{gu2018deep}. As code and text involves some form of natural language processing (NLP), it commonly starts with encoding words by a fixed size of vocabulary \cite{gu2018deep}. Afterwards, a DL model is used for embedding digitalized code/text into a vector space. 
However, the vocabulary built from training data usually suffers from the OOV issue on dynamic and growing size of testing data \cite{han2017learning,hellendoorn2017deep,liu2019automatic, xu2016predicting,ben2016testing}. 

The DL models used in SE are commonly based on three different types \cite{ha2019deepperf,zhang2019novel,zhao2019actionnet}. The first type is the traditional artificial neural network (ANN) that consists of fully connected neural networks, such as the multilayer perceptron (MLP) \cite{choi1992sensitivity,fine2006feedforward,goh1995back}, deep belief network (DBN) \cite{goh1995back}, etc. To enhance representation learning, researchers developed two popular DL types. One is the convolutional neural network (CNN) \cite{lawrence1997face,kalchbrenner2014convolutional} that can sense regional characteristics of a matrix data via special convolution functions. The other is the recurrent neural network (RNN) \cite{mikolov2010recurrent,mikolov2011extensions} that can capture the features of sequential data. The long short term memory (LSTM) structure is a popular variation of RNN, which can not only capture the long-term pattern of sequential data as RNN but also focus on the short-term features \cite{hochreiter1997long,sak2014long}.    

By leveraging a DL model to learn the representation from the vectorized code/text, the model can be optimized for a specific SE task \cite{bhatia2018neuro,guo2017semantically}. To train a DL model, the parameters within the model are usually initialized by a pseudo-random generator and optimized by randomly selected subsets of training data. The optimization stops when values of a loss function converge or after a fixed number of training iterations. Many DL studies show their effectiveness over traditional state-of-the-art models \cite{zhou2019deep}.

\subsection{Replicability and Reproducibility in SE}

Replicability and reproducibility are considered to be key aspects of the scientific method  \cite{louridas2012note,ince2012case,ince2012case}. This is because they help researchers confirm the merit of previous experimental findings and promote scientific innovations \cite{louridas2012note}. In the SE domain, replicability and reproducibility have been widely claimed as essential for all empirical studies due to the unsatisfactory quality of previous research findings \cite{juristo2011role,carver2014replications,gomez2014understanding,lung2008difficulty,gomez2010replications,da2014replication,amann2013software}. However, they are rarely investigated \cite{sjoberg2005survey,zannier2006success,gonzalez2012reproducibility,kitchenham2020meta}. 

Based on the findings of empirical studies, a large proportion of recent SE studies leveraged ML techniques to solve SE tasks \cite{malhotra2010application,al2013machine}. When verifying the validity of newly proposed ML models, the replication of previous state-of-the-art baselines are required. However, the replicability and reproducibility of these ML models are barely investigated in SE. Even for the most active SE research, defect prediction, a survey shows that very few ML models are replicated and  model reliability is unclear \cite{mahmood2018reproducibility}. To improve the replicability of ML models, it is suggested using replication infrastructures (e.g., OpenML\footnote{https://www.openml.org/}) to mitigate the replication efforts \cite{mahmood2018reproducibility}, leveraging the Docker container to save the experimental environment \cite{cito2016using}, and providing standard replication guidelines \cite{mahmood2018reproducibility}.

In recent years, many DL models have been successfully applied for diverse SE tasks \cite{wan2019multi,liu2018recommending}. Although DL models often show substantial outperformance over traditional ML models, we noticed that they often overlooked replicability and reproducibility, as done for many SE studies using traditional ML models. However, DL models have many differences with the traditional ML models. These include high complexity with many manually set parameters and the time-consuming optimization processes \cite{lecun2015deep,goodfellow2016deep}. Therefore, the requirement for DL replicability and reproducibility may  have many differences. Due to the lack of study of DL replicability and reproducibility, we conducted this study.

\section{Research Questions}\label{questions}
We present our study's seven key RQs on the replicability and reproducibility of DL models in SE, divided into two parts -- literature review and experiments. We summarise how we went about answering each RQ in this study.

\vspace{5pt}\noindent\textbf{Part I: Literature Review on DL Replicability and Reproducibility.}
The first part studies the prevalence of replicability and reproducibility issues via a literature review.

\RQ{1}{How are DL models used and evaluated in SE studies?}

Before investigating the replicability/reproducibility issues, we need to analyze how SE studies using DL models are reported in the SE literature and what are the characteristics of these studies. To do this we performed a literature review on studies published in twenty top SE journals/conferences in the last five years.

\RQ{2}{How often do SE studies provide replication packages to support the replicability of their DL models?}

To help others replicate work, replication packages are often released. We count the percentage of the reviewed studies in RQ1 that share accessible replication packages. A small percentage value implies a high prevalence of replicability issues in DL studies.

\RQ{3}{How often do SE studies investigate RQs affecting DL replicability/reproducibility?}

To further analyze the prevalence of replicability/reproducibility issues, we inspect the percentage of reviewed studies in RQ1 that discuss any RQ related to DL replicability or reproducibility. A lower ratio indicates low attention on DL replicability and reproducibility from the SE community.

\vspace{5pt}\noindent\textbf{Part II: Experiments on DL Replicability and Reproducibility.}

The second part of our study evaluates the importance of replicability/reproducibility in SE studies using DL models by a suite of experiments on four representative DL models. These are DeepCS \cite{gu2018deep}, RRGen \cite{gao2019automating}, RLNN \cite{liu2019automatic}, and ASTNN \cite{zhang2019novel}. These four models were selected because they cover four different SE tasks and  are published in the last two years in top SE venues with accessible replication packages.

\RQ{4}{How does DL model stability affect replicability?}

DL model stability is affected by the random initialization and the iterative optimization with randomly selected batches of training data. However, DL studies in SE often report one experimental result. It is unknown whether the reported result can be replicated with high probability. Therefore, we run the above four DL models multiple times and estimate their replicability.

\RQ{5}{How does DL model convergence affect reproducibility?}

Ideally, the random optimization of DL models stops when the iterative process converges or all the training data have been used. However, DL models in SE usually stop optimization after a limited number of iterations, leading to an uncertainty over their convergence level. To investigate the relationship between convergence and reproducibility, we analyze how DL model performance changes by training with more iterations. A low convergent model would generate highly turbulent performance, as the model is trained for a higher number of iterations. High turbulence indicates low reproducibility, because a good reported performance would be more difficult to be replicated.

\RQ{6}{How does the Out-of-Vocabulary Issue Affect Reproducibility?}

Although DL draws high popularity in code/text representation learning in SE, the OOV issue is rarely investigated. If a model suffers from the OOV issue, its performance would be sensitive to new words in testing data. Then the reported result would be hardly reproduced for different testing data, i.e., low reproducibility. To analyze whether the OOV issue is a high potential threat to reproducibility, we train the four DL models with different sizes of vocabulary and estimate the sensitivity degree of model performance.

\RQ{7}{How does testing data size affect reproducibility?}

The DL models in SE are commonly verified by testing data collected from a real-world environment. However, this data collection process is usually random, and the testing data size only covers a small fraction of the real world data that exists. It is also implicitly assumed that the reported good performance achieved from the testing data can be generalised and reproduced with larger test data. To investigate the relationship between testing data size and reproducibility, we analyze how a pre-trained DL model performs on different sizes of testing data. If the model performance varies largely at different scales of testing data, its reproducibility will be compromised.

\section{Literature Review}\label{review}
We present a literature review on DL models in SE to answer the first three RQs raised in Section \ref{questions}.

\subsection{RQ1. How are DL models used and evaluated in SE studies?}\label{review_dl}

\begin{table}
    \centering
    \footnotesize
    \caption{The selected SE conferences and journals for literature review.}
    \begin{tabular}{|c|l|l|}
        \toprule
        \textbf{No.} & \textbf{Venue} & \textbf{Full Name} \\
        \midrule
        1 & FSE & ACM SIGSOFT Symposium on the Foundation of Software Engineering/ European Software Engineering Conference\\
        2 & ICSE & International Conference on Software Engineering \\
        3 & ASE & International Conference on Automated Software Engineering \\
        4 & ISSTA & International Symposium on Software Testing and Analysis\\
        5 & ESEM &International Symposium on Empirical Software Engineering and Measurement\\
        6 & SANER & International Conference on Software Analysis, Evolution, and Reengineering \\
        7 & ICSME &International Conference on Software Maintenance and Evolution\\
        8 & ICPC &IEEE International Conference on Program Comprehension\\
        9 & ICST &IEEE International Conference on Software Testing, Verification and Validation\\
        10 & ISSRE &International Symposium on Software Reliability Engineering\\
        11 & TOSEM &ACM Transactions on Software Engineering and Methodology\\
        12 & TSE & IEEE Transactions on Software Engineering\\
        13 & JSS &Journal of Systems and Software\\
        14 & IST & Information and Software Systems\\
        15 & ASEJ &Automated Software Engineering\\
        16 & ESE &Empirical Software Engineering\\
        17 & IETS &IET Software\\
        18 & STVR &Software Testing, Verification and Reliability\\
        19 & JSEP &Journal of Software: Evolution and Process\\
        20 & SQJ &Software Quality Journal\\
        \bottomrule
    \end{tabular}
    \label{tab_review}
\end{table}

\vspace{5pt}\noindent\textbf{Motivation.}
Recently, DL techniques are frequently applied in the SE domain, therefore it is important to investigate how the DL models are used and evaluated for SE tasks. In this study we also investigate which SE tasks DL models work for, and which DL techniques are applied in SE tasks specifically. By answering these questions, we can perceive the overview of DL studies in SE and further analyze their replicability/reproducibility issues.

\vspace{5pt}\noindent\textbf{Method.}
To find DL studies in SE, we conducted a literature review using twenty widely read SE journals and conferences listed in Table \ref{tab_review}. We performed an automated search on three digital library portals, limiting to the past five years (2015-2019), including IEEE Xplore\footnote{https://ieeexplore.ieee.org}, ACM digital library\footnote{https://dl.acm.org}, and Web of Science\footnote{http://apps.webofknowledge.com} with terms "deep OR neural OR network" for literature title, abstract, and keywords. Afterward, we inspected the searched 676 papers one by one and excluded the ones that do not propose a DL model or are not a full research paper. We finally obtained 93 papers after this filtering.

\begin{figure}
    \centering
    \includegraphics[width=0.9\linewidth]{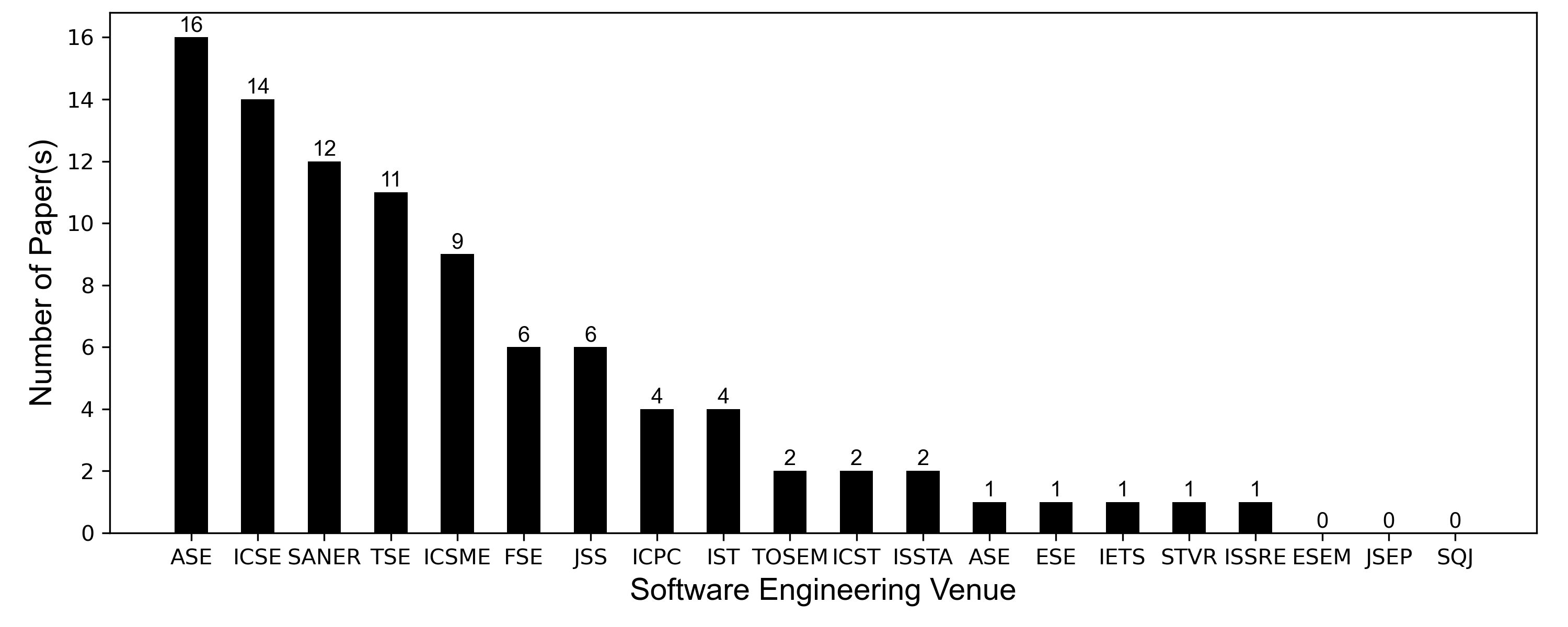}
    \caption{Number of papers published in twenty SE venues.}
    \label{fig_venues}
\end{figure}

\vspace{5pt}\noindent\textbf{Results.}
Table \ref{tab_paper_year} shows that \textbf{77.4\% of SE studies using SL} are published in these top SE venues in the last two years. The number of papers published in 2019 and 2018 nearly doubled compared with the previous three years. This result shows the growing popularity of the DL technique in SE. Table \ref{tab_paper_year} also shows that there are \textbf{far more DL studies published in SE conferences instead of journals} (66 vs. 27). This implies that researchers would like to present their new DL studies at top SE conferences. Fig. \ref{fig_venues} illustrates the total number of DL studies published in each SE venue. From the figure, we can observe that the \textbf{top-7 SE venues that published these SE studies using DL} are ASE, ICSE, SANER, TSE, ICSME, FSE, and JSS. These SE venues contribute 79.6\% of the total number of DL studies in the recent five years. 

\begin{table}
    \centering
    \small
    \caption{Number of papers published in the recent five years.}
    \setlength{\tabcolsep}{17pt}{
    \begin{tabular}{|l|ccccc|c|}
        \toprule
        \textbf{Year} & \textbf{2019} & \textbf{2018} & \textbf{2017} & \textbf{2016} & \textbf{2015} & \textbf{Total}\\
        \midrule
        \#Papers in SE conferences & 34 & 15 & 11 & 6 & 0 & 66\\
        \#Papers in SE journals & 13 & 10 & 1 & 2 & 1 & 27\\
        \#Papers in total  & 47 & 25 & 12 & 8 & 1 & 93\\
        \bottomrule
    \end{tabular}}
    \label{tab_paper_year}
\end{table}

\begin{figure}
    \centering
    \includegraphics[width=0.9\linewidth]{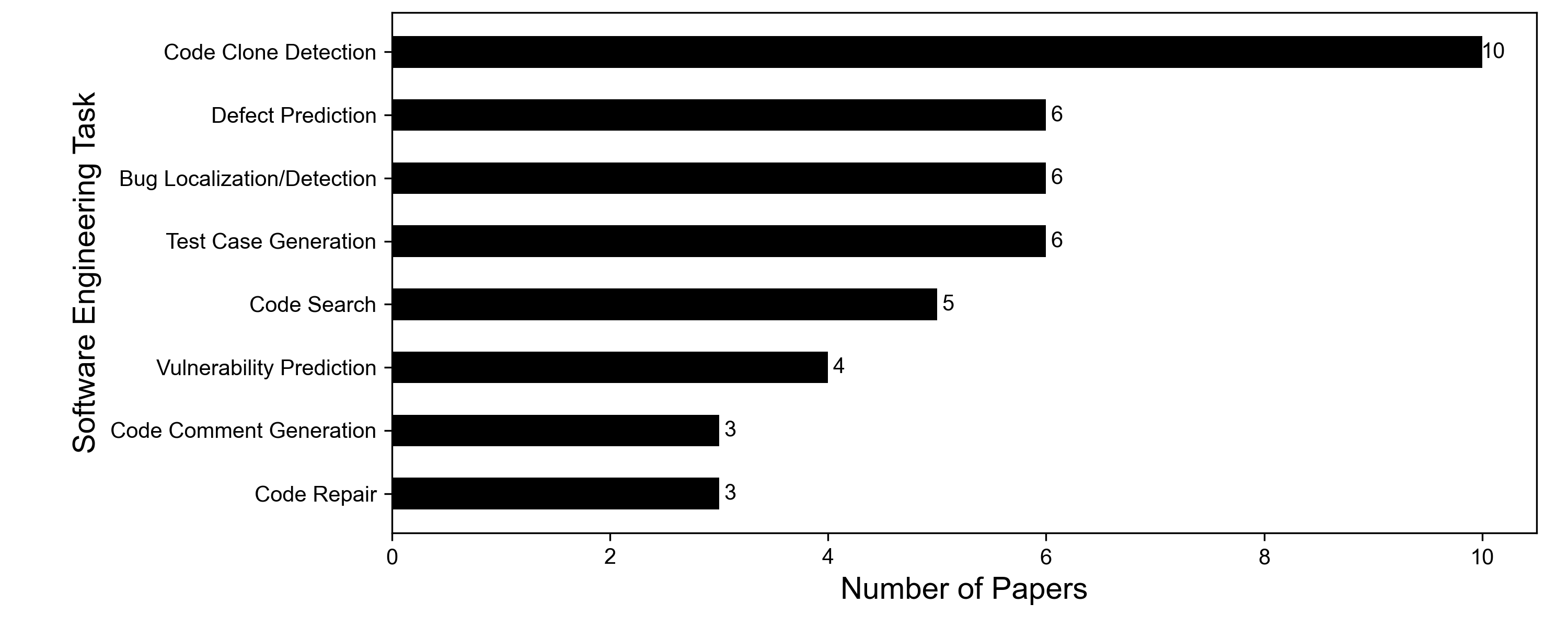}
    \caption{The top-8 popular SE tasks in terms of the number of published papers.}
    \label{fig_tasks}
\end{figure}

To understand the characteristics of these 93 reviewed studies, we classified them into 47 SE tasks and 10 study subjects as shown in Table \ref{tab_review_classify}. Note that as the study \cite{zhang2019novel} applied the proposed DL model on two different SE tasks (code classification and code clone detection), therefore we have 94 total number of studies in this table instead of 93. From Table \ref{tab_review_classify} we can observe that \textbf{63.8\% of the DL studies focus on the study of code and defects}, followed by GitHub, non-code software artifact, and Stack Overflow. Furthermore, we notice that \textbf{45.7\% of the proposed DL models work on eight SE tasks}, as illustrated in Fig. \ref{fig_tasks}. Here code clone detection \cite{zhang2019novel,gao2019teccd} and defect prediction \cite{loyola2017learning,wang2018deep} are the two most popular SE tasks for DL.

\begin{figure}
    \centering
    \includegraphics[width=0.85\linewidth]{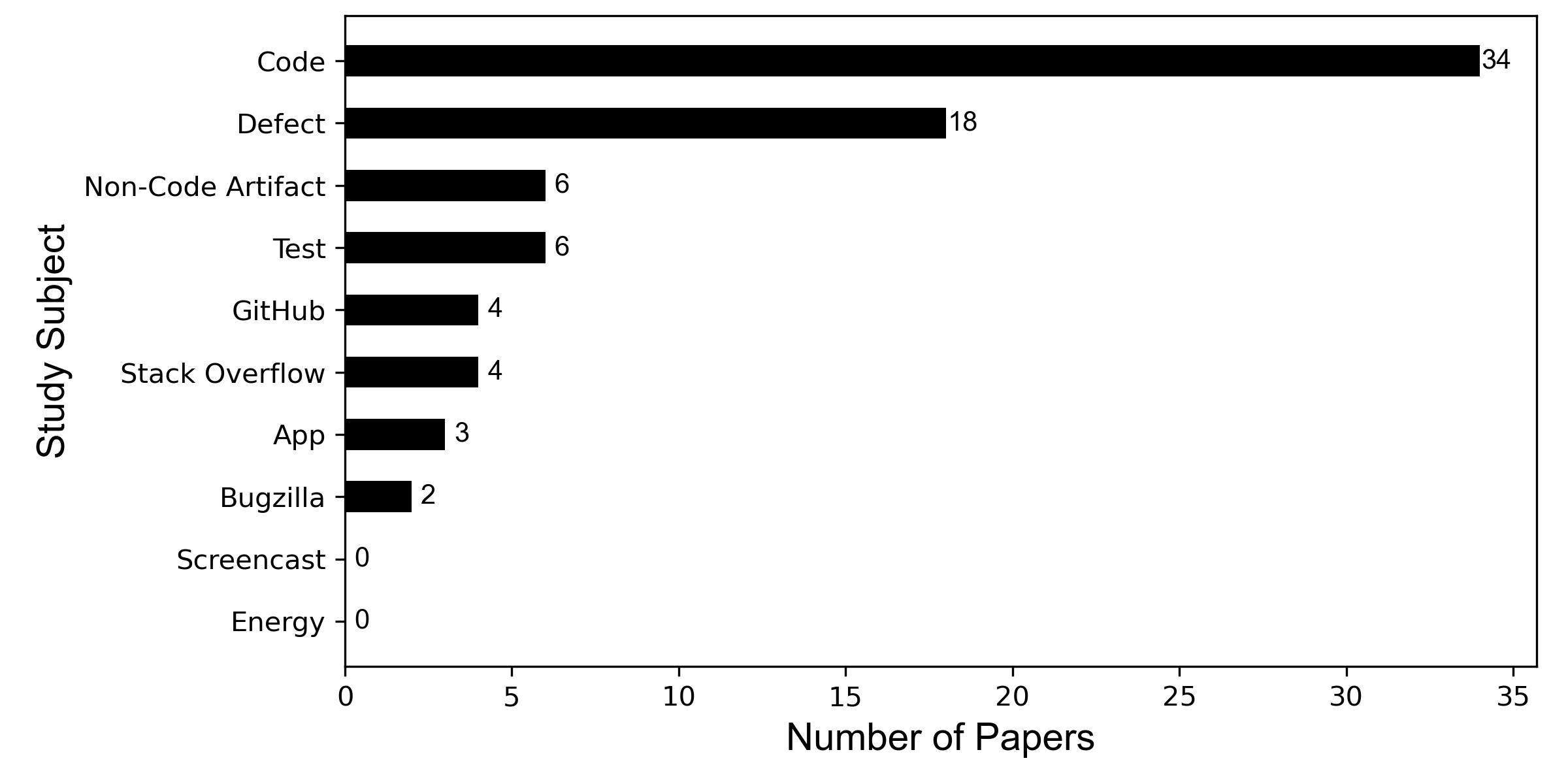}
    \caption{Number of papers addressing the semantics of code or text for different SE studies.}
    \label{fig_text}
\end{figure}

We observed that the reviewed DL models used are based on \textbf{five fundamental DL techniques}, including recurrent neural network (RNN) \cite{katz2018using}, convolutional neural network (CNN) \cite{xiao2019improving}, deep feed-forward network (DFFN) \cite{ilonen2003differential}, deep belief network (DBN) \cite{wang2016automatically}, and reinforcement learning (RL) \cite{liu2019automatic}. Note that a fundamental technique may contain many variations. For example, the long-short term memory (LSTM) belongs to the RNN model. Statistics in Table \ref{tab_review_model} show that \textbf{76.3\% 
of DL studies utilized the RNN, CNN, or their combination} for modeling SE tasks. We notice that most of DL studies leverage new DL techniques to overcome the challenges in understanding semantics of code and text and their transformation (e.g., code Summarization \cite{leclair2019neural,wan2018improving}, code change generation \cite{tufano2019learning}, etc.) in SE tasks, as illustrated in Fig. \ref{fig_text}. Furthermore, Table \ref{tab_review_model} also shows that 5.4\% of the DL studies have begun to leverage the RL technique to improve model performance. 

\vspace{5pt}\noindent\textbf{Implications.}
In the last two years, using DL techniques has gained increasing popularity in the SE community. The DL models are widely applied to various SE tasks (e.g., code clone detection, defect prediction, etc.) due to their better representation learning of code/text semantics. RNN and CNN based models and their combinations are the most frequently used DL techniques. Meanwhile, the RL technique shows its potential to further enhance the performance of existing DL models (e.g., RNN and CNN).

\begin{table}
    \centering
    \footnotesize
    \caption{Categories of ninety three DL studies in our literature review. There are ninety four studies in the table because the study \cite{zhang2019novel} performs predictions on two SE tasks (code clone detection and code classification).}
    \begin{tabular}{|l|L{0.67\linewidth}|C{22pt}|C{22pt}|}
        \toprule
        \textbf{Study Subject} & \textbf{Tasks} and \textbf{References} & \textbf{\#Task} & \textbf{\#Study}\\
        \midrule
        \multirow{7}{*}{Code} & Code clone detection \cite{zhang2019novel,gao2019teccd,li2017cclearner,zhao2018deepsim,nafi2019clcdsa,white2016deep,buch2019learning,yu2019neural,guo2019deep,xu2019ldfr} &\multirow{7}{*}{16} & \multirow{7}{*}{36}\\
        & Code search \cite{cambronero2019deep,wang2019domain,gu2018deep,wan2019multi,gu2016deep}, Code classification \cite{zhang2019novel,leclair2018adapting,dq2019bilateral} &&\\
        &code readability classification/prediction \cite{mi2018improving,rani2018neural}, Function type inferring \cite{malik2019nl2type,hellendoorn2018deep} &&\\
        & Code generation \cite{nguyen2018deep,gao2019neural}, Code Summarization \cite{leclair2019neural,wan2018improving}, Code Decompilation \cite{lacomis2019dire,katz2018using} &&\\
        & Code change generation \cite{tufano2019learning}, Data structure classification \cite{molina2019training}, Reverse execution \cite{mu2019renn} && \\
        & Design pattern detection \cite{thaller2019feature}, Technical debt detection \cite{ren2019neural}, Story points prediction \cite{choetkiertikul2018deep} &&\\
        & Inconsistent method name refactoring \cite{liu2019learning},
        Stable patch detection \cite{hoang2019patchnet} &&\\
        
        \midrule
        \multirow{4}{*}{Defect} & Defect prediction  \cite{loyola2017learning,wang2018deep,wen2018well,wang2016automatically,liu2018connecting,tong2018software}, Vulnerability prediction \cite{han2017learning,godefroid2017learn,dam2018automatic,yan2018new} & \multirow{4}{*}{8}&\multirow{4}{*}{24}\\
        &Bug localization/detection \cite{huo2019deep,xiao2019improving,lam2017bug,li2019deepfl,zhang2019cnn,wang2019textout} , Code repair \cite{bhatia2018neuro,white2019sorting,tufano2019empirical}&&\\
        &Code smell detection \cite{liu2018deep,liu2019deep}, Anti-pattern identification \cite{barbez2019deep}&&\\
        &Code naturalness prediction \cite{hellendoorn2017deep}, Static checker alarm classification \cite{lee2019classifying}  & & \\
        
        \midrule
        \multirow{2}{*}{GitHub} & Issue-commit link recovery \cite{xie2019deeplink,ruan2019deeplink}, Effort prediction \cite{kumar2016hybrid,bisi2016software}, Duration prediction \cite{lopez2015neural}&\multirow{2}{*}{5}&\multirow{2}{*}{7}\\
        & Commit message generation \cite{jiang2017automatically}, Pull request description generation \cite{liu2019automatic} &&\\
        
        \midrule
        \multirow{3}{*}{Non-Code Artifact} & Code comment generation \cite{hu2018deep,zhou2019augmenting,liu2019log}, Software resources tracking \cite{guo2017semantically} & \multirow{3}{*}{5}&\multirow{3}{*}{7}\\
        & Software configuration prediction \cite{ha2019deepperf}, Software incidents triage \cite{chen2019continuous} &&\\
        & requirement actor/action extraction \cite{al2018use} &&\\
        
        \midrule
        Test & Test case generation \cite{liu2017automatic,tufano2018learning,koo2019pyse,zheng2019wuji,ben2016testing,chen2018drlgencert} & 1 & 6\\
        
        \midrule
        \multirow{2}{*}{Stack Overflow} & Link classification \cite{xu2016predicting}, Question retrieval \cite{chen2016learning}, API extraction \cite{ma2019easy}&\multirow{2}{*}{5}&\multirow{2}{*}{5}\\
        &Tag recommendation \cite{liu2018fasttagrec}, Intention classification \cite{huang2018automating} &&\\
        
        \midrule
        \multirow{2}{*}{App} & GUI generation \cite{chen2018ui,moran2018machine}, Communication identification \cite{zhao2018neural}&\multirow{2}{*}{3}&\multirow{2}{*}{5}\\
        &Review response classification/generation \cite{guo2019systematic,gao2019automating} &&\\
        
        \midrule
        Bugzilla & Duplicate bug detection \cite{deshmukh2017towards}, Bug report summarization \cite{li2018unsupervised} & 2 & 2\\
        
        \midrule
        Screencast & Programming screencast action recognition \cite{zhao2019actionnet} & 1 & 1\\
        \midrule
        Energy & Software energy consumption prediction \cite{romansky2017deep} & 1 & 1\\
        \midrule
        Total & -& 47 & 94\\
        \bottomrule
    \end{tabular}
    \label{tab_review_classify}
\end{table}

\begin{table}
    \centering
    \small
    \caption{Statistics of basic model settings from ninety three DL studies in literature review. The basic DL models include recurrent neural network (RNN), convolutional neural network (CNN), deep belief network (DBN), deep feed-forward network (DFFN), and reinforcement learning (RL).}
    \begin{tabular}{|l|l|c|c|}
        \toprule
        \textbf{Configuration} & \textbf{Model Setting} & \textbf{\#Study} & \textbf{\%Study} \\
        \midrule
        \multirow{5}{*}{\tabincell{l}{The study\\ with individual\\ DL model}} & RNN & 36 & 38.7\% \\
        & CNN & 25 & 26.9\% \\
        & DFFN & 20 & 21.5\% \\
        & DBN & 2 & 02.2\% \\ 
        & RL & 0 & 00.0\% \\
        \midrule
        \multirow{6}{*}{\tabincell{l}{The study\\ with combined\\ DL models}} & RNN + CNN & 3 & 03.3\% \\
        & RNN + RL & 2 & 02.2\%\\
        & CNN + RL & 2 & 02.2\%\\
        & RNN + DFFN & 1 & 01.1\%\\
        & RNN + CNN + DFFN & 1 & 01.1\%\\
        & RNN + CNN + RL & 1 & 01.1\%\\
        \bottomrule
    \end{tabular}
    \label{tab_review_model}
\end{table}

\begin{table}
    \centering
    \small
    \caption{Statistics of ninety three studies that share links to their replication packages.}
    \begin{tabular}{|L{0.32\linewidth}|c|c|}
        \toprule
        \textbf{Item} & \textbf{\#Study} & \textbf{\%Study}  \\
        \midrule
        Link included & 27 & 29.0\%\\
        Accessible link & 24 & 25.8\%\\
        \bottomrule
    \end{tabular}
    \label{tab_review_link}
\end{table}

\subsection{RQ2. How Often do SE Studies Provide Replication Packages to Support the Replicability for DL Models?}\label{back_learning}

\vspace{5pt}\noindent\textbf{Motivation.}
Replicability is a fundamental requirement for DL models. DL replicability can however be compromised when other individual researchers re-implement a model by themselves with some careless errors or omissions. Sharing replication packages is a key to better support DL model replicability as it allows other researchers or practitioners to much more easily repeat results reported in a paper. Therefore, to analyze whether  replicability is a supported issue for DL studies in SE, we investigate the percentage of the reviewed DL studies that share links to replication packages and whether the links are accessible. 

\vspace{5pt}\noindent\textbf{Method.}
To investigate the prevalence of support for replicability issues in SE studies, we inspected the reviewed 93 DL studies in Section \ref{review_dl} one by one. For each study, we searched all the links within papers, and checked whether these links are related to replication packages. Finally, we counted the ratio of DL studies providing accessible replication packages. A lower ratio indicates lack of support for replication packages and hence replicability issues in SE studies using DL models.

\vspace{5pt}\noindent\textbf{Results.}
Table \ref{tab_review_link} shows the statistics of the reviewed DL studies that share links to their replication packages. We notice that \textbf{only 29\% of these SE studies using DL} provided a link to publicly shared source code and data. By entering the links with a browser, we found that three links are broken. Therefore, only 25.8\% of the 93 DL studies provide \emph{accessible} links to their replication packages. How long these remaining replication package links will remain available is unknown.

\vspace{5pt}\noindent\textbf{Implications.}
Most of our reviewed DL studies in SE do not provide replication packages that are publicly accessible (with the link given in the paper), leaving prevalent threats to low replicability. Under this circumstance, other researchers or practitioners need to replicate existing studies involving DL models from scratch and may miss important implementation details. 

\subsection{RQ3. How Often do SE Studies Investigate RQs Affecting DL Replicability/Reproducibility?}\label{sub_rqs}

\vspace{5pt}\noindent\textbf{Motivation.}
According to our observation, some internal factors within DL models may affect study replicability and reproducibility, although sharing replication packages can help support DL replicability. For example, a model with unstable performance will lead to low replicability. Model stability can be influenced by randomly initialized network weights for the DL model. Although some replication packages may provide initialized weights, the random optimization process with many manually set parameters may still influence model stability. Moreover, if the model performance is sensitive to the size of the testing data, DL-based study reproducibility would be unsatisfactory. Therefore, discussing these kind of internal factors in DL models can further strengthen the DL replicability and reproducibility. We analyzed how often DL studies in SE investigate RQs affecting DL replicability/reproducibility and further classify the related RQs. A work that discusses no RQ on DL replicability/reproducibility would possess potential threats to its validity.

\vspace{5pt}\noindent\textbf{Method.}
To investigate how DL studies in SE discuss internal factors (e.g., model stability, the performance sensitivity on testing data size, etc.) on DL-based study replicability/reproducibility, we inspected what the RQs in each paper aim to investigate. We then calculated the percentage of reviewed studies in Section \ref{review_dl} that conduct an RQ investigating DL replicability or reproducibility. If DL studies in SE rarely perform experiments on such internal factors, the DL replicability/reproducibility issue would be still prevalent even if some DL studies may provide replication packages.

\vspace{5pt}\noindent\textbf{Results.}
As shown in Table \ref{tab_review_useful}, we observe that the reviewed DL studies focus on three types of RQ, including effectiveness  -- whether the proposed DL model outperforms the state-of-the-art baseline; efficiency  -- whether the DL model works faster than the baseline; and replicability/reproducibility -- the degree to which that the reported model performance can be replicated/reproduced considering some model/experimental settings. Table \ref{tab_review_useful} shows that \textbf{all the 93 studies} investigate the effectiveness of the model e.g., in terms of precision, recall, F1, etc. This is because verifying model effectiveness is a basic requirement for DL studies. However, \textbf{only 40.9\% of the studies} provided runtime information that informs the efficiency of their proposed approach for training and prediction. As to the model replicability/reproducibility, \textbf{only 10.8\% of the total studies} investigated RQs related to these factors.

To understand the characteristics of RQs on DL replicability/reproducibility, we classify them into four RQ types as described in Table \ref{tab_rq_rr}. These RQ types include model randomness \cite{han2017learning,bisi2016software,tong2018software}, whether the randomness in DL model (e.g., random optimization process) affects model replicability; model convergence \cite{li2019deepfl,katz2018using}, whether the model optimization is divergent so that the turbulent performance leads to low replicability; out-of-vocabulary (OOV) issue \cite{han2017learning,hellendoorn2017deep,liu2019automatic, xu2016predicting,ben2016testing}, whether model performance is sensitive to the size of vocabulary for encoding words into vectors, resulting in low reproducibility; sensitivity to testing data size \cite{nafi2019clcdsa}, whether the model performance is sensitive to the size of testing data (i.e., low reproducibility). Table \ref{tab_rq_rr} shows that the OOV issue is the most frequently discussed RQs because this issue is widely considered in the domain of the natural language processing. However, \textbf{the number of papers that discuss this issueis low} (5 papers, 5.4\% of total studies), since 76.3\% of the reviewed 93 DL studies need to encode code/text by using a fixed size of vocabulary (as illustrated in Fig. \ref{fig_text}). The number of papers related to the other three RQ types is also very low.

Although some studies realized the importance of replicability/reproducibility, they just regarded them as some among many threats to validity and left them to future work \cite{tufano2019learning,leclair2019neural,gu2018deep,bhatia2018neuro}. One major obstacle to addressing DL replicability/reproducibility issue is the time-consuming nature of DL training. This is because researchers' limited computational resources have to be used for verifying the model effectiveness at first.

\vspace{5pt}\noindent\textbf{Implications.}
Some internal factors, such as model randomness, convergence, OOV issue, and sensitivity to testing data size in DL models, may threaten DL replicability and reproducibility even if the authors provide accessible replication packages. However, these factors are widely overlooked in DL-based SE studies. Therefore, there is a need for more studies on the impact of replicability and reproducibility issues in DL applications in SE. To promote such studies, reducing the time it takes to train DL models is also important because of limited computational resources.
    
\begin{table}
    \centering
    \footnotesize
    \caption{Statistics of studies that discuss RQs related to model effectiveness, efficiencey, or replicability/reproducibility.}
    \begin{tabular}{|l|L{0.59\linewidth}|c|c|}
        \toprule
        \textbf{Type of RQ} & \textbf{Description} & \textbf{\#Study} & \textbf{\%Study}\\
        \midrule
        Effectiveness & Whether the proposed new DL model outperforms some baselines in terms of some evaluation metrics. & 93 & 100.0\%\\
        \midrule
        
        Efficiency & Whether the proposed new DL model works faster than some baselines during model training and testing respectively. & 38 & 40.9\%\\
        \midrule
        
        Replicability/Reproducibility & The degree to which the reported model performance can be replicated/reproduced considering some model settings or experimental settings. & 10 & 10.8\%\\
        \bottomrule
    \end{tabular}
    \label{tab_review_useful}
\end{table}

\begin{table}
    \centering
    \footnotesize
    \caption{Classification of RQs related to replicability/reproducibility issues among reviewed DL studies in SE. There are ten DL studies in total, and one study discusses two RQ types.}
    \begin{tabular}{|L{94pt}|L{0.59\linewidth}|c|c|}
        \toprule
        \textbf{RQ Type} & \textbf{Description} & \textbf{\#Study} & \textbf{\%Study} \\
        \midrule
        Model randomness & Whether the randomness in model training (e.g., parameter initialization and random optimization process) affects DL replicability. & 3 & 3.2\%\\
        \midrule
        
        Model convergence & Whether the model optimization stops due to convergence instead of the limited size of training data; If so, whether it strongly influences the DL replicability. & 2& 2.2\%\\
        \midrule
        
        Out-of-vocabulary issue & If a DL model involves a vocabulary for encoding words, whether the fixed size of vocabulary leads to the reported model performance is unreproducible. & 5 & 5.4\%\\
        \midrule
        
        Sensitivity to testing data size & Whether the model performance is sensitive to the size of testing data, resulting in low reproducibility. & 1 & 1.1\%\\
        \bottomrule
    \end{tabular}
    \label{tab_rq_rr}
\end{table}

\section{Deep Learning Models}\label{models}
According to the literature review in Section \ref{review}, we know that the DL replicability/reproducibility issues are commonly overlooked and often regarded as minor threats in the SE domain. To analyze the importance of DL replicability/reproducibility, we designed four experiments, corresponding to the last four RQs in Section \ref{questions}, using four DL models for SE tasks. This section provides brief descriptions on these four DL models, which are used for different SE tasks including code search \cite{gu2018deep}, review response generation \cite{gao2019automating}, pull request description generation \cite{liu2019automatic}, and code classification \cite{zhang2019novel}. These DL models are selected as study subjects because they are published in top SE venues in the last two years with accessible replication packages. Table \ref{table:models} summarizes these four models and their details are briefly described as follows. 

\begin{table}
    \centering
    \small
    \caption{Summary of four DL models for different SE tasks.}
    \label{table:models}
    \begin{tabular}{|l|l|c|}
    \toprule
       \textbf{Task} & \textbf{Description} & \textbf{Model Name} \\
    \midrule
       Code search & Searching related code from codebase according to a user's query. & DeepCS \cite{gu2018deep} \\
       Review response generation & Generating high-quality response to a user's review of an App. & RRGen \cite{gao2019automating} \\
       Pull request description generation & Generating high-quality description for an empty pull request. & RLNN \cite{liu2019automatic} \\
       Code classification & Classifying code fragments according to their functionality. & ASTNN \cite{zhang2019novel} \\
    \bottomrule
    \end{tabular}
\end{table}

\subsection{Code Search by DeepCS}
\vspace{5pt}\noindent\textbf{Task and Solution.} Searching and reusing existing code can help developers accelerate software development. Code search research aims to provide developers a code search engine that returns some relevant code examples from a large scale codebase, e.g., GitHub, according to developers' search queries, such as "how to convert string to int". To solve this task, a DL model named DeepCS \cite{gu2018deep} first embeds a search query and all candidate code methods into a shared vector space, and then returns the top-10 methods relevant to the query in terms of the Cosine similarities.

\vspace{5pt}\noindent\textbf{Modeling and Optimization.} 
To learn the relationship between queries in natural language and code methods in a programming language, DeepCS leverages three recurrent neural networks (RNNs) \cite{mikolov2010recurrent} and one multilayer perceptron (MLP) \cite{choi1992sensitivity} to embeds code method ($c$) and query ($q$) into vectors respectively:

\begin{equation}
\label{eq_deepcs_embedding}
\left\{
    \begin{array}{lll}
        v_{c} & = RNN(m) + RNN(a) + MLP(t),&\\
        v_{q} & = RNN(q),&
    \end{array}
\right.
\end{equation}

\noindent
where $m$, $a$, $t$ are three components of a code method $c$ including method name, API sequence, and token set in method body; $v_{c}$ and $v_{q}$ are the vectorized method and query, respectively. 

Note that the token set $t$ is processed by a MLP instead of a RNN because the token set only considers the token frequency instead of the token order. With these two vectors, the relevant code methods to a query can calculated by the cosine similarity as $cos(v_{c}, v_{q})=(v_{c}^{T}v_{q})/({\|v_{c}\|\|v_{q}\|})$. To optimize parameters ($\theta$) in RNNs and MLP, DeepCS initialized them by a pseudo-random generator and trained by the loss function ($L$):

\begin{equation}
    \label{eq_loss_function}
    L(\theta)=\sum_{<C,Q+,Q->\in{P}}max(0,0.05-cos(v_{c},v_{q}^{+})+cos(v_{c},v_{q}^{-}),
\end{equation}

\noindent
where $v_{c}$$\in$$C$ is a code method randomly selected from training data ($P$); $v_{q}^{+}$$\in$$Q^{+}$ is the code related comment to stand for the related query; and $v_{q}^{-}$$\in$$Q^{-}$ is an irrelevant comment randomly selected from other methods. The objective of this loss function is to shorten the similarity between the matched method ($v_{q}^{+}$) and query ($v_{q}$) while enlarging the similarity for the unmatched pairs.

\subsection{App Review Response Generation by RRGen}
\vspace{5pt}\noindent\textbf{Task and Solution.}
Studies show that replying to users' reviews of an App largely increases the chances of a user updating their given rating, so that the App can maintain its user base and even attract more \cite{gao2019automating}. To automatically generate high-quality responses to user reviews, an RNN-based model called RRGen (review response generation) was developed to learn the relationship between review-response pairs \cite{gao2019automating}.

\vspace{5pt}\noindent\textbf{Modeling and Optimization.}
To generate a high-quality response ($y$), RRGen takes a review ($x$) as its input and builds an encoder-decoder model for $x$ and $y$ by leveraging RNN and MLP: 

\begin{equation}
    \label{eq_rrgen}
    y = RNN(x_{t}, MLP(x_{c},x_{l},x_{r},x_{s}))
\end{equation}

\noindent
where $x_{t}$ is a sequence of keywords in the review $x$; $x_{c}$, $x_{l}$, $x_{r}$, and $x_{s}$ are four high-level attributes of a review, which are App category, review length, user rating, and user's positive/negative sentiment respectively. 

To encode tokens in review keywords and responses, RRGen constructs a vocabulary with top-10k frequently occurred words in training data. Moreover, to optimize the randomly initialized parameters ($\theta$) in RRGen, the model utilized a loss function ($L$):

\begin{equation}
    \label{eq_rrgen_loss}
    L(\theta)=max_{\theta}\frac{1}{N}\sum_{i=1}^{N}logp_{\theta}(y'_{i}|y_{i},x_{c_{i}},x_{l_{i}},x_{r_{i}},x_{s_{i}})
\end{equation}

\noindent
where $p_{\theta}$ calculates the cross-entropy between the generated response ($y_{i}$) and the ground-truth ($y'_{i}$) for the $i$-th review; $N$ is the total number of review-response pairs; $x_{c_{i}}$, $x_{l_{i}}$, $x_{r_{i}}$, $x_{s_{i}}$ are the category, review length, user rating, and sentiment for the $i$-th review. This loss function intends to maximize cross-entropy between all pairs of generated responses and their ground-truth.

\subsection{Pull Request Description Generation by RLNN}
\vspace{5pt}\noindent\textbf{Task and Solution.}
Developers contribute to a project through a pull request (PR) with a description. The description helps reviewers and other developers understand their contributions. However, more than 34\% of real-world PR descriptions are empty \cite{liu2019automatic}. To help developers generate high-quality PR descriptions automatically, a reinforcement learning (RL) based RNN model was proposed \cite{liu2019automatic}. We call the model RLNN in this study. It takes a PR related context as model input, including commit messages, the added code comments, and then produces a summary as the PR description.

\vspace{5pt}\noindent\textbf{Modeling and Optimization.}
To generate the correct PR description ($g$), RLNN treats the commit message plus the related code comments as the model input ($s$). RLNN works in two phases. It first builds an RNN encoder-decoder model for the input-output pairs as $y=RNN(x)$ and it is pre-trained by a common loss function ($L_{ml}$):

\begin{equation}
    \label{eq_loss_ml}
    L_{ml} = -\frac{1}{|y|}\sum_{j=1}^{|y|}log p(y_{j}|\hat{y}_{0},\dots,\hat{y}_{j-1},w),
\end{equation}

\noindent
where $w$ the RNN parameters; $\hat{y}_{0},\dots,\hat{y}_{j-1}$ are the $0$-th to ($j$-1)-th tokens of the input; $y_{j}$ is the $j$-th generated token. The loss function intends to estimate the negative log-likelihood of the generated description.

In the second phase, all of the words are encoded by a fixed vocabulary with top-50 words shown in training data in terms of frequency. To overcome the (OOV) issue, RLNN integrates a pointer generator to select a token from the vocabulary or to copy one from the model decoding step. To further enhance the quality of the generated description, RLNN optimizes RNN by a special loss function ($L$):

\begin{equation}
    \label{eq_loss_combine}
    L = \gamma{}L_{rl} + (1-\gamma)L_{ml},
\end{equation}

\noindent
where $L_{ml}$ is the loss function defined in Eq. (\ref{eq_loss_ml}); $\gamma$ is a coefficient to balance two loss functions $L_{ml}$ and $L_{rl}$; $L_{rl}$ is an reinforcement loss function described as follows:

\begin{equation}
    \label{eq_loss_rl}
    L_{rl} = -(r(y^{s})-r(y)))\sum_{j=1}^{|y^{s}|}log p(y_{j}^{s}|y_{0}^{s},\dots,y_{j-1}^{s},w),
\end{equation}

\noindent
where $y^{s}$ is a sampled description by an Monte-Carlo method while $y$ is the ground-truth description; $r(y^{s})$ or $r(y)$ measures the generation accuracy between the generated description $y$ and $y^{s}$ or $y$ in terms of the ROUGE-L F1-score; $y_{0}^{s},\dots,y_{j-1}^{s}$ are the $0$-th to ($j$-1)-th tokens of the sampled description; $y_{j}^{s}$ is the $j$-th generated token for the sampled input. The major advantage of $L_{rl}$ is the incorporation of the automated evaluation ROUGE into the model optimization, training the model in a more natural and accurate way.

\subsection{Code Classification by ASTNN}
\vspace{5pt}\noindent\textbf{Task and Solution.}
Correctly classifying code fragments by their functionalities helps developers understand and maintain software projects. To improve this code classification accuracy, an AST-based neural network (ASTNN) was proposed \cite{zhang2019novel}, which represents a code by a sequence of ASTs and performs the classification by using a CNN based model.

\vspace{5pt}\noindent\textbf{Modeling and Optimization.}
To learn the representation of a code fragment ($x'$), ASTNN first parses the code ($x$) into a sequence of small statement trees. Each sequence is used to train a RNN based model that is optimized by the following loss function:

\begin{equation}
    \label{eq_loss_rnn}
    L_{RNN} = \|n_{1}-RNN(n_{1})\|_{2}^{2} + \|n_{2}-RNN(n_{2})\|_{2}^{2},
\end{equation}

\noindent
where $n_{1}$ and $n_{2}$ are two children nodes of a parent node in a parsed code ($x$) separately; $RNN(n_{1})$ and $RNN(n_{2})$ are the generated vectors by an RNN model corresponding to the nodes $n_{1}$ and $n_{2}$ respectively. To learn the representation of the AST, the loss function leverages the second normal form to assess the learning accuracy.

Furthermore, to apply the pre-trained ASTNN to the code classification, $\hat{x}=W_{0}x'+b_{0}$ is used to predict the its ground-truth category ($\hat{x}$), which is optimized by the following loss function:

\begin{equation}
    \label{eq_loss_astnn}
    L(\theta,\hat{x},y) = \sum{}\left(-log\frac{exp(\hat{x}_{y})}{\sum{}_{j}exp(\hat{x_{j}})}\right),
\end{equation}

\noindent
where $W_{0}$ is a weight matrix and $b_{0}$ is a bias term; $\hat{x}$ is the predicted category while $y$ is the ground-truth. The loss function $L$ intends to increase the accuracy between the generated category and the ground-truth via the cross-entropy measurement.

\section{Experimental Setup}\label{experiment}

We describe the experimental setup for the four DL models summarised in Section \ref{models}. Table \ref{table:configurations} provides their important settings related to the last four RQs in Section \ref{questions}. We run the public replication packages shared by authors on two servers with eight GPUs in total. The detailed experiment setup is presented as follows.

\begin{table*}[t]
    \centering
    \small
    \caption{Experiment setup for four DL models.}
    \label{table:configurations}
    \setlength{\tabcolsep}{11.5pt}{
    \begin{tabular}{|c|cccc|}
    \toprule
        \textbf{Model} & \textbf{Training Iterations} & \textbf{Vocabulary Size} & \textbf{Testing Data Size} & \textbf{Evaluation Metric} \\
    \midrule
        DeepCS & 500 epochs & 10,000 & 16,262,602 methods & MRR\\
        RRGen & 3 epochs & 10,000 & 14,727 reviews & BLEU-4\\
        RLNN & 22,000 iterations & 50,000 & 4,100 pull requests & F1-Score of ROUGE-L\\
        ASTNN & 15 epochs & 8,182 & 10,401 code fragments & Accuracy\\
    \bottomrule
    \end{tabular}}
\end{table*}

\vspace{5pt}\noindent\textbf{DeepCS.} 
Following the original study \cite{gu2018deep}, DeepCS is trained using around 18 million commented Java methods with 10k fixed size of vocabulary, and the optimization stops at 500 epochs. Afterward, the model is tested using about 16 million Java code methods with top-50 search queries extracted from Stack Overflow as developers' search requirements. For each query, DeepCS returns top-10 relevant code. The performance of code search is estimated by a widely used metric MRR (mean reciprocal rank). $MRR$ is defined as $Q^{-1}\sum_{q=1}^{Q}FRank_{q}^{-1}$, where $Q$ is the total number of queries; $FRank$ is the rank of the first correct search to a query. We re-ran DeepCS by using the authors' replication package\footnote{https://github.com/guxd/deep-code-search} from GitHub.

\vspace{5pt}\noindent\textbf{RRGen.}
For App review response generation, RRGen encodes words of reviews and responses in training and testing data by vocabulary with top-10k words appeared in training data. Subsequently, RRGen is trained using 279,792 review-response pairs with three epochs and tested using 14,727 reviews \cite{gao2019automating}. Note that the three epochs are long enough for model optimization, as each epoch takes more than 40 hours due to the large size of training data \cite{gao2019automating}. To assess the generation accuracy, the textual similarity between the generated responses ($\hat{y}$) and the ground-truth ($y$) is measured by the BLEU (bilingual evaluation understudy) score \cite{bahdanau2014neural}. Generally, the BLEU score analyzes the co-occurrences of $n$-grams between $y$ and $\hat{y}$, where $n$ is set to 4 because it is demonstrated to be more correlated with human judgments than other settings \cite{liu2016not}. We re-ran RRGen by using the replication package\footnote{https://github.com/armor-ai/RRGen} shared by the authors in GitHub.
    
\vspace{5pt}\noindent\textbf{RLNN.} 
To generate a high-quality pull request description, RLNN is optimized using training data with 32.8k PRs. In the encoding phase, RLNN uses a fixed vocabulary with 50k words frequently occurred in training data. Note that RLNN is based on a pre-trained RNN model with 12,000 iterations and then performed an RL optimization on it with 22,000 iterations. To better understand the replicability of RLNN, we pre-trained the RNN one time and mainly investigated RQs on the RL part. Moreover, the validity of RLNN is verified on testing data involving 4.1k PRs. Model performance is evaluated by using ROUGE (recall-oriented understudy for gisting evaluation) metrics, which highly correlate with human assessment of summarized text quality \cite{lin2004looking}. Specifically, the F1-score of ROUGE-L is adopted in the model evaluation and its detailed definition can be found in the RLNN study \cite{liu2019automatic}. We re-ran their public replication package\footnote{https://github.com/Tbabm/PRSummarizer} shared on GitHub.
    
\vspace{5pt}\noindent\textbf{ASTNN.} 
For code classification, the validity of ASTNN is verified by a public dataset built by Mou et al. \cite{mou2016convolutional}\footnote{https://sites.google.com/site/treebasedcnn/}. The dataset contains 46,887 code fragments with training data encoded using a vocabulary containing 8,182 words. It involves 10,402 code fragments in the testing dataset. The model training stops at 15 epochs. The classification accuracy is computed as $Accuracy = M^{-1}\sum_{m=1}^{M}f(x_{m},y_{m})$, where $f$ is an indicator function that returns 1 if the predicted category of $m$-th code fragment ($x_{m}$) equals to the ground-truth ($y_{m}$), otherwise it returns 0; $M$ denotes the total number of code fragments in the testing data. We re-ran the authors' replication package\footnote{https://s3.us-east-2.amazonaws.com/icse2018/index.html.} that are shared publicly.

Note that the units (epoch or iteration) and size (3 or 500) of the model iterations are kept the same as the original studies, but their actual running time could be substantially different.

\section{Experimental Results}\label{results}
We present our experimental results to help answer the last four RQs described in Section \ref{questions}, investigating the importance of DL replicability and reproducibility. The RQs aim to analyze how the internal factors (i.e., model stability, convergence, OOV issue, and testing data size) in DLs affect DL-based SE study replicability and reproducibility, and whether this influence is strong enough to threaten the validity of the studies.

\subsection{RQ4. How does Model Stability Affect  Replicability?}\label{result_stability}

\vspace{5pt}\noindent\textbf{Motivation.}
DL replicability can be largely supported by sharing a replication package. Ideally, re-running the source code and data provided can replicate the reported model performance. However, a publicly shared replication package cannot guarantee that the models trained by different researchers are the same and produce the same experimental result as the reported one, due to the randomness in model initialization and optimization \cite{li2019random}. Therefore, the DL model performance could be varied for different researchers, so that a model with unstable performance would be of low replicability. This is the reason why the DL replicability requires that the reported performance by a DL study can be approximately reproduced (although not identically) with high probability. Thus, it is important to investigate whether DL models are stable enough and whether the stability level strongly affects the DL replicability.

\vspace{5pt}\noindent\textbf{Method.}
To estimate the four DL models' stability, we re-run each DL model ten times and analyze the statistics of our experimental results (e.g., mean, standard deviation, etc.). To assess the model replicability, we utilize the mean value of ten experimental results as the performance that the DL model likely produces with high probability. Hence, if the mean value is far from the result reported by the authors of the original studies, then the reported study results are difficult to reproduce i.e., it has low replicability. 

\begin{figure}
    \centering
    \includegraphics[width=\linewidth]{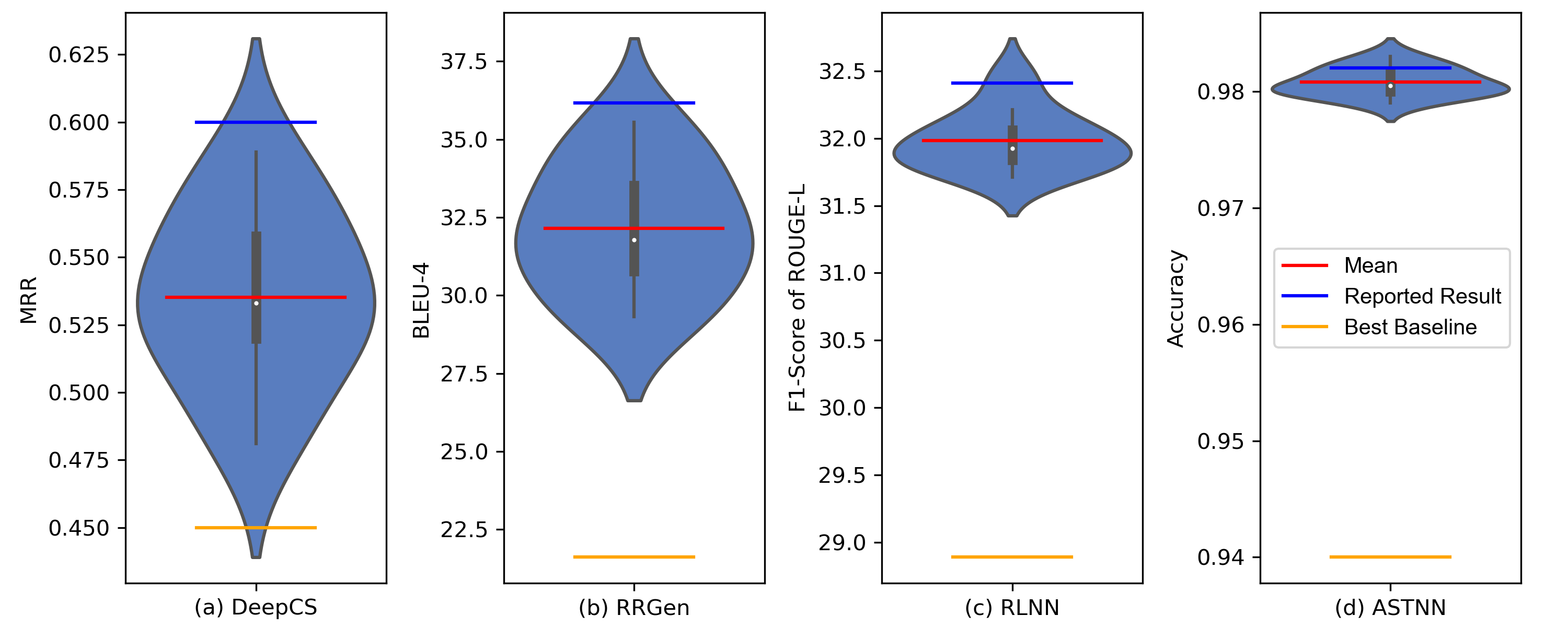}
    \caption{Violin plots of four DL models with ten repeated experimental results, where white dot and red line indicate median and mean values respectively; blue and orange lines denote the performance of four models and the best baseline reported in the original studies, respectively.}
    \label{fig_random}
\end{figure}

\vspace{5pt}\noindent\textbf{Results.}
Fig. \ref{fig_random} illustrates the ten experimental results of four DL models (i.e., DeepCS, RRGen, RLNN, and ASTNn), where white dot and red line indicate the median and mean values, respectively; blue and orange lines show the performance of the DL model and the best baseline reported in the original study, respectively.  The important statistics of Fig. \ref{fig_random} are listed in Table \ref{tab_random}. From the table, we notice that the experimental performance of DeepCS obtains a mean 0.535 and standard deviation (std) 0.033 with the minimum 0.481 (min) and the maximum (max) 0.589. The mean performance of RRGen, RLNN, ASTNN are 32.144 (std = 0.032, min = 0.481, max = 0.589), 31.982 (std = 0.213, Min = 31.710, max = 32.457), 0.981 (std = 0.001, min = 0.979, max = 0.983) respectively. 

To assess the relative stability between these four DL models, we calculated the coefficient of variation ($c_{v}$), equal to the standard deviation normalized by its mean. Table \ref{tab_random} shows that the $c_{v}$ values of four models are 5.92\%, 6.27\%, 0.67\%, and 0.1\% respectively. Thus, \textbf{the DeepCS and RRGen DL models possess much lower stability} than the other two. We also observe that all of the reported results are larger or equal to the mean plus standard deviation of our multiple re-runs. This observation implies that \textbf{the reported performance of all four DL models can be achieved by other researchers with low probability}, i.e., low replicability. Therefore, a good reported DL performance without multiple runs could be achieved by chance.

\begin{table}
    \centering
    \small
    \caption{Statistics of four DL models with ten repeated experimental results, where 'reported' and 'baseline' are the reported result of a model and the compared best baseline in author's study, respectively; Std is the standard deviation; $c_{v}$ is the coefficient of variance equaling to std/mean.}
    \setlength{\tabcolsep}{12.5pt}{
    \begin{tabular}{|c|rrrr|c|c|c|}
        \toprule
        \textbf{Model} & \textbf{Min} & \textbf{Max} & \textbf{Mean} & \textbf{Std} & \textbf{Reported (vs. Mean)} & \textbf{Baseline} & \bm{$c_{v}$} \\
        \midrule
        DeepCS & 0.481 & 0.589 & 0.535 & 0.032 & 0.600 (+12.2\%) & 0.450 & 5.9\%\\
        RRGen & 29.314 & 35.549 & 32.144 & 2.016 & 36.170 (+12.5\%) & 21.610 & 6.3\%\\
        RLNN & 31.710 & 32.457 & 31.982 & 0.213 & 32.410  (+1.3\%) & 28.890 & 0.7\%\\
        ASTNN & 0.979 & 0.983 & 0.981 & 0.001 & 0.982 (+0.1\%) & 0.940 & 0.1\%\\
        \bottomrule
    \end{tabular}}
    \label{tab_random}
\end{table}

Compared with the generated mean performance of the four DL models, the results reported in the original studies increased by 12.2\%, 12.5\%, 1.3\%, and 0.1\% respectively. This case suggests that \textbf{the performance of four models in the original studies are overestimated} to different degrees, where first two cases are more substantial ($>$12\%) than the others ($<$1.4\%). Thus, \textbf{a DL model with low replicability may strongly affect the overall validity of the study}. To analyze the correlation between the stability and replicability, we performed a Pearson correlation test \cite{benesty2009pearson} at a 5\% significance level on the $c_{v}$ values and the overestimation rates (12.2\%, 12.5\%, 1.3\%, and 0.1\%) between four models. This statistical test shows that DL stability and replicability are strongly correlated (cor = 1, p-value = 1). Our test result implies that \textbf{an unstable DL model leads to low study replicability}. The premise is that the reported performance is far from the mean, especially larger than the mean plus standard deviation. 

To further analyze the impacts of this overestimation, we compare the obtained mean values of multiple experiments with the performance of the best baselines reported in the DL-based SE studies. We find that the mean results of four models outperform the best baselines reported by four DL studies by 18.9\%, 48.8\%, 10.7\%, and 4.4\%, respectively. However, due to overestimation, the advantages of the four DL models over the best baselines are reduced by 43.3\%, 27.6\%, 12.2\%, and 2.5\% correspondingly. Therefore,\textbf{only reporting a good DL performance may threaten the experimental conclusions}.

\vspace{5pt}\noindent\textbf{Implications.}
DL models possess randomness, so that only reporting one good experimental result is not enough. If the reported performance is far from the mean values of multiple runs (especially larger than the mean plus standard deviation), \textbf{the reported result could be replicated with low probability}. In this situation, low DL replicability will produce greater negative effects on an unstable model. Therefore, it is important to estimate and improve the DL stability and replicability, instead of only reporting one good experimental result. Otherwise, the model validity could be largely threatened.

\subsection{RQ5. How does Model Convergence Affect Reproducibility?}

\vspace{5pt}\noindent\textbf{Motivation.}
DL models in SE are optimized with many manually set parameters (e.g., learning rate, the total number of training iterations, convergence coefficient, etc.), hypothesizing that the optimization is convergent or training model with more iterations would gain better performance. However, if this unverified assumption does not hold, the DL-based study reproducibility will not be high because the manually set parameters cannot adapt to data in different new experiments. Thus, the goal is to estimate the DL model convergence and investigate whether the convergence level is an influential factor for DL study reproducibility.

\vspace{5pt}\noindent\textbf{Method.}
To analyze the model convergence level, we investigate whether training DL models with more iterations can achieve better and more stable performance. We again use the four DL models, which have already trained ten times in Section \ref{result_stability}, as our study subject. Thus, we have forty pre-trained models. For each DL model, we continue to optimize them with one-third more iterations/epochs and compare the change rate (CR) of performance before and after the new training iterations. A divergent model would produce highly turbulent performance, i.e., a large standard deviation of change rates ($CR_{std}$). Note that one-third more epochs or iterations are determined because this extended training is enough to analyze the turbulence level, and more iterations/epochs will substantially increase the total training time for forty DL models but our computation resource is limited. Besides, training DL models with too many iterations/epochs may not be always better \cite{nakkiran2019deep}. The overfitting issue is one major reason \cite{poggio2017theory}. 

\begin{figure}[ht]
    \centering
    \includegraphics[width=0.9\linewidth]{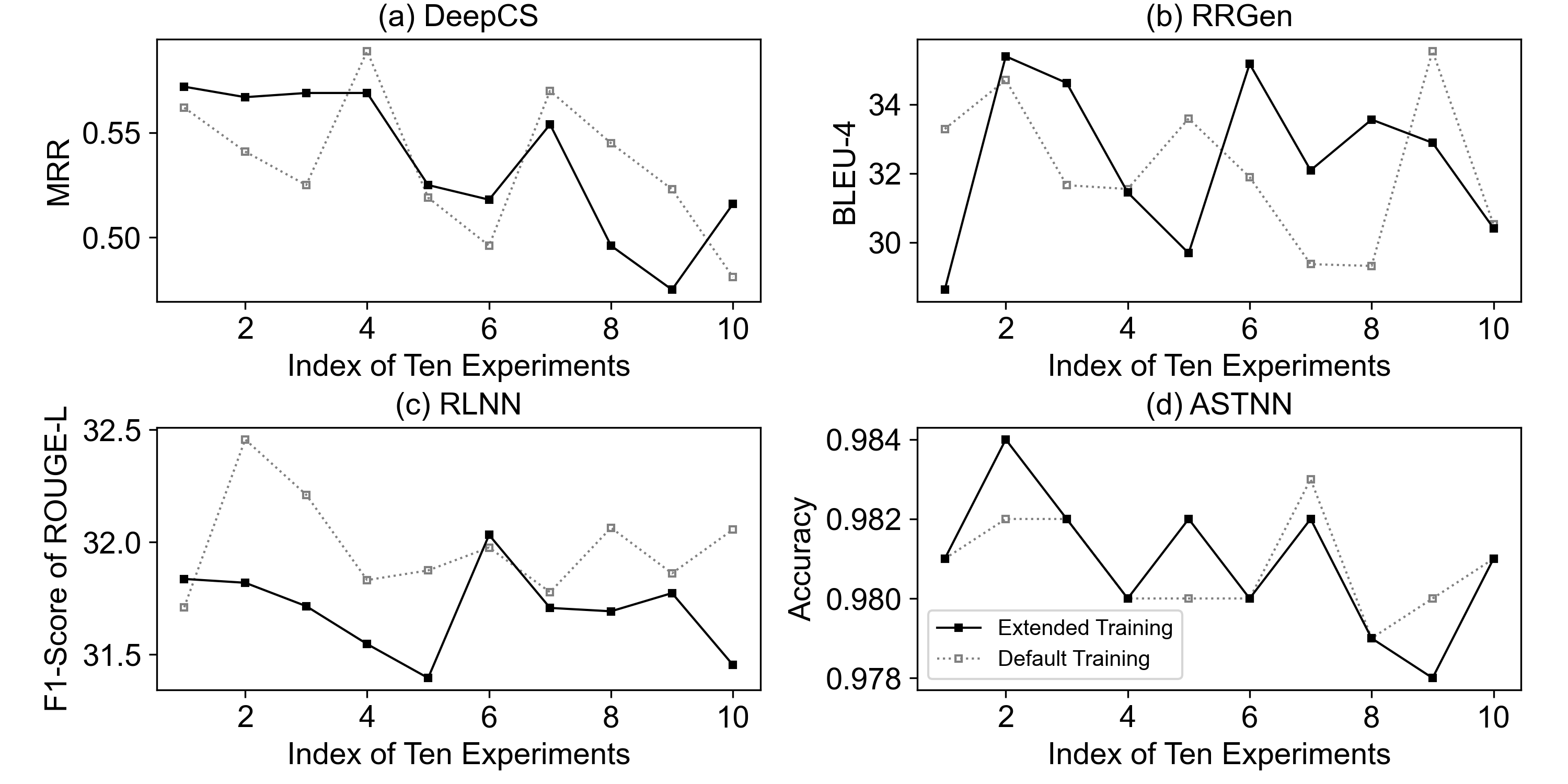}
    \caption{Plots of four DL models with default training (grey squares) vs. extended training (black squares).}
    \label{fig_convergence}
\end{figure}

\vspace{5pt}\noindent\textbf{Results.}
Fig. \ref{fig_convergence} shows our experimental results from the four DL models with default training iterations (black squares) vs. extended iterations (grey squares), whose statistics are listed in Table \ref{tab_random} and \ref{tab_convergence_cmp} respectively. From the Table \ref{tab_convergence_cmp}, we can notice that, after the extended training, the mean performance of DeepCS, RRGen, RLNN, and ASTNN are 0.536 (min = 0.475, max = 0.572, and std = 0.033), 32.391 (min = 28.625, max = 35.399, std = 2.232), 31.697 (min = 31.395, max = 32.032, std = 0.181), and 0.981 (min = 0.978, max = 0.984, std = 0.002) respectively. Compared with our generated mean values, the model performance reported in the original study 
increased by 11.9\%, 11.7\%, 2.2\%, and 0.1\% respectively. These results indicate that \textbf{the reported results for these four DL models cannot be reproduced to different degrees} for experiments with extended model training. The reproducibility issue of the first two models, DeepCS and RRGen, show much higher negative effects over the other two.

For each DL model, we performed a Wilcoxon signed-rank test \cite{wilcoxon1945individual} at a 5\% significance level between the experimental results before and after extended training. The tests show that all the statistical differences are not significant (p-value > 0.05), as shown in Table \ref{tab_convergence_cmp}. This statistical analysis result implies that training all four DL models does not produce better performance. Also, we can notice that, comparing with the values in Table \ref{tab_random}, Table \ref{tab_convergence_cmp} shows little difference, where the reported performance of four models cannot be better reproduced with the extended training (the reported performance outperforms the mean value by 11.9\%, 11.7\%, 2.2\%, 0.2\%); the relative model stability is not changed substantially ($c_{v}$ values of four models are 6.2\%, 6.9\%, 0.6\%, 0.2\% respectively). Thus, \textbf{the extended training did not improve the model performance} due to the unchanged degree of model stability. The unchanged stability indicates that model training tends to be not convergent for a model with lower stability.

\begin{table}[ht]
    \centering
    \small
    \caption{Statistics of four DL models trained with extended iterations/epochs for ten repeated experimental results, where 'reported' is the reported result of a model in the author's study, respectively; Std is the standard deviation; $c_{v}$ is the coefficient of variance equaling to std/mean; p-value is the result of the Wilcoxon signed-rank test, at a 5\% significance level, on the model performance before and after the extended training.}
    \setlength{\tabcolsep}{12.5pt}{
    \begin{tabular}{|c|rrrr|c|c|c|}
        \toprule
        \textbf{Model} & \textbf{Min} & \textbf{Max} & \textbf{Mean} & \textbf{Std} & \textbf{Reported (vs. Mean)} & \bm{$c_{v}$} & \textbf{p-value} \\
        \midrule
        DeepCS & 0.475 & 0.572 & 0.536 & 0.033 & 0.600 (+11.9\%) & 6.2\% & >0.05\\
        RRGen & 28.625 & 35.399 & 32.391 & 2.232 & 36.170 (+11.7\%) & 6.9\% & >0.05\\
        RLNN & 31.395 & 32.032 & 31.697 & 0.181 & 32.410 (+2.2\%) & 0.6\% & >0.05\\
        ASTNN & 0.978 & 0.984 & 0.981 & 0.002 & 0.982 (+0.1\%) & 0.2\% & >0.05\\
        \bottomrule
    \end{tabular}}
    \label{tab_convergence_cmp}
\end{table}

\begin{figure}
    \centering
    \includegraphics[width=\linewidth]{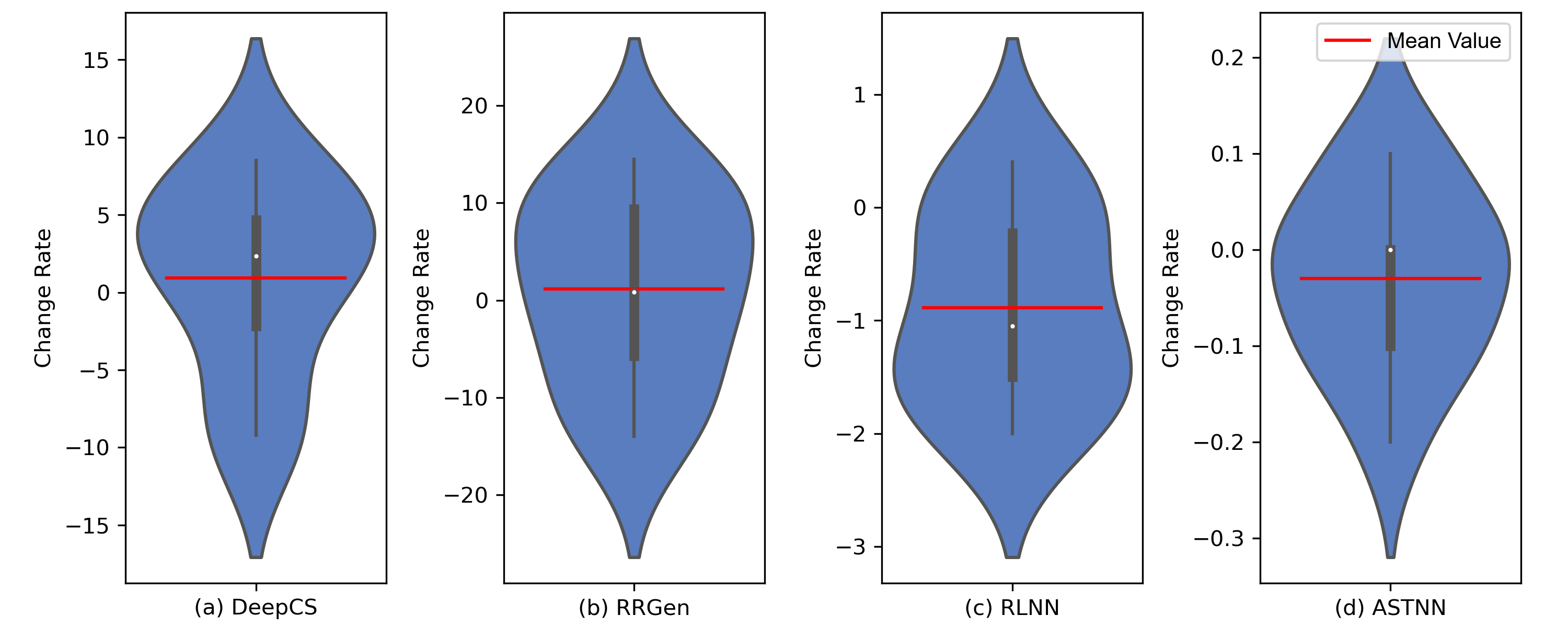}
    \caption{Violin plots of the performance change rates after training models with extended iterations/epochs, where the red horzontal line indicates the average change rate.}
    \label{fig_change}
\end{figure}

To estimate the convergence levels of these four DL models and analyze their effects, we calculated the change rates (CR) of model performance before and after training with more iterations/epochs. Fig. \ref{fig_change} illustrates violin plots of the change rates for each DL model, where the red horizontal line indicates the mean value of the change rate. The statistics of change rates are listed in Table \ref{tab_convergence}, including the minimum (min) and maximum (max) values, mean, and standard deviation (std). We can observe that the convergence levels of four models are 6\%, 9.3\%, 0.8\%, and 0.1\% respectively in terms of the $CR_{std}$ values, where the convergence level of the first two models is lower than the others. By comparing the min and max values, we can notice that a model with a lower convergence level may improve its performance by 14.5\% or even reduce the performance by 9.1\% after training the model with more iterations/epochs. In this case, \textbf{the variation caused by poor convergence levels may lead to different experimental conclusions}. Therefore, DL model convergence is an important factor for the DL study reproducibility issue.

\vspace{5pt}\noindent\textbf{Implications.}
We cannot assume that a DL model is convergent or the model performance can be improved with further training unless we provide any verification. This is because a model with a low convergence level may produce highly turbulent and unstable performance under different experimental data. In other words, DL model convergence is of high importance for DL study reproducibility. To enhance  confidence in any SE study with reported good effectiveness, it is necessary to justify the parameters in the used DL model training and verify the convergence level of the used DL model.

\begin{table}
    \centering
    \small
    \caption{Statistics of change rates (CR) between ten experimental results on four DL models, where 'CR' indicates the change rate before and after more iterations or epochs of training; min, max, mean, and std indicate the minimim, maximum, average, standard deviation of the change rates respectively.}
    \setlength{\tabcolsep}{8pt}{
    \begin{tabular}{|c|cccc|}
        \toprule
        \textbf{Model} & \bm{$CR_{min}$} & \bm{$CR_{max}$} & \bm{$CR_{mean}$} & \bm{$CR_{std}$}\\
        \midrule
        DeepCS & -9.1\% & +8.5\% & +0.3\% & +6.0\%\\
        RRGen & -4.0\% & +14.5\% & +1.2\% & +9.3\%\\
        RLNN & -2.0\% & +0.4\% & -0.9\% & +0.8\%\\
        ASTNN & -0.2\% & +0.1\% & -0.0\% & +0.1\%\\
        \bottomrule
    \end{tabular}}
    \label{tab_convergence}
\end{table}

\subsection{RQ6. How does the Out-of-Vocabulary Issue Affect Reproducibility?}\label{sub_vocab}

\vspace{5pt}\noindent\textbf{Motivation.}
As described in Section \ref{sub_rqs}, most of the DL-based studies in SE (76.3\% of reviewed literature) address code/text by using a limited vocabulary size. Usually, the vocabulary only contains the top frequently occurred words. However, when addressing code/text representation, DL models commonly meet the out-of-vocabulary (OOV) issue where the vocabulary built from training data does not cover the words in testing data. This is because their testing data is collected from real-world environments e.g., GitHub and Stack Overflow, and is thus growing and changing frequently and new words will increase substantially as time goes by. Under this situation, a DL model can not capture the semantics of new words and may misunderstand the semantics of code/text. Therefore, DL models may substantially suffer from the OOV issue with different experimental settings, leading to low reproducibility.

\vspace{5pt}\noindent\textbf{Method.}
To investigate the impact of this OOV issue, we trained four DL models (DeepCS, RRGen, RLNN, and ASTNN) with different sizes of vocabulary and performed predictions on their corresponding testing data. If the impact is low, it is expected that the model performance shows little variation as the vocabulary size increases. Otherwise, one good reported model performance may not be reproduced in new experiments. This is because the DL model is likely misunderstand the semantics of code/text with new words, resulting in substantially poorer performance.

Specifically, the vocabulary size is determined by ten scales of the default vocabulary used in the original studies. The division scale ranges from 10\% to 100\% with a step 10\%. The full scale (100\%) indicates that we used the vocabulary with the default setting in the original studies. Table \ref{table:configurations} shows that four DL models contain 10k, 10k, 50k, and 8k most frequently occurred words in their training data, respectively. However, we note that the original DL studies do not use all words in the training data to build their vocabularies, hence even the 100\% scale does not mean that the OOV issue does not occur. Besides, when we trained a model with the vocabulary at a lower scale (e.g., 90\%), we only excluded the (10\%) words that have the least frequency in the vocabulary. If a vocabulary does not contain word frequency information, we collected the information from the training data by ourselves. In this way, we can avoid overestimating the effect of the OOV issue by suddenly dropping some frequently appearing words. After we obtained the vocabularies with lower scales, we addressed the encoded vectors in training and testing data one by one. For each vector, we removed the word that does not appear in a new vocabulary. Moreover, to mitigate the influence of model stability, we ran each model ten times for different vocabulary settings and reported the mean values of model performance following the setting described in Section \ref{result_stability}. 

\vspace{5pt}\noindent\textbf{Results.}
Table \ref{tab_oov} lists the average performance of four DL models trained with different vocabulary scales in ten repeated runs. The 'improve' rows show the performance improvement from a certain scale (e.g., 90\%) to the initial scale 10\%. Table \ref{tab_oov} shows that the performance of DeepCS keeps improving from 0.262 to 0.5 as the vocabulary scale increases, where the relative improvement over the 10\% scale ranges from 3.4\% to 104.2\%. For the RRGen model, when increasing the vocabulary scale from 10\% to 20\%, the model performance reduces by 1.2\%. However, the RRGen gains improvements (6.8\% to 17.6\%) when the vocabulary scale keeps increasing from 30\% to 90\%, where the performance ranges from 32.144 to 35.394. The maximum performance occurs at the 70\% scale of vocabulary size. The performance of the RLNN model ranges from 31.685 to 32.877 with no substantial improvement or deterioration. The performance of ASTNN (around 0.98) did not change with the vocabulary scale ranging from 20\% to 100\%.

To analyze the trend -- increasing, decreasing, or neither -- of the DL model performance when increasing the vocabulary scales, we performed a Cox Stuart trend test \cite{cox1955some} at a 5\% significance level. Table \ref{tab_oov_trend} illustrates the trendlines and statistical results, where '$\uparrow$' indicates an increasing trend while '-' denotes no increasing/decreasing trend. $c_{v}$ is the coefficient of variation (i.e., the standard deviation normalized by the mean) to measure the relative variation between model performance. 'scale(s)' means that for each listed vocabulary scale the model performance shows the significant difference compared with the performance at the previous scale,  where the difference is tested by the Wilcoxon signed-rank test \cite{wilcoxon1945individual} at a 5\% significance level.

The statistical results in Table \ref{tab_oov_trend} show that the performance of DeepCS is increasing (p-value < 0.05) for larger-scale vocabulary scales. \textbf{This means that DeepCS would suffer from the OOV issue with a limited size of vocabulary}, unless we enlarge the size as much as possible. The $c_{v}$ value indicates that when increasing the vocabulary scale from 10\% to 100\%, the mean performance can be improved by more than 23.5\%, where the significant improvements occur at the scales of 30\%, 40\%, and 90\%. Therefore, \textbf{the OOV issue could strongly affect model performance, resulting in low study reproducibility}. 

The performance of RRGen tends to be neither increasing nor decreasing (p-value > 0.05). Comparing with DeepCS, the performance variation ($c_{v}$ = 5.9\%) of RRGen is 74.9\% smaller. A significant change only happened at the 30\% scale. By excluding the scales 10\% and 20\%, the $c_{v}$ drops to only 2.9\%. Thus, \textbf{the new out-of-vocabulary words do not affect the performance of RRGen anywhere near as severely as DeepCS}. The performance of RLNN also shows no upward/downward trend (p-value > 0.05) as RRGen, with the variation in the degree of performance is smaller ($c_{v}$ = 1.2\%). There is some significant performance turbulence at the vocabulary scales 30\%, 40\%, 50\%, and 70\%. Therefore, \textbf{the OOV issue generates a less negative effect on the RLNN, compared with DeepCS and RRGen}. The performance of ASTNN is proportional to the vocabulary scale (p-value < 0.05) with the smallest variation ($c_{v}$ = 0.2\%). A significant improvement only takes place at the 20\% scale. For the vocabulary scales larger than 10\%, the $c_{v}$ is reduced to no more than 0.01\%. Thus, the performance of ASTNN converges to 0.981 for higher vocabulary scales. \textbf{This means that the ASTNN is not affected by the OOV issue and the reported performance of ASTNN can be reproduced for datasets with different scales of new words}. We observed that the reproducibility of ASTNN is not influenced by the OOV issue because it represents code as an abstract syntax tree (AST). In this way, the code classification task is not only affected by the code semantics but also strongly affected by the code structure. Moreover, RLNN can largely mitigate the effect of the OOV issue since it leverages the pointer generator technique \cite{see2017get} to replace the OOV words with the ones in training data in an appropriate way.

\begin{table}
    \centering
    \small
    \caption{Performance of four DL models trained with different sizes of vocabulary, where 'average' indicate the mean performance of a model with one vocabulary scale running in ten times; 'improve' indicates the percent of improvement comparing to the average performance with the 10\% scale.}
    \setlength{\tabcolsep}{5.8pt}{
    \begin{tabular}{|c|c|cccccccccc|}
        \toprule
        \textbf{Model} & \textbf{Type} & \textbf{10\%} & \textbf{20\%} & \textbf{30\%} & \textbf{40\%} & \textbf{50\%} & \textbf{60\%} & \textbf{70\%} & \textbf{80\%} & \textbf{90\%} & \textbf{100\%}\\
        \midrule
        \multirow{2}{*}{DeepCS} & Average & 0.262 &0.271 &0.324 &0.421 &0.446 &0.445 &0.470 &0.463 &0.520 &0.540\\
        & Improve & -& +3.4\% & +23.7\% & +60.7\% & +70.2\% & +69.8\% & +79.4\% & +76.7\% & +98.5\% & +104.2\%\\
        \midrule
        \multirow{2}{*}{RRGen} & Average & 30.085 &29.714 &33.232 &34.753 &34.039 &35.394 &34.846 &33.238 &34.146 &32.144\\
        & Improve & - & -1.2\% & +15.5\% & +15.5\% & +13.1\% & +17.6\% & +15.8\% & +10.5\% & +13.5\% & +6.8\%\\
        \midrule
        \multirow{2}{*}{RLNN} & Average & 32.766 & 32.031 & 31.685 & 32.064 & 32.393 & 32.307 & 32.877 & 32.465 & 32.668 & 31.982\\
        & Improve & - & -2.2\% & -3.3\% & -2.1\% & -1.1\% & -1.4\% & +0.3\% & -0.9\% & -0.3\% & -2.4\%\\
        \midrule
        \multirow{2}{*}{ASTNN} & Average & 0.975 & 0.980 & 0.980 & 0.980 & 0.980 & 0.980 & 0.981 & 0.981 & 0.981 & 0.981\\
        & Improve & - & +0.5\% & +0.5\% & +0.5\% & +0.5\% & +0.5\% & +0.5\% & +0.5\% & +0.5\% & +0.5\%\\
        \bottomrule
    \end{tabular}}
    \label{tab_oov}
\end{table}

\begin{table}
    \centering
    \small
    \caption{Performance Trends of the four DL models trained with difference scales of vocabulary. The trend is determined by the Cox Stuart trend test at a 5\% significance level, where '$\uparrow$' indicates an increasing trend while '$-$' indicates no trend. For the trendlines, X axes indicate the vocabulary scale (ranging from 10\% to 100\% with a step 10\%) while Y axes indicate the model performance. $c_{v}$ is the coefficient of variation on model performance at different vocabulary scales, equaling to mean divided by standard deviation. 'Scale(s)' indicates that at each listed vocabulary scale the model performance is significantly different from the one with the previous scale, where the difference is tested by the Wilcoxon signed-rank test at a 5\% significance level.}
    \setlength{\tabcolsep}{16pt}{
    \begin{tabular}{|c|c|c|c|c|}
        \toprule
        \textbf{Model} & \textbf{Trend (p-value)} & \textbf{Trendline} & \bm{$c_{v}$} & \textbf{Scale(s)}\\
        \midrule
        DeepCS & $\uparrow$ (0.03*) & \includegraphics[width=0.25\linewidth]{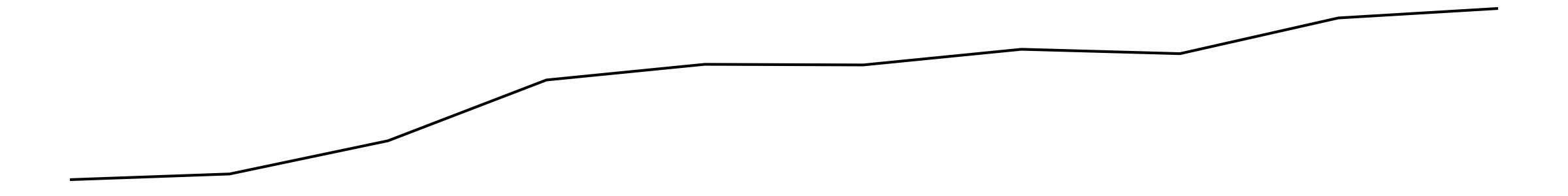} & 23.5\% & 30\%,40\%,90\%\\ 
        \midrule
        RRGen & $-$ (0.50)& \includegraphics[width=0.25\linewidth]{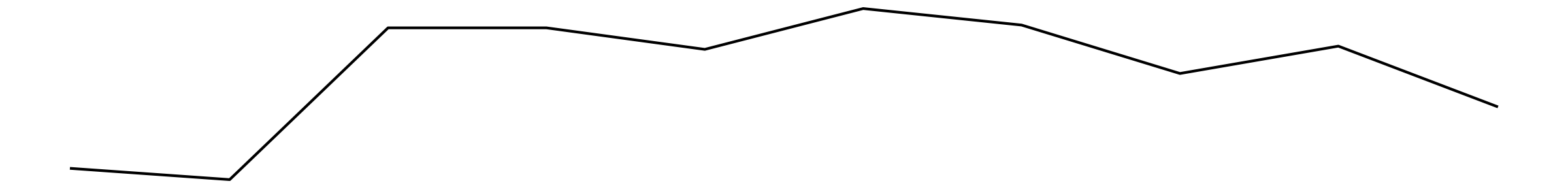} & 5.9\%& 30\%\\ 
        \midrule
        RLNN & $-$ (0.50) & \includegraphics[width=0.25\linewidth]{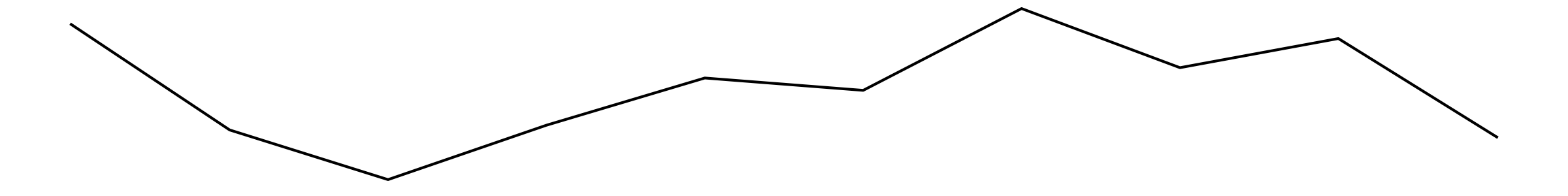} & 1.2\%& 30\%,40\%,50\%,70\%\\ 
        \midrule
        ASTNN & $\uparrow$ (0.03*) & \includegraphics[width=0.25\linewidth]{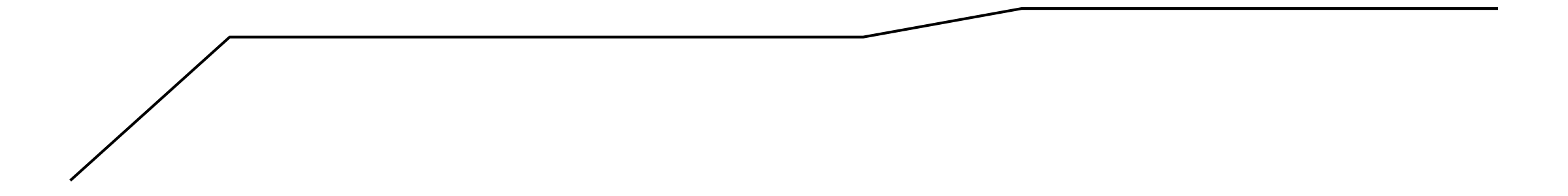} & 0.2\%& 20\%\\ 
        \bottomrule
    \end{tabular}}
    \label{tab_oov_trend}
\end{table}

\vspace{5pt}\noindent\textbf{Implications.}
The OOV issue is prevalent when modeling code or text in SE tasks. This issue can strongly influence the DL model performance for new settings involving occurrences of new words that do not appear in the training data. Thus, the sensitivity of DL model performance to vocabulary size can lead to low study reproducibility. Therefore, DL-based studies should address the OOV issue and mitigate its negative effect as much as possible.

\subsection{RQ7. How does Testing Data Size Affect Reproducibility?}

\vspace{5pt}\noindent\textbf{Motivation.}
It is common to verify the validity of a DL model by collecting a subset of data from the real-world environment, e.g., GitHub and Stack Overflow. However, researchers usually start an experiment with a limited size of testing data and assume that the model performance can be reproduced for a larger scale of testing data. However this assumption is rarely investigated. Therefore, our goal is to analyze how our four SE study DL models perform with different sizes of testing data, and further investigate how the testing data size affects DL-based study reproducibility.

\vspace{5pt}\noindent\textbf{Method.}
To analyze the effect of testing data size, we trained each DL model as default (namely, the training data does not change) but performed predictions on different sizes of testing data. Specifically, for the models for DeepCS, RRGen, RLNN, and ASTNN, they contain around 16m, 14k, 4k, and 10k testing data in default as shown in Table \ref{table:configurations}. We sampled a subset of testing data at ten different scales, ranging from 10\% to 100\% with a step 10\%. As the sampling process for each scale is random, we repeated the sampling and testing ten times to mitigate the effect of randomness. If the model performance is sensitive to the testing data scale,  DL-based SE study reproducibility will be compromised. Therefore, it is expected that the performance of a DL model will not decrease substantially when the scale of testing data increases.

\vspace{5pt}\noindent\textbf{Results.}
Table \ref{tab_testing} lists the average performance of DL models tested on different scales of testing data in ten repeated runs. The 'improve' rows show the performance improvement from a certain scale (e.g., 80\% or 90\%) to the initial scale 10\%. Table \ref{tab_testing} shows that when the testing data size is increased, the DeepCS performs the worst (0.531) at the 30\% scale and the best (0.581) at the 80\% scale. The performance of RRGen did not change largely from the scale 10\% to 80\%, ranging from 29.128 to 29.404. However its performance is largely improved at scales 90\% and 100\% (improved by 6.0\% and 9.7\%, compared with its performance at the 10\% scale.) For the models, RLNN and ASTNN, Table \ref{tab_testing} shows that their performance is hardly affected by the scale of their testing data, where the performance of RLNN is varied by no more than 1.3\% while ASTNN shows negligible performance difference (<0.2\%) for different scales of testing data.

Similar to Section \ref{sub_vocab}, we performed a Cox Stuart trend test \cite{cox1955some} at a 5\% significance level on the DL model performance at different scales of testing data to analyze their performance trend -- increasing, decreasing, or neither. We also calculated the $c_{v}$ values to measure the variation of performance. We performed a Wilcoxon signed-rank test at a 5\% significance level and recorded the scales at which the model performance is significantly different from the one in the previous scale.

The trendlines and statistical results in Table \ref{tab_test_trend} indicate that \textbf{all four models (DeepCS, RRGen, RLNN, and ASTNN) show neither increasing nor decreasing trend with statistical significance} (p-values > 0.05). However their performance variation does differ. Specifically, the $c_{v}$ of DeepCS is 4\% with significant changes at the scales 50\% and 90\%. Compared with the other three DL models, DeepCS obtains a higher variation. We observe that this variation is mainly caused by the quality of the testing data used. This is because for a search query the number of ground-truth may be substantially changed due to random sampling.
RRGen shows smaller performance variation with $c_{v}$ = 3.5\% with no significant change at any scale. Similarly to DeepCS,  RRGen achieves varied performance at different testing data scales because the difficulty of review response generation could be largely varied for different testing data. Thus \textbf{the reproducibility of studies using a DeepCS or RRGen DL model may be low with different sizes of testing data used}. In contrast, the performance turbulence of RLNN is reduced by 80\% in terms of $c_{v}$ (0.7\%) even if there is a significant improvement at the 100\% scale. We observed that although RLNN is used for a generation task similar to RRGen, \textbf{its performance is not as sensitive to the testing data size as RRGen}, as RRGen leverages a reinforcement learning technique to optimize the generation result. Similarly, the $c_{v}$ of ASTNN is 0\%, even if there is a significant improvement at 20\%. It means that the model performance converges. The model performance ASTNN is not sensitive to the testing data size because its performance is very high with accuracy 98.1\%. Hence, no matter how we sampled a subset of testing data, the prediction accuracy would not differ. Therefore, if the characteristics of the testing data do not change, \textbf{the reproducibility of studies using ASTNN should be high}.

\begin{table}
    \centering
    \small
    \caption{Performance of four DL models tested on different sizes of testing data, where 'average' indicate the mean performance of a model with one testing data scale running in ten times; 'improve' indicates the percent of improvement comparing to the average performance with the 10\% scale.}
    \setlength{\tabcolsep}{6.6pt}{
    \begin{tabular}{|c|c|cccccccccc|}
        \toprule
        \textbf{Model} & \textbf{Type} & \textbf{10\%} & \textbf{20\%} & \textbf{30\%} & \textbf{40\%} & \textbf{50\%} & \textbf{60\%} & \textbf{70\%} & \textbf{80\%} & \textbf{90\%} & \textbf{100\%}\\
        \midrule
        \multirow{2}{*}{DeepCS} & Average & 0.540 & 0.523 & 0.531 & 0.520 & 0.564 & 0.550 & 0.569 & 0.581 & 0.566 & 0.535\\
        & Improve & -&-3.1\% & -1.7\% & -3.7\% & +4.4\% & +1.9\% & +5.4\% & +7.6\% & +0.0\% &-0.9\% \\
        \midrule
        \multirow{2}{*}{RRGen} & Average & 29.290 & 29.218 & 29.396 & 29.373 & 29.316 & 29.128 & 29.326 & 29.404 & 31.305 & 32.144\\
        & Improve & - & -0.2\% & +0.4\% & +0.3\% & +0.1\% & -0.6\% & +0.1\% & +0.4\% & +6.9\% & +9.7\%\\
        \midrule
        \multirow{2}{*}{RLNN} & Average & 31.556 & 31.528 & 31.086 & 31.611 & 31.554 & 31.385 & 31.447 & 31.532 & 31.554 & 31.982\\
        & Improve & - & -0.1\% & -1.5\% & +0.2\% & +0.0\% & -0.5\% & -0.3\% & -0.1\% & +0.0\% & +1.3\%\\
        \midrule
        \multirow{2}{*}{ASTNN} & Average & 0.980 & 0.981 & 0.981 & 0.981 & 0.981 & 0.981 & 0.981 & 0.981 & 0.981 & 0.981\\
        & Improve & - & +0.1\% & +0.1\% & +0.1\% & +0.1\% & +0.1\% & +0.1\% & +0.1\% & +0.1\% & +0.1\%\\
        \bottomrule
    \end{tabular}}
    \label{tab_testing}
\end{table}

\begin{table}
    \centering
    \small
    \caption{Performance Trends of the four DL models tested on difference scales of testing data. The trend is determined by the Cox Stuart trend test at a 5\% significance level, where '$\uparrow$' indicates an increasing trend while '$-$' indicates no trend. For the trendlines, X axes indicate the testing data scale (ranging from 10\% to 100\% with a step 10\%) while Y axes indicate the model performance. $c_{v}$ is the coefficient of variation on model performance at different scales of testing data, equaling to mean divided by standard deviation. 'Scale(s)' indicates that at each listed scale of testing data the model performance is significantly different from the one with the previous scale, where the difference is tested by the Wilcoxon signed-rank test at a 5\% significance level.}
    \setlength{\tabcolsep}{19.5pt}{
    \begin{tabular}{|c|c|c|c|c|}
        \toprule
        \textbf{Model} & \textbf{Trend (p-value)} & \textbf{Trendline} & \bm{$c_{v}$} & \textbf{Scale(s)} \\
        \midrule
        DeepCS & $-$ (0.19) & \includegraphics[width=0.25\linewidth]{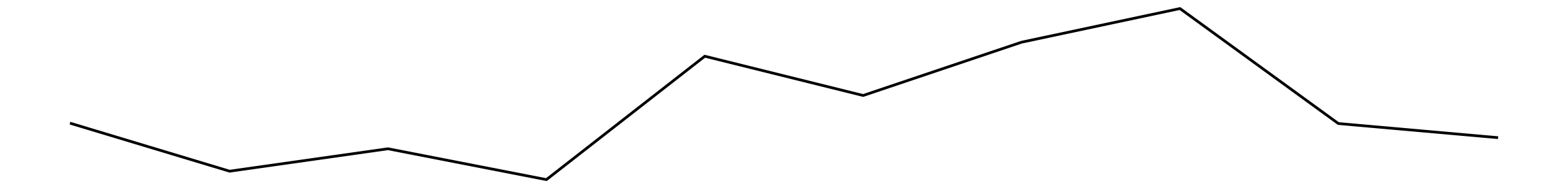} & 4.0\% & 50\%,90\%\\ 
        \midrule
        RRGen & $-$ (0.50)& \includegraphics[width=0.25\linewidth]{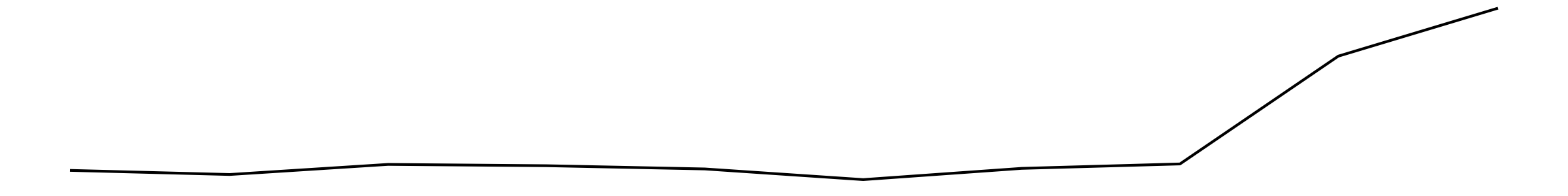} & 3.5\% & -\\ 
        \midrule
        RLNN & $-$ (0.50) & \includegraphics[width=0.25\linewidth]{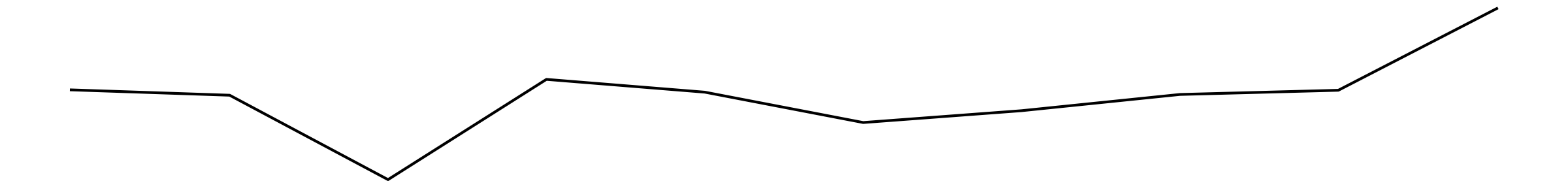} & 0.7\% & 100\%\\ 
        \midrule
        ASTNN & $-$ (0.50) & \includegraphics[width=0.25\linewidth]{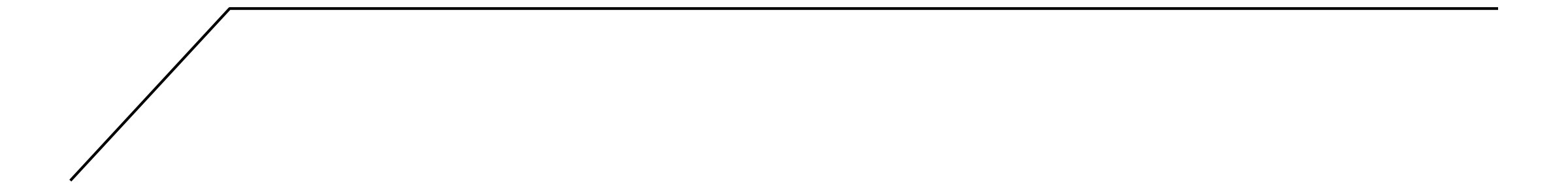} & 0.0\% & 20\%\\ 
        \bottomrule
    \end{tabular}}
    \label{tab_test_trend}
\end{table}

\vspace{5pt}\noindent\textbf{Implications.}
Testing DL-based models with different sizes of testing data may achieve substantially different model performance. In this case, a reported performance can be difficult to reproduce in new experiments when using newly sampled testing data. Therefore, mitigating the negative effect of the testing data size should be fully considered in DL-based SE studies.

\section{Threats to Validity}\label{threats}

This validity of this study could be affected by the following threats.

\vspace{5pt}\noindent\textbf{Manual Efforts in Literature Review.}
Like any manual evaluation, our literature review is subject to personal bias and subjectivity. As our research topic is on DL-based studies in SE, we performed our literature search only on twenty selected SE journals or conferences. Any DL-based SE studies outside of this search scope would be excluded. We believe this threat should be minor because these selections are the most commonly published SE venues. To analyze how often DL studies provide replication packages, we checked all the links in the paper one by one. We may have missed replication packages not mentioned in the paper but existing on the Internet, e.g., GitHub. To analyze the characteristics of the reviewed DL studies, we manually classified them into ten different categories, e.g., study subjects, SE tasks, basic DL techniques, and types of RQs discussed in each paper. These manual efforts may involve some interpretation or mis-classification errors that threaten the validity of our findings.

\vspace{5pt}\noindent\textbf{Limited Experiments.}
To estimate the importance of the DL replicability/reproducibility, we conducted experiments on four representative DL models used in four representative SE tasks. Our experiments involve four prevalent factors that could strongly affect the DL replicability/reproducibility --  model stability, convergence, OOV issue, and the testing data size. This investigation on DL replicability/reproducibility may not be exhaustive, but they are the most commonly discussed factors in SE as described in Section \ref{sub_rqs}, and hence are enough to show the importance of the DL replicability/reproducibility issue. We selected only four DL models for use in our experiments. Running more DL models would help researchers better understand the DL replicability/reproducibility issues for other SE tasks and other DL- based models. However, running DL models are time-consuming and we have limited computational resources ourselves. We therefore presented an exploratory meta-analysis in this study to help researchers understand the prevalence and importance of replicability/reproducibility issues. Due to the same reason, we ran each DL model ten times for each research question. In the future, we plan to replicate more DL studies and investigate more influential factors relating to DL replicability/reproducibility.

\vspace{5pt}\noindent\textbf{Manual Evaluation.}
When we re-ran the code search model DeepCS \cite{gu2018deep}, the relevance of top-10 returned code methods to a query requires manual identification, which could suffer from subjectivity bias. To mitigate this threat, the manual analysis was performed by two independent experienced developers. If a conflict occurred, it was resolved by an open discussion. To save manual efforts, the relevancy of code-query pair was labeled by a script if that relevancy has been identified before.

\section{Discussion}\label{discussion}

Our experimental findings in Sections \ref{review} and \ref{results} imply that replicability and reproducibility issues are prevalent and important for DL-based studies in SE. To mitigate these replicability and replicability issues, DL studies in SE should consider the following guidelines:

\textbf{Provide a long-lasting link for a replication package.} An accessible replication package with source code and data can substantially facilitate study replication, help other researchers have a deeper understanding of the DL model used, and largely mitigate manual errors in any replication study. 

\textbf{Estimating stability and convergence of DL models.} Reporting only one good experimental result likely threatens the DL model replicability due to its unverified stability and convergence. It is recommended to estimate the model stability by running it multiple times and assess the model convergence by reporting model performance at different optimization iterations or epochs.

\textbf{Enhance stability and convergence of DL models.} DL model stability and convergence can be improved by choosing an appropriate DL initialization method and training the model until convergence, instead of a fixed number of optimization iterations. 

\textbf{Use an automated evaluation approach to avoid the effect of human bias and subjectivity.} The replicability of a SE study using a DL model could be compromised if the evaluation involves any manual process. Although many DL studies leverage some methods to mitigate the effect of this issue, a better way is to design an automated evaluation approach. 

\textbf{Mitigate the effect of OOV issue for code/text representation learning.} The OOV issue is prevalent in many SE tasks and the prevalent solutions are to replace new words by existing words in the vocabulary \cite{liu2019automatic}, transforming new words \cite{habash2008four}, or build a character level vocabulary instead of the word level one \cite{kim2016character}. 

\textbf{Measure DL-based study reproducibility on different testing data.} Study reproducibility requires that an experimental finding can be reproduced using different sampled testing data. Thus, one reported result on only one testing dataset will be not enough. It is suggested to test the model using different sizes of testing data and analyze the model performance.

\textbf{Mitigate DL-based SE study reproducibility issues on different testing data.} DL model performance commonly varies for different testing data. To mitigate this issue, it is recommended to incorporate some more robust features for representation learning, such as using abstract syntax trees \cite{zhang2019novel}.  

\textbf{Improve DL model efficiency.} Time-consuming model optimization is a large obstacle when DL-based studies investigate model replicability and reproducibility. Therefore, reducing model complexity and accelerating model optimization will substantially promote the replicability and reproducibility studies.

\section{Conclusion}\label{conclusion}
Replicability and reproducibility are important for scientific research but DL-based studies in SE often ignore or minimalise them. To investigate the merit of SE study replicability/reproducibility using DL-based models, we first analyzed the characteristics of DL-based studies in SE. Specifically, we conducted a literature review on DL studies that are published on twenty prevalent SE venues in the recent five years. The observation from 93 reviewed literature shows that more than 70\% of DL-based studies were published in the last two years. Most of the studies leverage DL techniques to better learn  semantics in code/text, and CNN and RNN based models are the most frequently used DL techniques. Reinforcement learning is begining to show its potential to further improve the performance of existing DL-based models.

To assess the prevalence of replicability/replicability issues in SE studies using DL, we checked all the links in the reviewed DL-based studies and found that only 25.8\% of studies provided accessible links to their replication packages. Thus, the DL replicability cannot be ensured because other researchers need to replicate existing studies from scratch and may miss important implementation details. We also counted the percentage of DL studies that investigated any research questions on the replicability/replicability issues, and found that only 10.8\% of studies considered these issues. Therefore, the replicability/replicability issue is prevalent, and it is suggested for further DL-based studies in SE to share their source code and data publicly and better analyze their DL-based study replicability/replicability issues to better support their DL model validity.

To investigate the importance of replicability/replicability, we performed four experiments on four representative DL models used for four representative SE tasks. These studies were published in top SE venues in the last two years with accessible replication packages. Our experiments considered how the model stability affects DL replicability, and analyzed how model convergence, the out-of-vocabulary issue, and testing data size all impact on DL-based SE study reproducibility. Our experimental results show that DL models possess randomness by nature so that study replicability can be compromised for an unstable model. A DL model with low convergence level can produce highly turbulent and unstable performance, leading to low study reproducibility. The performance of a DL model can be highly sensitive to the size of vocabulary for training and the scale of testing data used. Thus a good reported model performance can be hard to reproduce for a new experimental dataset. Therefore, it is recommended for the SE community to pay more attention to these influential factors that may cause low DL-based study replicability/replicability. We also need to investigate viable solutions to strengthen study DL-based model validity instead of just presenting the advantages of chosen DL model effectiveness.

\bibliographystyle{ACM-Reference-Format}
\bibliography{reference.bib}


\begin{thebibliography}{162}


\ifx \showCODEN    \undefined \def \showCODEN     #1{\unskip}     \fi
\ifx \showDOI      \undefined \def \showDOI       #1{#1}\fi
\ifx \showISBNx    \undefined \def \showISBNx     #1{\unskip}     \fi
\ifx \showISBNxiii \undefined \def \showISBNxiii  #1{\unskip}     \fi
\ifx \showISSN     \undefined \def \showISSN      #1{\unskip}     \fi
\ifx \showLCCN     \undefined \def \showLCCN      #1{\unskip}     \fi
\ifx \shownote     \undefined \def \shownote      #1{#1}          \fi
\ifx \showarticletitle \undefined \def \showarticletitle #1{#1}   \fi
\ifx \showURL      \undefined \def \showURL       {\relax}        \fi
\providecommand\bibfield[2]{#2}
\providecommand\bibinfo[2]{#2}
\providecommand\natexlab[1]{#1}
\providecommand\showeprint[2][]{arXiv:#2}

\bibitem[\protect\citeauthoryear{Al-Hroob, Imam, and Al-Heisa}{Al-Hroob
  et~al\mbox{.}}{2018}]%
        {al2018use}
\bibfield{author}{\bibinfo{person}{Aysh Al-Hroob}, \bibinfo{person}{Ayad~Tareq
  Imam}, {and} \bibinfo{person}{Rawan Al-Heisa}.}
  \bibinfo{year}{2018}\natexlab{}.
\newblock \showarticletitle{The use of artificial neural networks for
  extracting actions and actors from requirements document}.
\newblock \bibinfo{journal}{\emph{Information and Software Technology}}
  \bibinfo{volume}{101} (\bibinfo{year}{2018}), \bibinfo{pages}{1--15}.
\newblock


\bibitem[\protect\citeauthoryear{Al-Jamimi and Ahmed}{Al-Jamimi and
  Ahmed}{2013}]%
        {al2013machine}
\bibfield{author}{\bibinfo{person}{Hamdi~A Al-Jamimi} {and}
  \bibinfo{person}{Moataz Ahmed}.} \bibinfo{year}{2013}\natexlab{}.
\newblock \showarticletitle{Machine learning-based software quality prediction
  models: state of the art}. In \bibinfo{booktitle}{\emph{2013 International
  Conference on Information Science and Applications (ICISA)}}. IEEE,
  \bibinfo{pages}{1--4}.
\newblock


\bibitem[\protect\citeauthoryear{Amann, Beyer, Kevic, and Gall}{Amann
  et~al\mbox{.}}{2013}]%
        {amann2013software}
\bibfield{author}{\bibinfo{person}{Sven Amann}, \bibinfo{person}{Stefanie
  Beyer}, \bibinfo{person}{Katja Kevic}, {and} \bibinfo{person}{Harald Gall}.}
  \bibinfo{year}{2013}\natexlab{}.
\newblock \showarticletitle{Software mining studies: Goals, approaches,
  artifacts, and replicability}.
\newblock In \bibinfo{booktitle}{\emph{Software Engineering}}.
  \bibinfo{publisher}{Springer}, \bibinfo{pages}{121--158}.
\newblock


\bibitem[\protect\citeauthoryear{Anda, Sj{\o}berg, and Mockus}{Anda
  et~al\mbox{.}}{2008}]%
        {anda2008variability}
\bibfield{author}{\bibinfo{person}{Bente~CD Anda}, \bibinfo{person}{Dag~IK
  Sj{\o}berg}, {and} \bibinfo{person}{Audris Mockus}.}
  \bibinfo{year}{2008}\natexlab{}.
\newblock \showarticletitle{Variability and reproducibility in software
  engineering: A study of four companies that developed the same system}.
\newblock \bibinfo{journal}{\emph{IEEE Transactions on Software Engineering}}
  \bibinfo{volume}{35}, \bibinfo{number}{3} (\bibinfo{year}{2008}),
  \bibinfo{pages}{407--429}.
\newblock


\bibitem[\protect\citeauthoryear{Arpteg, Brinne, Crnkovic-Friis, and
  Bosch}{Arpteg et~al\mbox{.}}{2018}]%
        {arpteg2018software}
\bibfield{author}{\bibinfo{person}{Anders Arpteg}, \bibinfo{person}{Bj{\"o}rn
  Brinne}, \bibinfo{person}{Luka Crnkovic-Friis}, {and} \bibinfo{person}{Jan
  Bosch}.} \bibinfo{year}{2018}\natexlab{}.
\newblock \showarticletitle{Software engineering challenges of deep learning}.
  In \bibinfo{booktitle}{\emph{2018 44th Euromicro Conference on Software
  Engineering and Advanced Applications (SEAA)}}. IEEE,
  \bibinfo{pages}{50--59}.
\newblock


\bibitem[\protect\citeauthoryear{Bahdanau, Cho, and Bengio}{Bahdanau
  et~al\mbox{.}}{2014}]%
        {bahdanau2014neural}
\bibfield{author}{\bibinfo{person}{Dzmitry Bahdanau},
  \bibinfo{person}{Kyunghyun Cho}, {and} \bibinfo{person}{Yoshua Bengio}.}
  \bibinfo{year}{2014}\natexlab{}.
\newblock \showarticletitle{Neural machine translation by jointly learning to
  align and translate}.
\newblock \bibinfo{journal}{\emph{arXiv preprint arXiv:1409.0473}}
  (\bibinfo{year}{2014}).
\newblock


\bibitem[\protect\citeauthoryear{Barbez, Khomh, and Gu{\'e}h{\'e}neuc}{Barbez
  et~al\mbox{.}}{2019}]%
        {barbez2019deep}
\bibfield{author}{\bibinfo{person}{Antoine Barbez}, \bibinfo{person}{Foutse
  Khomh}, {and} \bibinfo{person}{Yann-Ga{\"e}l Gu{\'e}h{\'e}neuc}.}
  \bibinfo{year}{2019}\natexlab{}.
\newblock \showarticletitle{Deep Learning Anti-patterns from Code Metrics
  History}. In \bibinfo{booktitle}{\emph{2019 IEEE International Conference on
  Software Maintenance and Evolution (ICSME)}}. IEEE,
  \bibinfo{pages}{114--124}.
\newblock


\bibitem[\protect\citeauthoryear{Ben~Abdessalem, Nejati, Briand, and
  Stifter}{Ben~Abdessalem et~al\mbox{.}}{2016}]%
        {ben2016testing}
\bibfield{author}{\bibinfo{person}{Raja Ben~Abdessalem}, \bibinfo{person}{Shiva
  Nejati}, \bibinfo{person}{Lionel~C Briand}, {and} \bibinfo{person}{Thomas
  Stifter}.} \bibinfo{year}{2016}\natexlab{}.
\newblock \showarticletitle{Testing advanced driver assistance systems using
  multi-objective search and neural networks}. In
  \bibinfo{booktitle}{\emph{Proceedings of the 31st IEEE/ACM International
  Conference on Automated Software Engineering}}. \bibinfo{pages}{63--74}.
\newblock


\bibitem[\protect\citeauthoryear{Benesty, Chen, Huang, and Cohen}{Benesty
  et~al\mbox{.}}{2009}]%
        {benesty2009pearson}
\bibfield{author}{\bibinfo{person}{Jacob Benesty}, \bibinfo{person}{Jingdong
  Chen}, \bibinfo{person}{Yiteng Huang}, {and} \bibinfo{person}{Israel Cohen}.}
  \bibinfo{year}{2009}\natexlab{}.
\newblock \showarticletitle{Pearson correlation coefficient}.
\newblock In \bibinfo{booktitle}{\emph{Noise reduction in speech processing}}.
  \bibinfo{publisher}{Springer}, \bibinfo{pages}{1--4}.
\newblock


\bibitem[\protect\citeauthoryear{Bhatia, Kohli, and Singh}{Bhatia
  et~al\mbox{.}}{2018}]%
        {bhatia2018neuro}
\bibfield{author}{\bibinfo{person}{Sahil Bhatia}, \bibinfo{person}{Pushmeet
  Kohli}, {and} \bibinfo{person}{Rishabh Singh}.}
  \bibinfo{year}{2018}\natexlab{}.
\newblock \showarticletitle{Neuro-symbolic program corrector for introductory
  programming assignments}. In \bibinfo{booktitle}{\emph{2018 IEEE/ACM 40th
  International Conference on Software Engineering (ICSE)}}. IEEE,
  \bibinfo{pages}{60--70}.
\newblock


\bibitem[\protect\citeauthoryear{Bisi and Goyal}{Bisi and Goyal}{2016}]%
        {bisi2016software}
\bibfield{author}{\bibinfo{person}{Manjubala Bisi} {and}
  \bibinfo{person}{Neeraj~Kumar Goyal}.} \bibinfo{year}{2016}\natexlab{}.
\newblock \showarticletitle{Software development efforts prediction using
  artificial neural network}.
\newblock \bibinfo{journal}{\emph{IET Software}} \bibinfo{volume}{10},
  \bibinfo{number}{3} (\bibinfo{year}{2016}), \bibinfo{pages}{63--71}.
\newblock


\bibitem[\protect\citeauthoryear{Boylan, Goodwin, Mohammadipour, and
  Syntetos}{Boylan et~al\mbox{.}}{2015}]%
        {boylan2015reproducibility}
\bibfield{author}{\bibinfo{person}{John~E Boylan}, \bibinfo{person}{Paul
  Goodwin}, \bibinfo{person}{Maryam Mohammadipour}, {and}
  \bibinfo{person}{Aris~A Syntetos}.} \bibinfo{year}{2015}\natexlab{}.
\newblock \showarticletitle{Reproducibility in forecasting research}.
\newblock \bibinfo{journal}{\emph{International Journal of Forecasting}}
  \bibinfo{volume}{31}, \bibinfo{number}{1} (\bibinfo{year}{2015}),
  \bibinfo{pages}{79--90}.
\newblock


\bibitem[\protect\citeauthoryear{Branco, Cohen, Vossen, Ide, and
  Calzolari}{Branco et~al\mbox{.}}{2017}]%
        {branco2017replicability}
\bibfield{author}{\bibinfo{person}{Ant{\'o}nio Branco},
  \bibinfo{person}{Kevin~Bretonnel Cohen}, \bibinfo{person}{Piek Vossen},
  \bibinfo{person}{Nancy Ide}, {and} \bibinfo{person}{Nicoletta Calzolari}.}
  \bibinfo{year}{2017}\natexlab{}.
\newblock \bibinfo{title}{Replicability and reproducibility of research results
  for human language technology: Introducing an LRE special section}.
\newblock
\newblock


\bibitem[\protect\citeauthoryear{B{\"u}ch and Andrzejak}{B{\"u}ch and
  Andrzejak}{2019}]%
        {buch2019learning}
\bibfield{author}{\bibinfo{person}{Lutz B{\"u}ch} {and} \bibinfo{person}{Artur
  Andrzejak}.} \bibinfo{year}{2019}\natexlab{}.
\newblock \showarticletitle{Learning-based recursive aggregation of abstract
  syntax trees for code clone detection}. In \bibinfo{booktitle}{\emph{2019
  IEEE 26th International Conference on Software Analysis, Evolution and
  Reengineering (SANER)}}. IEEE, \bibinfo{pages}{95--104}.
\newblock


\bibitem[\protect\citeauthoryear{Cambronero, Li, Kim, Sen, and
  Chandra}{Cambronero et~al\mbox{.}}{2019}]%
        {cambronero2019deep}
\bibfield{author}{\bibinfo{person}{Jose Cambronero}, \bibinfo{person}{Hongyu
  Li}, \bibinfo{person}{Seohyun Kim}, \bibinfo{person}{Koushik Sen}, {and}
  \bibinfo{person}{Satish Chandra}.} \bibinfo{year}{2019}\natexlab{}.
\newblock \showarticletitle{When deep learning met code search}. In
  \bibinfo{booktitle}{\emph{Proceedings of the 2019 27th ACM Joint Meeting on
  European Software Engineering Conference and Symposium on the Foundations of
  Software Engineering}}. \bibinfo{pages}{964--974}.
\newblock


\bibitem[\protect\citeauthoryear{Carver, Juristo, Baldassarre, and
  Vegas}{Carver et~al\mbox{.}}{2014}]%
        {carver2014replications}
\bibfield{author}{\bibinfo{person}{Jeffrey~C Carver}, \bibinfo{person}{Natalia
  Juristo}, \bibinfo{person}{Maria~Teresa Baldassarre}, {and}
  \bibinfo{person}{Sira Vegas}.} \bibinfo{year}{2014}\natexlab{}.
\newblock \bibinfo{title}{Replications of software engineering experiments}.
\newblock
\newblock


\bibitem[\protect\citeauthoryear{Chen, Diao, Zeng, Guo, and Hu}{Chen
  et~al\mbox{.}}{2018a}]%
        {chen2018drlgencert}
\bibfield{author}{\bibinfo{person}{Chao Chen}, \bibinfo{person}{Wenrui Diao},
  \bibinfo{person}{Yingpei Zeng}, \bibinfo{person}{Shanqing Guo}, {and}
  \bibinfo{person}{Chengyu Hu}.} \bibinfo{year}{2018}\natexlab{a}.
\newblock \showarticletitle{DRLgencert: Deep learning-based automated testing
  of certificate verification in SSL/TLS implementations}. In
  \bibinfo{booktitle}{\emph{2018 IEEE International Conference on Software
  Maintenance and Evolution (ICSME)}}. IEEE, \bibinfo{pages}{48--58}.
\newblock


\bibitem[\protect\citeauthoryear{Chen, Su, Meng, Xing, and Liu}{Chen
  et~al\mbox{.}}{2018b}]%
        {chen2018ui}
\bibfield{author}{\bibinfo{person}{Chunyang Chen}, \bibinfo{person}{Ting Su},
  \bibinfo{person}{Guozhu Meng}, \bibinfo{person}{Zhenchang Xing}, {and}
  \bibinfo{person}{Yang Liu}.} \bibinfo{year}{2018}\natexlab{b}.
\newblock \showarticletitle{From UI design image to GUI skeleton: a neural
  machine translator to bootstrap mobile GUI implementation}. In
  \bibinfo{booktitle}{\emph{Proceedings of the 40th International Conference on
  Software Engineering}}. \bibinfo{pages}{665--676}.
\newblock


\bibitem[\protect\citeauthoryear{Chen, Chen, Xing, and Xu}{Chen
  et~al\mbox{.}}{2016}]%
        {chen2016learning}
\bibfield{author}{\bibinfo{person}{Guibin Chen}, \bibinfo{person}{Chunyang
  Chen}, \bibinfo{person}{Zhenchang Xing}, {and} \bibinfo{person}{Bowen Xu}.}
  \bibinfo{year}{2016}\natexlab{}.
\newblock \showarticletitle{Learning a dual-language vector space for
  domain-specific cross-lingual question retrieval}. In
  \bibinfo{booktitle}{\emph{2016 31st IEEE/ACM International Conference on
  Automated Software Engineering (ASE)}}. IEEE, \bibinfo{pages}{744--755}.
\newblock


\bibitem[\protect\citeauthoryear{Chen, He, Lin, Zhang, Hao, Gao, Xu, Dang, and
  Zhang}{Chen et~al\mbox{.}}{2019}]%
        {chen2019continuous}
\bibfield{author}{\bibinfo{person}{Junjie Chen}, \bibinfo{person}{Xiaoting He},
  \bibinfo{person}{Qingwei Lin}, \bibinfo{person}{Hongyu Zhang},
  \bibinfo{person}{Dan Hao}, \bibinfo{person}{Feng Gao},
  \bibinfo{person}{Zhangwei Xu}, \bibinfo{person}{Yingnong Dang}, {and}
  \bibinfo{person}{Dongmei Zhang}.} \bibinfo{year}{2019}\natexlab{}.
\newblock \showarticletitle{Continuous incident triage for large-scale online
  service systems}. In \bibinfo{booktitle}{\emph{2019 34th IEEE/ACM
  International Conference on Automated Software Engineering (ASE)}}. IEEE,
  \bibinfo{pages}{364--375}.
\newblock


\bibitem[\protect\citeauthoryear{Choetkiertikul, Dam, Tran, Pham, Ghose, and
  Menzies}{Choetkiertikul et~al\mbox{.}}{2018}]%
        {choetkiertikul2018deep}
\bibfield{author}{\bibinfo{person}{Morakot Choetkiertikul},
  \bibinfo{person}{Hoa~Khanh Dam}, \bibinfo{person}{Truyen Tran},
  \bibinfo{person}{Trang Pham}, \bibinfo{person}{Aditya Ghose}, {and}
  \bibinfo{person}{Tim Menzies}.} \bibinfo{year}{2018}\natexlab{}.
\newblock \showarticletitle{A deep learning model for estimating story points}.
\newblock \bibinfo{journal}{\emph{IEEE Transactions on Software Engineering}}
  \bibinfo{volume}{45}, \bibinfo{number}{7} (\bibinfo{year}{2018}),
  \bibinfo{pages}{637--656}.
\newblock


\bibitem[\protect\citeauthoryear{Choi and Choi}{Choi and Choi}{1992}]%
        {choi1992sensitivity}
\bibfield{author}{\bibinfo{person}{Jin~Young Choi} {and}
  \bibinfo{person}{Chong-Ho Choi}.} \bibinfo{year}{1992}\natexlab{}.
\newblock \showarticletitle{Sensitivity analysis of multilayer perceptron with
  differentiable activation functions}.
\newblock \bibinfo{journal}{\emph{IEEE Transactions on Neural Networks}}
  \bibinfo{volume}{3}, \bibinfo{number}{1} (\bibinfo{year}{1992}),
  \bibinfo{pages}{101--107}.
\newblock


\bibitem[\protect\citeauthoryear{Cito, Ferme, and Gall}{Cito
  et~al\mbox{.}}{2016}]%
        {cito2016using}
\bibfield{author}{\bibinfo{person}{J{\"u}rgen Cito}, \bibinfo{person}{Vincenzo
  Ferme}, {and} \bibinfo{person}{Harald~C Gall}.}
  \bibinfo{year}{2016}\natexlab{}.
\newblock \showarticletitle{Using docker containers to improve reproducibility
  in software and web engineering research}. In
  \bibinfo{booktitle}{\emph{International Conference on Web Engineering}}.
  Springer, \bibinfo{pages}{609--612}.
\newblock


\bibitem[\protect\citeauthoryear{Collobert, Weston, Bottou, Karlen,
  Kavukcuoglu, and Kuksa}{Collobert et~al\mbox{.}}{2011}]%
        {collobert2011natural}
\bibfield{author}{\bibinfo{person}{Ronan Collobert}, \bibinfo{person}{Jason
  Weston}, \bibinfo{person}{L{\'e}on Bottou}, \bibinfo{person}{Michael Karlen},
  \bibinfo{person}{Koray Kavukcuoglu}, {and} \bibinfo{person}{Pavel Kuksa}.}
  \bibinfo{year}{2011}\natexlab{}.
\newblock \showarticletitle{Natural language processing (almost) from scratch}.
\newblock \bibinfo{journal}{\emph{Journal of machine learning research}}
  \bibinfo{volume}{12}, \bibinfo{number}{Aug} (\bibinfo{year}{2011}),
  \bibinfo{pages}{2493--2537}.
\newblock


\bibitem[\protect\citeauthoryear{Cox and Stuart}{Cox and Stuart}{1955}]%
        {cox1955some}
\bibfield{author}{\bibinfo{person}{David~Roxbee Cox} {and}
  \bibinfo{person}{Alan Stuart}.} \bibinfo{year}{1955}\natexlab{}.
\newblock \showarticletitle{Some quick sign tests for trend in location and
  dispersion}.
\newblock \bibinfo{journal}{\emph{Biometrika}} \bibinfo{volume}{42},
  \bibinfo{number}{1/2} (\bibinfo{year}{1955}), \bibinfo{pages}{80--95}.
\newblock


\bibitem[\protect\citeauthoryear{Da~Silva, Suassuna, Fran{\c{c}}a, Grubb,
  Gouveia, Monteiro, and dos Santos}{Da~Silva et~al\mbox{.}}{2014}]%
        {da2014replication}
\bibfield{author}{\bibinfo{person}{Fabio~QB Da~Silva}, \bibinfo{person}{Marcos
  Suassuna}, \bibinfo{person}{A~C{\'e}sar~C Fran{\c{c}}a},
  \bibinfo{person}{Alicia~M Grubb}, \bibinfo{person}{Tatiana~B Gouveia},
  \bibinfo{person}{Cleviton~VF Monteiro}, {and} \bibinfo{person}{Igor~Ebrahim
  dos Santos}.} \bibinfo{year}{2014}\natexlab{}.
\newblock \showarticletitle{Replication of empirical studies in software
  engineering research: a systematic mapping study}.
\newblock \bibinfo{journal}{\emph{Empirical Software Engineering}}
  \bibinfo{volume}{19}, \bibinfo{number}{3} (\bibinfo{year}{2014}),
  \bibinfo{pages}{501--557}.
\newblock


\bibitem[\protect\citeauthoryear{Dam, Tran, Pham, Ng, Grundy, and Ghose}{Dam
  et~al\mbox{.}}{2018}]%
        {dam2018automatic}
\bibfield{author}{\bibinfo{person}{Hoa~Khanh Dam}, \bibinfo{person}{Truyen
  Tran}, \bibinfo{person}{Trang Thi~Minh Pham}, \bibinfo{person}{Shien~Wee Ng},
  \bibinfo{person}{John Grundy}, {and} \bibinfo{person}{Aditya Ghose}.}
  \bibinfo{year}{2018}\natexlab{}.
\newblock \showarticletitle{Automatic feature learning for predicting
  vulnerable software components}.
\newblock \bibinfo{journal}{\emph{IEEE Transactions on Software Engineering}}
  (\bibinfo{year}{2018}).
\newblock


\bibitem[\protect\citeauthoryear{Deselaers, Hasan, Bender, and Ney}{Deselaers
  et~al\mbox{.}}{2009}]%
        {deselaers2009deep}
\bibfield{author}{\bibinfo{person}{Thomas Deselaers},
  \bibinfo{person}{Sa{\v{s}}a Hasan}, \bibinfo{person}{Oliver Bender}, {and}
  \bibinfo{person}{Hermann Ney}.} \bibinfo{year}{2009}\natexlab{}.
\newblock \showarticletitle{A deep learning approach to machine
  transliteration}. In \bibinfo{booktitle}{\emph{Proceedings of the Fourth
  Workshop on Statistical Machine Translation}}. Association for Computational
  Linguistics, \bibinfo{pages}{233--241}.
\newblock


\bibitem[\protect\citeauthoryear{Deshmukh, Podder, Sengupta, Dubash,
  et~al\mbox{.}}{Deshmukh et~al\mbox{.}}{2017}]%
        {deshmukh2017towards}
\bibfield{author}{\bibinfo{person}{Jayati Deshmukh}, \bibinfo{person}{Sanjay
  Podder}, \bibinfo{person}{Shubhashis Sengupta}, \bibinfo{person}{Neville
  Dubash}, {et~al\mbox{.}}} \bibinfo{year}{2017}\natexlab{}.
\newblock \showarticletitle{Towards accurate duplicate bug retrieval using deep
  learning techniques}. In \bibinfo{booktitle}{\emph{2017 IEEE International
  conference on software maintenance and evolution (ICSME)}}. IEEE,
  \bibinfo{pages}{115--124}.
\newblock


\bibitem[\protect\citeauthoryear{DQ, Yu, and Jiang}{DQ et~al\mbox{.}}{2019}]%
        {dq2019bilateral}
\bibfield{author}{\bibinfo{person}{Bui~Nghi DQ}, \bibinfo{person}{Yijun Yu},
  {and} \bibinfo{person}{Lingxiao Jiang}.} \bibinfo{year}{2019}\natexlab{}.
\newblock \showarticletitle{Bilateral dependency neural networks for
  cross-language algorithm classification}. In \bibinfo{booktitle}{\emph{2019
  IEEE 26th International Conference on Software Analysis, Evolution and
  Reengineering (SANER)}}. IEEE, \bibinfo{pages}{422--433}.
\newblock


\bibitem[\protect\citeauthoryear{Fine}{Fine}{2006}]%
        {fine2006feedforward}
\bibfield{author}{\bibinfo{person}{Terrence~L Fine}.}
  \bibinfo{year}{2006}\natexlab{}.
\newblock \bibinfo{booktitle}{\emph{Feedforward neural network methodology}}.
\newblock \bibinfo{publisher}{Springer Science \& Business Media}.
\newblock


\bibitem[\protect\citeauthoryear{Fischer and Krauss}{Fischer and
  Krauss}{2018}]%
        {fischer2018deep}
\bibfield{author}{\bibinfo{person}{Thomas Fischer} {and}
  \bibinfo{person}{Christopher Krauss}.} \bibinfo{year}{2018}\natexlab{}.
\newblock \showarticletitle{Deep learning with long short-term memory networks
  for financial market predictions}.
\newblock \bibinfo{journal}{\emph{European Journal of Operational Research}}
  \bibinfo{volume}{270}, \bibinfo{number}{2} (\bibinfo{year}{2018}),
  \bibinfo{pages}{654--669}.
\newblock


\bibitem[\protect\citeauthoryear{Forsyth and Ponce}{Forsyth and Ponce}{2002}]%
        {forsyth2002computer}
\bibfield{author}{\bibinfo{person}{David~A Forsyth} {and} \bibinfo{person}{Jean
  Ponce}.} \bibinfo{year}{2002}\natexlab{}.
\newblock \bibinfo{booktitle}{\emph{Computer vision: a modern approach}}.
\newblock \bibinfo{publisher}{Prentice Hall Professional Technical Reference}.
\newblock


\bibitem[\protect\citeauthoryear{Gao, Zeng, Xia, Lo, Lyu, and King}{Gao
  et~al\mbox{.}}{2019b}]%
        {gao2019automating}
\bibfield{author}{\bibinfo{person}{Cuiyun Gao}, \bibinfo{person}{Jichuan Zeng},
  \bibinfo{person}{Xin Xia}, \bibinfo{person}{David Lo},
  \bibinfo{person}{Michael~R Lyu}, {and} \bibinfo{person}{Irwin King}.}
  \bibinfo{year}{2019}\natexlab{b}.
\newblock \showarticletitle{Automating App Review Response Generation}. In
  \bibinfo{booktitle}{\emph{2019 34th IEEE/ACM International Conference on
  Automated Software Engineering (ASE)}}. IEEE, \bibinfo{pages}{163--175}.
\newblock


\bibitem[\protect\citeauthoryear{Gao, Chen, Xing, Ma, Song, and Lin}{Gao
  et~al\mbox{.}}{2019a}]%
        {gao2019neural}
\bibfield{author}{\bibinfo{person}{Sa Gao}, \bibinfo{person}{Chunyang Chen},
  \bibinfo{person}{Zhenchang Xing}, \bibinfo{person}{Yukun Ma},
  \bibinfo{person}{Wen Song}, {and} \bibinfo{person}{Shang-Wei Lin}.}
  \bibinfo{year}{2019}\natexlab{a}.
\newblock \showarticletitle{A neural model for method name generation from
  functional description}. In \bibinfo{booktitle}{\emph{2019 IEEE 26th
  International Conference on Software Analysis, Evolution and Reengineering
  (SANER)}}. IEEE, \bibinfo{pages}{414--421}.
\newblock


\bibitem[\protect\citeauthoryear{Gao, Wang, Liu, Yang, Sang, and Cai}{Gao
  et~al\mbox{.}}{[n.d.]}]%
        {gao2019teccd}
\bibfield{author}{\bibinfo{person}{Yi Gao}, \bibinfo{person}{Zan Wang},
  \bibinfo{person}{Shuang Liu}, \bibinfo{person}{Lin Yang},
  \bibinfo{person}{Wei Sang}, {and} \bibinfo{person}{Yuanfang Cai}.}
  \bibinfo{year}{[n.d.]}\natexlab{}.
\newblock \showarticletitle{TECCD: A Tree Embedding Approach for Code Clone
  Detection}. In \bibinfo{booktitle}{\emph{2019 IEEE International Conference
  on Software Maintenance and Evolution (ICSME)}}. IEEE,
  \bibinfo{pages}{145--156}.
\newblock


\bibitem[\protect\citeauthoryear{Ge, Chen, Liu, Chen, Huang, and Wang}{Ge
  et~al\mbox{.}}{2018}]%
        {ge2018deep}
\bibfield{author}{\bibinfo{person}{Yongxin Ge}, \bibinfo{person}{Min Chen},
  \bibinfo{person}{Chao Liu}, \bibinfo{person}{Feiyi Chen},
  \bibinfo{person}{Sheng Huang}, {and} \bibinfo{person}{Hongxing Wang}.}
  \bibinfo{year}{2018}\natexlab{}.
\newblock \showarticletitle{Deep metric learning for software change-proneness
  prediction}. In \bibinfo{booktitle}{\emph{International Conference on
  Intelligent Science and Big Data Engineering}}. Springer,
  \bibinfo{pages}{287--300}.
\newblock


\bibitem[\protect\citeauthoryear{Godefroid, Peleg, and Singh}{Godefroid
  et~al\mbox{.}}{2017}]%
        {godefroid2017learn}
\bibfield{author}{\bibinfo{person}{Patrice Godefroid}, \bibinfo{person}{Hila
  Peleg}, {and} \bibinfo{person}{Rishabh Singh}.}
  \bibinfo{year}{2017}\natexlab{}.
\newblock \showarticletitle{Learn\&fuzz: Machine learning for input fuzzing}.
  In \bibinfo{booktitle}{\emph{2017 32nd IEEE/ACM International Conference on
  Automated Software Engineering (ASE)}}. IEEE, \bibinfo{pages}{50--59}.
\newblock


\bibitem[\protect\citeauthoryear{Goh}{Goh}{1995}]%
        {goh1995back}
\bibfield{author}{\bibinfo{person}{Anthony~TC Goh}.}
  \bibinfo{year}{1995}\natexlab{}.
\newblock \showarticletitle{Back-propagation neural networks for modeling
  complex systems}.
\newblock \bibinfo{journal}{\emph{Artificial Intelligence in Engineering}}
  \bibinfo{volume}{9}, \bibinfo{number}{3} (\bibinfo{year}{1995}),
  \bibinfo{pages}{143--151}.
\newblock


\bibitem[\protect\citeauthoryear{G{\'o}mez, Juristo, and Vegas}{G{\'o}mez
  et~al\mbox{.}}{2010}]%
        {gomez2010replications}
\bibfield{author}{\bibinfo{person}{Omar~S G{\'o}mez}, \bibinfo{person}{Natalia
  Juristo}, {and} \bibinfo{person}{Sira Vegas}.}
  \bibinfo{year}{2010}\natexlab{}.
\newblock \showarticletitle{Replications types in experimental disciplines}. In
  \bibinfo{booktitle}{\emph{Proceedings of the 2010 ACM-IEEE international
  symposium on empirical software engineering and measurement}}.
  \bibinfo{pages}{1--10}.
\newblock


\bibitem[\protect\citeauthoryear{G{\'o}mez, Juristo, and Vegas}{G{\'o}mez
  et~al\mbox{.}}{2014}]%
        {gomez2014understanding}
\bibfield{author}{\bibinfo{person}{Omar~S G{\'o}mez}, \bibinfo{person}{Natalia
  Juristo}, {and} \bibinfo{person}{Sira Vegas}.}
  \bibinfo{year}{2014}\natexlab{}.
\newblock \showarticletitle{Understanding replication of experiments in
  software engineering: A classification}.
\newblock \bibinfo{journal}{\emph{Information and Software Technology}}
  \bibinfo{volume}{56}, \bibinfo{number}{8} (\bibinfo{year}{2014}),
  \bibinfo{pages}{1033--1048}.
\newblock


\bibitem[\protect\citeauthoryear{Gonz{\'a}lez-Barahona and
  Robles}{Gonz{\'a}lez-Barahona and Robles}{2012}]%
        {gonzalez2012reproducibility}
\bibfield{author}{\bibinfo{person}{Jes{\'u}s~M Gonz{\'a}lez-Barahona} {and}
  \bibinfo{person}{Gregorio Robles}.} \bibinfo{year}{2012}\natexlab{}.
\newblock \showarticletitle{On the reproducibility of empirical software
  engineering studies based on data retrieved from development repositories}.
\newblock \bibinfo{journal}{\emph{Empirical Software Engineering}}
  \bibinfo{volume}{17}, \bibinfo{number}{1-2} (\bibinfo{year}{2012}),
  \bibinfo{pages}{75--89}.
\newblock


\bibitem[\protect\citeauthoryear{Goodfellow, Bengio, and Courville}{Goodfellow
  et~al\mbox{.}}{2016}]%
        {goodfellow2016deep}
\bibfield{author}{\bibinfo{person}{Ian Goodfellow}, \bibinfo{person}{Yoshua
  Bengio}, {and} \bibinfo{person}{Aaron Courville}.}
  \bibinfo{year}{2016}\natexlab{}.
\newblock \bibinfo{booktitle}{\emph{Deep learning}}.
\newblock \bibinfo{publisher}{MIT press}.
\newblock


\bibitem[\protect\citeauthoryear{Gu, Zhang, and Kim}{Gu et~al\mbox{.}}{2018}]%
        {gu2018deep}
\bibfield{author}{\bibinfo{person}{Xiaodong Gu}, \bibinfo{person}{Hongyu
  Zhang}, {and} \bibinfo{person}{Sunghun Kim}.}
  \bibinfo{year}{2018}\natexlab{}.
\newblock \showarticletitle{Deep code search}. In
  \bibinfo{booktitle}{\emph{2018 IEEE/ACM 40th International Conference on
  Software Engineering (ICSE)}}. IEEE, \bibinfo{pages}{933--944}.
\newblock


\bibitem[\protect\citeauthoryear{Gu, Zhang, Zhang, and Kim}{Gu
  et~al\mbox{.}}{2016}]%
        {gu2016deep}
\bibfield{author}{\bibinfo{person}{Xiaodong Gu}, \bibinfo{person}{Hongyu
  Zhang}, \bibinfo{person}{Dongmei Zhang}, {and} \bibinfo{person}{Sunghun
  Kim}.} \bibinfo{year}{2016}\natexlab{}.
\newblock \showarticletitle{Deep API learning}. In
  \bibinfo{booktitle}{\emph{Proceedings of the 2016 24th ACM SIGSOFT
  International Symposium on Foundations of Software Engineering}}.
  \bibinfo{pages}{631--642}.
\newblock


\bibitem[\protect\citeauthoryear{Guo, Huang, Dong, Ye, Xu, Fan, Yang, and
  Xu}{Guo et~al\mbox{.}}{2019a}]%
        {guo2019deep}
\bibfield{author}{\bibinfo{person}{Chenkai Guo}, \bibinfo{person}{Dengrong
  Huang}, \bibinfo{person}{Naipeng Dong}, \bibinfo{person}{Quanqi Ye},
  \bibinfo{person}{Jing Xu}, \bibinfo{person}{Yaqing Fan}, \bibinfo{person}{Hui
  Yang}, {and} \bibinfo{person}{Yifan Xu}.} \bibinfo{year}{2019}\natexlab{a}.
\newblock \showarticletitle{Deep review sharing}. In
  \bibinfo{booktitle}{\emph{2019 IEEE 26th International Conference on Software
  Analysis, Evolution and Reengineering (SANER)}}. IEEE,
  \bibinfo{pages}{61--72}.
\newblock


\bibitem[\protect\citeauthoryear{Guo, Wang, Wu, Dong, Ye, Xu, and Zhang}{Guo
  et~al\mbox{.}}{2019b}]%
        {guo2019systematic}
\bibfield{author}{\bibinfo{person}{Chenkai Guo}, \bibinfo{person}{Weijing
  Wang}, \bibinfo{person}{Yanfeng Wu}, \bibinfo{person}{Naipeng Dong},
  \bibinfo{person}{Quanqi Ye}, \bibinfo{person}{Jing Xu}, {and}
  \bibinfo{person}{Sen Zhang}.} \bibinfo{year}{2019}\natexlab{b}.
\newblock \showarticletitle{Systematic comprehension for developer reply in
  mobile system forum}. In \bibinfo{booktitle}{\emph{2019 IEEE 26th
  International Conference on Software Analysis, Evolution and Reengineering
  (SANER)}}. IEEE, \bibinfo{pages}{242--252}.
\newblock


\bibitem[\protect\citeauthoryear{Guo, Cheng, and Cleland-Huang}{Guo
  et~al\mbox{.}}{2017}]%
        {guo2017semantically}
\bibfield{author}{\bibinfo{person}{Jin Guo}, \bibinfo{person}{Jinghui Cheng},
  {and} \bibinfo{person}{Jane Cleland-Huang}.} \bibinfo{year}{2017}\natexlab{}.
\newblock \showarticletitle{Semantically enhanced software traceability using
  deep learning techniques}. In \bibinfo{booktitle}{\emph{2017 IEEE/ACM 39th
  International Conference on Software Engineering (ICSE)}}. IEEE,
  \bibinfo{pages}{3--14}.
\newblock


\bibitem[\protect\citeauthoryear{Ha and Zhang}{Ha and Zhang}{2019}]%
        {ha2019deepperf}
\bibfield{author}{\bibinfo{person}{Huong Ha} {and} \bibinfo{person}{Hongyu
  Zhang}.} \bibinfo{year}{2019}\natexlab{}.
\newblock \showarticletitle{DeepPerf: performance prediction for configurable
  software with deep sparse neural network}. In \bibinfo{booktitle}{\emph{2019
  IEEE/ACM 41st International Conference on Software Engineering (ICSE)}}.
  IEEE, \bibinfo{pages}{1095--1106}.
\newblock


\bibitem[\protect\citeauthoryear{Habash}{Habash}{2008}]%
        {habash2008four}
\bibfield{author}{\bibinfo{person}{Nizar Habash}.}
  \bibinfo{year}{2008}\natexlab{}.
\newblock \showarticletitle{Four techniques for online handling of
  out-of-vocabulary words in Arabic-English statistical machine translation}.
  In \bibinfo{booktitle}{\emph{Proceedings of ACL-08: HLT, Short Papers}}.
  \bibinfo{pages}{57--60}.
\newblock


\bibitem[\protect\citeauthoryear{Han, Shihab, Wan, Deng, and Xia}{Han
  et~al\mbox{.}}{[n.d.]}]%
        {hanprogrammers}
\bibfield{author}{\bibinfo{person}{Junxiao Han}, \bibinfo{person}{Emad Shihab},
  \bibinfo{person}{Zhiyuan Wan}, \bibinfo{person}{Shuiguang Deng}, {and}
  \bibinfo{person}{Xin Xia}.} \bibinfo{year}{[n.d.]}\natexlab{}.
\newblock \showarticletitle{What do Programmers Discuss about Deep Learning
  Frameworks}.
\newblock  (\bibinfo{year}{[n.\,d.]}).
\newblock


\bibitem[\protect\citeauthoryear{Han, Li, Xing, Liu, and Feng}{Han
  et~al\mbox{.}}{2017}]%
        {han2017learning}
\bibfield{author}{\bibinfo{person}{Zhuobing Han}, \bibinfo{person}{Xiaohong
  Li}, \bibinfo{person}{Zhenchang Xing}, \bibinfo{person}{Hongtao Liu}, {and}
  \bibinfo{person}{Zhiyong Feng}.} \bibinfo{year}{2017}\natexlab{}.
\newblock \showarticletitle{Learning to predict severity of software
  vulnerability using only vulnerability description}. In
  \bibinfo{booktitle}{\emph{2017 IEEE International Conference on Software
  Maintenance and Evolution (ICSME)}}. IEEE, \bibinfo{pages}{125--136}.
\newblock


\bibitem[\protect\citeauthoryear{Hellendoorn, Bird, Barr, and
  Allamanis}{Hellendoorn et~al\mbox{.}}{2018}]%
        {hellendoorn2018deep}
\bibfield{author}{\bibinfo{person}{Vincent~J Hellendoorn},
  \bibinfo{person}{Christian Bird}, \bibinfo{person}{Earl~T Barr}, {and}
  \bibinfo{person}{Miltiadis Allamanis}.} \bibinfo{year}{2018}\natexlab{}.
\newblock \showarticletitle{Deep learning type inference}. In
  \bibinfo{booktitle}{\emph{Proceedings of the 2018 26th acm joint meeting on
  european software engineering conference and symposium on the foundations of
  software engineering}}. \bibinfo{pages}{152--162}.
\newblock


\bibitem[\protect\citeauthoryear{Hellendoorn and Devanbu}{Hellendoorn and
  Devanbu}{2017}]%
        {hellendoorn2017deep}
\bibfield{author}{\bibinfo{person}{Vincent~J Hellendoorn} {and}
  \bibinfo{person}{Premkumar Devanbu}.} \bibinfo{year}{2017}\natexlab{}.
\newblock \showarticletitle{Are deep neural networks the best choice for
  modeling source code?}. In \bibinfo{booktitle}{\emph{Proceedings of the 2017
  11th Joint Meeting on Foundations of Software Engineering}}.
  \bibinfo{pages}{763--773}.
\newblock


\bibitem[\protect\citeauthoryear{Hoang, Lawall, Tian, Oentaryo, and Lo}{Hoang
  et~al\mbox{.}}{2019}]%
        {hoang2019patchnet}
\bibfield{author}{\bibinfo{person}{Thong Hoang}, \bibinfo{person}{Julia
  Lawall}, \bibinfo{person}{Yuan Tian}, \bibinfo{person}{Richard~J Oentaryo},
  {and} \bibinfo{person}{David Lo}.} \bibinfo{year}{2019}\natexlab{}.
\newblock \showarticletitle{PatchNet: Hierarchical Deep Learning-Based Stable
  Patch Identification for the Linux Kernel}.
\newblock \bibinfo{journal}{\emph{IEEE Transactions on Software Engineering}}
  (\bibinfo{year}{2019}).
\newblock


\bibitem[\protect\citeauthoryear{Hochreiter and Schmidhuber}{Hochreiter and
  Schmidhuber}{1997}]%
        {hochreiter1997long}
\bibfield{author}{\bibinfo{person}{Sepp Hochreiter} {and}
  \bibinfo{person}{J{\"u}rgen Schmidhuber}.} \bibinfo{year}{1997}\natexlab{}.
\newblock \showarticletitle{Long short-term memory}.
\newblock \bibinfo{journal}{\emph{Neural computation}} \bibinfo{volume}{9},
  \bibinfo{number}{8} (\bibinfo{year}{1997}), \bibinfo{pages}{1735--1780}.
\newblock


\bibitem[\protect\citeauthoryear{Hu, Li, Xia, Lo, and Jin}{Hu
  et~al\mbox{.}}{2018}]%
        {hu2018deep}
\bibfield{author}{\bibinfo{person}{Xing Hu}, \bibinfo{person}{Ge Li},
  \bibinfo{person}{Xin Xia}, \bibinfo{person}{David Lo}, {and}
  \bibinfo{person}{Zhi Jin}.} \bibinfo{year}{2018}\natexlab{}.
\newblock \showarticletitle{Deep code comment generation}. In
  \bibinfo{booktitle}{\emph{Proceedings of the 26th Conference on Program
  Comprehension}}. \bibinfo{pages}{200--210}.
\newblock


\bibitem[\protect\citeauthoryear{Huang, Xia, Lo, and Murphy}{Huang
  et~al\mbox{.}}{2018}]%
        {huang2018automating}
\bibfield{author}{\bibinfo{person}{Qiao Huang}, \bibinfo{person}{Xin Xia},
  \bibinfo{person}{David Lo}, {and} \bibinfo{person}{Gail~C Murphy}.}
  \bibinfo{year}{2018}\natexlab{}.
\newblock \showarticletitle{Automating intention mining}.
\newblock \bibinfo{journal}{\emph{IEEE Transactions on Software Engineering}}
  (\bibinfo{year}{2018}).
\newblock


\bibitem[\protect\citeauthoryear{Huo, Thung, Li, Lo, and Shi}{Huo
  et~al\mbox{.}}{2019}]%
        {huo2019deep}
\bibfield{author}{\bibinfo{person}{Xuan Huo}, \bibinfo{person}{Ferdian Thung},
  \bibinfo{person}{Ming Li}, \bibinfo{person}{David Lo}, {and}
  \bibinfo{person}{Shu-Ting Shi}.} \bibinfo{year}{2019}\natexlab{}.
\newblock \showarticletitle{Deep transfer bug localization}.
\newblock \bibinfo{journal}{\emph{IEEE Transactions on Software Engineering}}
  (\bibinfo{year}{2019}).
\newblock


\bibitem[\protect\citeauthoryear{Ilonen, Kamarainen, and Lampinen}{Ilonen
  et~al\mbox{.}}{2003}]%
        {ilonen2003differential}
\bibfield{author}{\bibinfo{person}{Jarmo Ilonen},
  \bibinfo{person}{Joni-Kristian Kamarainen}, {and} \bibinfo{person}{Jouni
  Lampinen}.} \bibinfo{year}{2003}\natexlab{}.
\newblock \showarticletitle{Differential evolution training algorithm for
  feed-forward neural networks}.
\newblock \bibinfo{journal}{\emph{Neural Processing Letters}}
  \bibinfo{volume}{17}, \bibinfo{number}{1} (\bibinfo{year}{2003}),
  \bibinfo{pages}{93--105}.
\newblock


\bibitem[\protect\citeauthoryear{Ince, Hatton, and Graham-Cumming}{Ince
  et~al\mbox{.}}{2012}]%
        {ince2012case}
\bibfield{author}{\bibinfo{person}{Darrel~C Ince}, \bibinfo{person}{Leslie
  Hatton}, {and} \bibinfo{person}{John Graham-Cumming}.}
  \bibinfo{year}{2012}\natexlab{}.
\newblock \showarticletitle{The case for open computer programs}.
\newblock \bibinfo{journal}{\emph{Nature}} \bibinfo{volume}{482},
  \bibinfo{number}{7386} (\bibinfo{year}{2012}), \bibinfo{pages}{485--488}.
\newblock


\bibitem[\protect\citeauthoryear{Jiang, Armaly, and McMillan}{Jiang
  et~al\mbox{.}}{2017}]%
        {jiang2017automatically}
\bibfield{author}{\bibinfo{person}{Siyuan Jiang}, \bibinfo{person}{Ameer
  Armaly}, {and} \bibinfo{person}{Collin McMillan}.}
  \bibinfo{year}{2017}\natexlab{}.
\newblock \showarticletitle{Automatically generating commit messages from diffs
  using neural machine translation}. In \bibinfo{booktitle}{\emph{2017 32nd
  IEEE/ACM International Conference on Automated Software Engineering (ASE)}}.
  IEEE, \bibinfo{pages}{135--146}.
\newblock


\bibitem[\protect\citeauthoryear{Juristo and G{\'o}mez}{Juristo and
  G{\'o}mez}{2010}]%
        {juristo2010replication}
\bibfield{author}{\bibinfo{person}{Natalia Juristo} {and}
  \bibinfo{person}{Omar~S G{\'o}mez}.} \bibinfo{year}{2010}\natexlab{}.
\newblock \showarticletitle{Replication of software engineering experiments}.
\newblock In \bibinfo{booktitle}{\emph{Empirical software engineering and
  verification}}. \bibinfo{publisher}{Springer}, \bibinfo{pages}{60--88}.
\newblock


\bibitem[\protect\citeauthoryear{Juristo and Vegas}{Juristo and Vegas}{2011}]%
        {juristo2011role}
\bibfield{author}{\bibinfo{person}{Natalia Juristo} {and} \bibinfo{person}{Sira
  Vegas}.} \bibinfo{year}{2011}\natexlab{}.
\newblock \showarticletitle{The role of non-exact replications in software
  engineering experiments}.
\newblock \bibinfo{journal}{\emph{Empirical Software Engineering}}
  \bibinfo{volume}{16}, \bibinfo{number}{3} (\bibinfo{year}{2011}),
  \bibinfo{pages}{295--324}.
\newblock


\bibitem[\protect\citeauthoryear{Kalchbrenner, Grefenstette, and
  Blunsom}{Kalchbrenner et~al\mbox{.}}{2014}]%
        {kalchbrenner2014convolutional}
\bibfield{author}{\bibinfo{person}{Nal Kalchbrenner}, \bibinfo{person}{Edward
  Grefenstette}, {and} \bibinfo{person}{Phil Blunsom}.}
  \bibinfo{year}{2014}\natexlab{}.
\newblock \showarticletitle{A convolutional neural network for modelling
  sentences}.
\newblock \bibinfo{journal}{\emph{arXiv preprint arXiv:1404.2188}}
  (\bibinfo{year}{2014}).
\newblock


\bibitem[\protect\citeauthoryear{Katz, Ruchti, and Schulte}{Katz
  et~al\mbox{.}}{2018}]%
        {katz2018using}
\bibfield{author}{\bibinfo{person}{Deborah~S Katz}, \bibinfo{person}{Jason
  Ruchti}, {and} \bibinfo{person}{Eric Schulte}.}
  \bibinfo{year}{2018}\natexlab{}.
\newblock \showarticletitle{Using recurrent neural networks for decompilation}.
  In \bibinfo{booktitle}{\emph{2018 IEEE 25th International Conference on
  Software Analysis, Evolution and Reengineering (SANER)}}. IEEE,
  \bibinfo{pages}{346--356}.
\newblock


\bibitem[\protect\citeauthoryear{Kim, Jernite, Sontag, and Rush}{Kim
  et~al\mbox{.}}{2016}]%
        {kim2016character}
\bibfield{author}{\bibinfo{person}{Yoon Kim}, \bibinfo{person}{Yacine Jernite},
  \bibinfo{person}{David Sontag}, {and} \bibinfo{person}{Alexander~M Rush}.}
  \bibinfo{year}{2016}\natexlab{}.
\newblock \showarticletitle{Character-aware neural language models}. In
  \bibinfo{booktitle}{\emph{Thirtieth AAAI Conference on Artificial
  Intelligence}}.
\newblock


\bibitem[\protect\citeauthoryear{Kitchenham, Madeyski, and Brereton}{Kitchenham
  et~al\mbox{.}}{2020}]%
        {kitchenham2020meta}
\bibfield{author}{\bibinfo{person}{Barbara Kitchenham}, \bibinfo{person}{Lech
  Madeyski}, {and} \bibinfo{person}{Pearl Brereton}.}
  \bibinfo{year}{2020}\natexlab{}.
\newblock \showarticletitle{Meta-analysis for families of experiments in
  software engineering: a systematic review and reproducibility and validity
  assessment}.
\newblock \bibinfo{journal}{\emph{Empirical Software Engineering}}
  \bibinfo{volume}{25}, \bibinfo{number}{1} (\bibinfo{year}{2020}),
  \bibinfo{pages}{353--401}.
\newblock


\bibitem[\protect\citeauthoryear{Koo, Saumya, Kulkarni, and Bagchi}{Koo
  et~al\mbox{.}}{2019}]%
        {koo2019pyse}
\bibfield{author}{\bibinfo{person}{Jinkyu Koo}, \bibinfo{person}{Charitha
  Saumya}, \bibinfo{person}{Milind Kulkarni}, {and} \bibinfo{person}{Saurabh
  Bagchi}.} \bibinfo{year}{2019}\natexlab{}.
\newblock \showarticletitle{PySE: Automatic Worst-Case Test Generation by
  Reinforcement Learning}. In \bibinfo{booktitle}{\emph{2019 12th IEEE
  Conference on Software Testing, Validation and Verification (ICST)}}. IEEE,
  \bibinfo{pages}{136--147}.
\newblock


\bibitem[\protect\citeauthoryear{Kumar and Rath}{Kumar and Rath}{2016}]%
        {kumar2016hybrid}
\bibfield{author}{\bibinfo{person}{Lov Kumar} {and} \bibinfo{person}{Santanu~Ku
  Rath}.} \bibinfo{year}{2016}\natexlab{}.
\newblock \showarticletitle{Hybrid functional link artificial neural network
  approach for predicting maintainability of object-oriented software}.
\newblock \bibinfo{journal}{\emph{Journal of Systems and Software}}
  \bibinfo{volume}{121} (\bibinfo{year}{2016}), \bibinfo{pages}{170--190}.
\newblock


\bibitem[\protect\citeauthoryear{Lacomis, Yin, Schwartz, Allamanis, Le~Goues,
  Neubig, and Vasilescu}{Lacomis et~al\mbox{.}}{2019}]%
        {lacomis2019dire}
\bibfield{author}{\bibinfo{person}{Jeremy Lacomis}, \bibinfo{person}{Pengcheng
  Yin}, \bibinfo{person}{Edward Schwartz}, \bibinfo{person}{Miltiadis
  Allamanis}, \bibinfo{person}{Claire Le~Goues}, \bibinfo{person}{Graham
  Neubig}, {and} \bibinfo{person}{Bogdan Vasilescu}.}
  \bibinfo{year}{2019}\natexlab{}.
\newblock \showarticletitle{Dire: A neural approach to decompiled identifier
  naming}. In \bibinfo{booktitle}{\emph{2019 34th IEEE/ACM International
  Conference on Automated Software Engineering (ASE)}}. IEEE,
  \bibinfo{pages}{628--639}.
\newblock


\bibitem[\protect\citeauthoryear{Lam, Nguyen, Nguyen, and Nguyen}{Lam
  et~al\mbox{.}}{2017}]%
        {lam2017bug}
\bibfield{author}{\bibinfo{person}{An~Ngoc Lam}, \bibinfo{person}{Anh~Tuan
  Nguyen}, \bibinfo{person}{Hoan~Anh Nguyen}, {and} \bibinfo{person}{Tien~N
  Nguyen}.} \bibinfo{year}{2017}\natexlab{}.
\newblock \showarticletitle{Bug localization with combination of deep learning
  and information retrieval}. In \bibinfo{booktitle}{\emph{2017 IEEE/ACM 25th
  International Conference on Program Comprehension (ICPC)}}. IEEE,
  \bibinfo{pages}{218--229}.
\newblock


\bibitem[\protect\citeauthoryear{Lawrence, Giles, Tsoi, and Back}{Lawrence
  et~al\mbox{.}}{1997}]%
        {lawrence1997face}
\bibfield{author}{\bibinfo{person}{Steve Lawrence}, \bibinfo{person}{C~Lee
  Giles}, \bibinfo{person}{Ah~Chung Tsoi}, {and} \bibinfo{person}{Andrew~D
  Back}.} \bibinfo{year}{1997}\natexlab{}.
\newblock \showarticletitle{Face recognition: A convolutional neural-network
  approach}.
\newblock \bibinfo{journal}{\emph{IEEE transactions on neural networks}}
  \bibinfo{volume}{8}, \bibinfo{number}{1} (\bibinfo{year}{1997}),
  \bibinfo{pages}{98--113}.
\newblock


\bibitem[\protect\citeauthoryear{LeClair, Eberhart, and McMillan}{LeClair
  et~al\mbox{.}}{2018}]%
        {leclair2018adapting}
\bibfield{author}{\bibinfo{person}{Alexander LeClair}, \bibinfo{person}{Zachary
  Eberhart}, {and} \bibinfo{person}{Collin McMillan}.}
  \bibinfo{year}{2018}\natexlab{}.
\newblock \showarticletitle{Adapting neural text classification for improved
  software categorization}. In \bibinfo{booktitle}{\emph{2018 IEEE
  International Conference on Software Maintenance and Evolution (ICSME)}}.
  IEEE, \bibinfo{pages}{461--472}.
\newblock


\bibitem[\protect\citeauthoryear{LeClair, Jiang, and McMillan}{LeClair
  et~al\mbox{.}}{2019}]%
        {leclair2019neural}
\bibfield{author}{\bibinfo{person}{Alexander LeClair}, \bibinfo{person}{Siyuan
  Jiang}, {and} \bibinfo{person}{Collin McMillan}.}
  \bibinfo{year}{2019}\natexlab{}.
\newblock \showarticletitle{A neural model for generating natural language
  summaries of program subroutines}. In \bibinfo{booktitle}{\emph{2019 IEEE/ACM
  41st International Conference on Software Engineering (ICSE)}}. IEEE,
  \bibinfo{pages}{795--806}.
\newblock


\bibitem[\protect\citeauthoryear{LeCun, Bengio, and Hinton}{LeCun
  et~al\mbox{.}}{2015}]%
        {lecun2015deep}
\bibfield{author}{\bibinfo{person}{Yann LeCun}, \bibinfo{person}{Yoshua
  Bengio}, {and} \bibinfo{person}{Geoffrey Hinton}.}
  \bibinfo{year}{2015}\natexlab{}.
\newblock \showarticletitle{Deep learning}.
\newblock \bibinfo{journal}{\emph{nature}} \bibinfo{volume}{521},
  \bibinfo{number}{7553} (\bibinfo{year}{2015}), \bibinfo{pages}{436--444}.
\newblock


\bibitem[\protect\citeauthoryear{Lee, Hong, Yi, Kim, Kim, and Yoo}{Lee
  et~al\mbox{.}}{2019}]%
        {lee2019classifying}
\bibfield{author}{\bibinfo{person}{Seongmin Lee}, \bibinfo{person}{Shin Hong},
  \bibinfo{person}{Jungbae Yi}, \bibinfo{person}{Taeksu Kim},
  \bibinfo{person}{Chul-Joo Kim}, {and} \bibinfo{person}{Shin Yoo}.}
  \bibinfo{year}{2019}\natexlab{}.
\newblock \showarticletitle{Classifying False Positive Static Checker Alarms in
  Continuous Integration Using Convolutional Neural Networks}. In
  \bibinfo{booktitle}{\emph{2019 12th IEEE Conference on Software Testing,
  Validation and Verification (ICST)}}. IEEE, \bibinfo{pages}{391--401}.
\newblock


\bibitem[\protect\citeauthoryear{Li, Feng, Zhuang, Meng, and Ryder}{Li
  et~al\mbox{.}}{2017}]%
        {li2017cclearner}
\bibfield{author}{\bibinfo{person}{Liuqing Li}, \bibinfo{person}{He Feng},
  \bibinfo{person}{Wenjie Zhuang}, \bibinfo{person}{Na Meng}, {and}
  \bibinfo{person}{Barbara Ryder}.} \bibinfo{year}{2017}\natexlab{}.
\newblock \showarticletitle{Cclearner: A deep learning-based clone detection
  approach}. In \bibinfo{booktitle}{\emph{2017 IEEE International Conference on
  Software Maintenance and Evolution (ICSME)}}. IEEE,
  \bibinfo{pages}{249--260}.
\newblock


\bibitem[\protect\citeauthoryear{Li and Talwalkar}{Li and Talwalkar}{2019}]%
        {li2019random}
\bibfield{author}{\bibinfo{person}{Liam Li} {and} \bibinfo{person}{Ameet
  Talwalkar}.} \bibinfo{year}{2019}\natexlab{}.
\newblock \showarticletitle{Random search and reproducibility for neural
  architecture search}.
\newblock \bibinfo{journal}{\emph{arXiv preprint arXiv:1902.07638}}
  (\bibinfo{year}{2019}).
\newblock


\bibitem[\protect\citeauthoryear{Li, Jiang, Liu, Ren, and Li}{Li
  et~al\mbox{.}}{2018}]%
        {li2018unsupervised}
\bibfield{author}{\bibinfo{person}{Xiaochen Li}, \bibinfo{person}{He Jiang},
  \bibinfo{person}{Dong Liu}, \bibinfo{person}{Zhilei Ren}, {and}
  \bibinfo{person}{Ge Li}.} \bibinfo{year}{2018}\natexlab{}.
\newblock \showarticletitle{Unsupervised deep bug report summarization}. In
  \bibinfo{booktitle}{\emph{Proceedings of the 26th Conference on Program
  Comprehension}}. \bibinfo{pages}{144--155}.
\newblock


\bibitem[\protect\citeauthoryear{Li, Li, Zhang, and Zhang}{Li
  et~al\mbox{.}}{2019}]%
        {li2019deepfl}
\bibfield{author}{\bibinfo{person}{Xia Li}, \bibinfo{person}{Wei Li},
  \bibinfo{person}{Yuqun Zhang}, {and} \bibinfo{person}{Lingming Zhang}.}
  \bibinfo{year}{2019}\natexlab{}.
\newblock \showarticletitle{Deepfl: Integrating multiple fault diagnosis
  dimensions for deep fault localization}. In
  \bibinfo{booktitle}{\emph{Proceedings of the 28th ACM SIGSOFT International
  Symposium on Software Testing and Analysis}}. \bibinfo{pages}{169--180}.
\newblock


\bibitem[\protect\citeauthoryear{Lin and Och}{Lin and Och}{2004}]%
        {lin2004looking}
\bibfield{author}{\bibinfo{person}{Chin-Yew Lin} {and} \bibinfo{person}{FJ
  Och}.} \bibinfo{year}{2004}\natexlab{}.
\newblock \showarticletitle{Looking for a few good metrics: ROUGE and its
  evaluation}. In \bibinfo{booktitle}{\emph{Ntcir Workshop}}.
\newblock


\bibitem[\protect\citeauthoryear{Litjens, Kooi, Bejnordi, Setio, Ciompi,
  Ghafoorian, Van Der~Laak, Van~Ginneken, and S{\'a}nchez}{Litjens
  et~al\mbox{.}}{2017}]%
        {litjens2017survey}
\bibfield{author}{\bibinfo{person}{Geert Litjens}, \bibinfo{person}{Thijs
  Kooi}, \bibinfo{person}{Babak~Ehteshami Bejnordi}, \bibinfo{person}{Arnaud
  Arindra~Adiyoso Setio}, \bibinfo{person}{Francesco Ciompi},
  \bibinfo{person}{Mohsen Ghafoorian}, \bibinfo{person}{Jeroen~Awm Van
  Der~Laak}, \bibinfo{person}{Bram Van~Ginneken}, {and}
  \bibinfo{person}{Clara~I S{\'a}nchez}.} \bibinfo{year}{2017}\natexlab{}.
\newblock \showarticletitle{A survey on deep learning in medical image
  analysis}.
\newblock \bibinfo{journal}{\emph{Medical image analysis}}
  \bibinfo{volume}{42} (\bibinfo{year}{2017}), \bibinfo{pages}{60--88}.
\newblock


\bibitem[\protect\citeauthoryear{Liu, Yang, Xia, Yan, and Zhang}{Liu
  et~al\mbox{.}}{2018c}]%
        {liu2018cross}
\bibfield{author}{\bibinfo{person}{Chao Liu}, \bibinfo{person}{Dan Yang},
  \bibinfo{person}{Xin Xia}, \bibinfo{person}{Meng Yan}, {and}
  \bibinfo{person}{Xiaohong Zhang}.} \bibinfo{year}{2018}\natexlab{c}.
\newblock \showarticletitle{Cross-project change-proneness prediction}. In
  \bibinfo{booktitle}{\emph{2018 IEEE 42nd Annual Computer Software and
  Applications Conference (COMPSAC)}}, Vol.~\bibinfo{volume}{1}. IEEE,
  \bibinfo{pages}{64--73}.
\newblock


\bibitem[\protect\citeauthoryear{Liu, Yang, Xia, Yan, and Zhang}{Liu
  et~al\mbox{.}}{2019d}]%
        {liu2019two}
\bibfield{author}{\bibinfo{person}{Chao Liu}, \bibinfo{person}{Dan Yang},
  \bibinfo{person}{Xin Xia}, \bibinfo{person}{Meng Yan}, {and}
  \bibinfo{person}{Xiaohong Zhang}.} \bibinfo{year}{2019}\natexlab{d}.
\newblock \showarticletitle{A two-phase transfer learning model for
  cross-project defect prediction}.
\newblock \bibinfo{journal}{\emph{Information and Software Technology}}
  \bibinfo{volume}{107} (\bibinfo{year}{2019}), \bibinfo{pages}{125--136}.
\newblock


\bibitem[\protect\citeauthoryear{Liu, Yang, Zhang, Hu, Barson, and Ray}{Liu
  et~al\mbox{.}}{2018d}]%
        {liu2018recommender}
\bibfield{author}{\bibinfo{person}{Chao Liu}, \bibinfo{person}{Dan Yang},
  \bibinfo{person}{Xiaohong Zhang}, \bibinfo{person}{Haibo Hu},
  \bibinfo{person}{Jed Barson}, {and} \bibinfo{person}{Baishakhi Ray}.}
  \bibinfo{year}{2018}\natexlab{d}.
\newblock \showarticletitle{A recommender system for developer onboarding}. In
  \bibinfo{booktitle}{\emph{Proceedings of the 40th International Conference on
  Software Engineering: Companion Proceeedings}}. \bibinfo{pages}{319--320}.
\newblock


\bibitem[\protect\citeauthoryear{Liu, Yang, Zhang, Ray, and Rahman}{Liu
  et~al\mbox{.}}{2018e}]%
        {liu2018recommending}
\bibfield{author}{\bibinfo{person}{Chao Liu}, \bibinfo{person}{Dan Yang},
  \bibinfo{person}{Xiaohong Zhang}, \bibinfo{person}{Baishakhi Ray}, {and}
  \bibinfo{person}{Md~Masudur Rahman}.} \bibinfo{year}{2018}\natexlab{e}.
\newblock \showarticletitle{Recommending GitHub Projects for Developer
  Onboarding}.
\newblock \bibinfo{journal}{\emph{IEEE Access}}  \bibinfo{volume}{6}
  (\bibinfo{year}{2018}), \bibinfo{pages}{52082--52094}.
\newblock


\bibitem[\protect\citeauthoryear{Liu, Lowe, Serban, Noseworthy, Charlin, and
  Pineau}{Liu et~al\mbox{.}}{2016}]%
        {liu2016not}
\bibfield{author}{\bibinfo{person}{Chia-Wei Liu}, \bibinfo{person}{Ryan Lowe},
  \bibinfo{person}{Iulian~V Serban}, \bibinfo{person}{Michael Noseworthy},
  \bibinfo{person}{Laurent Charlin}, {and} \bibinfo{person}{Joelle Pineau}.}
  \bibinfo{year}{2016}\natexlab{}.
\newblock \showarticletitle{How not to evaluate your dialogue system: An
  empirical study of unsupervised evaluation metrics for dialogue response
  generation}.
\newblock \bibinfo{journal}{\emph{arXiv preprint arXiv:1603.08023}}
  (\bibinfo{year}{2016}).
\newblock


\bibitem[\protect\citeauthoryear{Liu, Jin, Xu, Bu, Zou, and Zhang}{Liu
  et~al\mbox{.}}{2019a}]%
        {liu2019deep}
\bibfield{author}{\bibinfo{person}{Hui Liu}, \bibinfo{person}{Jiahao Jin},
  \bibinfo{person}{Zhifeng Xu}, \bibinfo{person}{Yifan Bu},
  \bibinfo{person}{Yanzhen Zou}, {and} \bibinfo{person}{Lu Zhang}.}
  \bibinfo{year}{2019}\natexlab{a}.
\newblock \showarticletitle{Deep learning based code smell detection}.
\newblock \bibinfo{journal}{\emph{IEEE Transactions on Software Engineering}}
  (\bibinfo{year}{2019}).
\newblock


\bibitem[\protect\citeauthoryear{Liu, Xu, and Zou}{Liu et~al\mbox{.}}{2018b}]%
        {liu2018deep}
\bibfield{author}{\bibinfo{person}{Hui Liu}, \bibinfo{person}{Zhifeng Xu},
  {and} \bibinfo{person}{Yanzhen Zou}.} \bibinfo{year}{2018}\natexlab{b}.
\newblock \showarticletitle{Deep learning based feature envy detection}. In
  \bibinfo{booktitle}{\emph{Proceedings of the 33rd ACM/IEEE International
  Conference on Automated Software Engineering}}. \bibinfo{pages}{385--396}.
\newblock


\bibitem[\protect\citeauthoryear{Liu, Zhou, Yang, Liu, and Grundy}{Liu
  et~al\mbox{.}}{2018f}]%
        {liu2018fasttagrec}
\bibfield{author}{\bibinfo{person}{Jin Liu}, \bibinfo{person}{Pingyi Zhou},
  \bibinfo{person}{Zijiang Yang}, \bibinfo{person}{Xiao Liu}, {and}
  \bibinfo{person}{John Grundy}.} \bibinfo{year}{2018}\natexlab{f}.
\newblock \showarticletitle{FastTagRec: fast tag recommendation for software
  information sites}.
\newblock \bibinfo{journal}{\emph{Automated Software Engineering}}
  \bibinfo{volume}{25}, \bibinfo{number}{4} (\bibinfo{year}{2018}),
  \bibinfo{pages}{675--701}.
\newblock


\bibitem[\protect\citeauthoryear{Liu, Kim, Bissyand{\'e}, Kim, Kim, Koyuncu,
  Kim, and Le~Traon}{Liu et~al\mbox{.}}{2019b}]%
        {liu2019learning}
\bibfield{author}{\bibinfo{person}{Kui Liu}, \bibinfo{person}{Dongsun Kim},
  \bibinfo{person}{Tegawend{\'e}~F Bissyand{\'e}}, \bibinfo{person}{Taeyoung
  Kim}, \bibinfo{person}{Kisub Kim}, \bibinfo{person}{Anil Koyuncu},
  \bibinfo{person}{Suntae Kim}, {and} \bibinfo{person}{Yves Le~Traon}.}
  \bibinfo{year}{2019}\natexlab{b}.
\newblock \showarticletitle{Learning to spot and refactor inconsistent method
  names}. In \bibinfo{booktitle}{\emph{2019 IEEE/ACM 41st International
  Conference on Software Engineering (ICSE)}}. IEEE, \bibinfo{pages}{1--12}.
\newblock


\bibitem[\protect\citeauthoryear{Liu, Zhang, Pistoia, Zheng, Marques, and
  Zeng}{Liu et~al\mbox{.}}{2017}]%
        {liu2017automatic}
\bibfield{author}{\bibinfo{person}{Peng Liu}, \bibinfo{person}{Xiangyu Zhang},
  \bibinfo{person}{Marco Pistoia}, \bibinfo{person}{Yunhui Zheng},
  \bibinfo{person}{Manoel Marques}, {and} \bibinfo{person}{Lingfei Zeng}.}
  \bibinfo{year}{2017}\natexlab{}.
\newblock \showarticletitle{Automatic text input generation for mobile
  testing}. In \bibinfo{booktitle}{\emph{2017 IEEE/ACM 39th International
  Conference on Software Engineering (ICSE)}}. IEEE, \bibinfo{pages}{643--653}.
\newblock


\bibitem[\protect\citeauthoryear{Liu, Li, Guo, Zhou, and Xu}{Liu
  et~al\mbox{.}}{2018a}]%
        {liu2018connecting}
\bibfield{author}{\bibinfo{person}{Yibin Liu}, \bibinfo{person}{Yanhui Li},
  \bibinfo{person}{Jianbo Guo}, \bibinfo{person}{Yuming Zhou}, {and}
  \bibinfo{person}{Baowen Xu}.} \bibinfo{year}{2018}\natexlab{a}.
\newblock \showarticletitle{Connecting software metrics across versions to
  predict defects}. In \bibinfo{booktitle}{\emph{2018 IEEE 25th International
  Conference on Software Analysis, Evolution and Reengineering (SANER)}}. IEEE,
  \bibinfo{pages}{232--243}.
\newblock


\bibitem[\protect\citeauthoryear{Liu, Xia, Treude, Lo, and Li}{Liu
  et~al\mbox{.}}{2019c}]%
        {liu2019automatic}
\bibfield{author}{\bibinfo{person}{Zhongxin Liu}, \bibinfo{person}{Xin Xia},
  \bibinfo{person}{Christoph Treude}, \bibinfo{person}{David Lo}, {and}
  \bibinfo{person}{Shanping Li}.} \bibinfo{year}{2019}\natexlab{c}.
\newblock \showarticletitle{Automatic Generation of Pull Request Descriptions}.
\newblock \bibinfo{journal}{\emph{arXiv preprint arXiv:1909.06987}}
  (\bibinfo{year}{2019}).
\newblock


\bibitem[\protect\citeauthoryear{L{\'o}pez-Mart{\'\i}n and
  Abran}{L{\'o}pez-Mart{\'\i}n and Abran}{2015}]%
        {lopez2015neural}
\bibfield{author}{\bibinfo{person}{Cuauht{\'e}moc L{\'o}pez-Mart{\'\i}n} {and}
  \bibinfo{person}{Alain Abran}.} \bibinfo{year}{2015}\natexlab{}.
\newblock \showarticletitle{Neural networks for predicting the duration of new
  software projects}.
\newblock \bibinfo{journal}{\emph{Journal of Systems and Software}}
  \bibinfo{volume}{101} (\bibinfo{year}{2015}), \bibinfo{pages}{127--135}.
\newblock


\bibitem[\protect\citeauthoryear{Louridas and Gousios}{Louridas and
  Gousios}{2012}]%
        {louridas2012note}
\bibfield{author}{\bibinfo{person}{Panos Louridas} {and}
  \bibinfo{person}{Georgios Gousios}.} \bibinfo{year}{2012}\natexlab{}.
\newblock \showarticletitle{A note on rigour and replicability}.
\newblock \bibinfo{journal}{\emph{ACM SIGSOFT Software Engineering Notes}}
  \bibinfo{volume}{37}, \bibinfo{number}{5} (\bibinfo{year}{2012}),
  \bibinfo{pages}{1--4}.
\newblock


\bibitem[\protect\citeauthoryear{Loyola and Matsuo}{Loyola and Matsuo}{2017}]%
        {loyola2017learning}
\bibfield{author}{\bibinfo{person}{Pablo Loyola} {and} \bibinfo{person}{Yutaka
  Matsuo}.} \bibinfo{year}{2017}\natexlab{}.
\newblock \showarticletitle{Learning feature representations from change
  dependency graphs for defect prediction}. In \bibinfo{booktitle}{\emph{2017
  IEEE 28th International Symposium on Software Reliability Engineering
  (ISSRE)}}. IEEE, \bibinfo{pages}{361--372}.
\newblock


\bibitem[\protect\citeauthoryear{Lung, Aranda, Easterbrook, and Wilson}{Lung
  et~al\mbox{.}}{2008}]%
        {lung2008difficulty}
\bibfield{author}{\bibinfo{person}{Jonathan Lung}, \bibinfo{person}{Jorge
  Aranda}, \bibinfo{person}{Steve Easterbrook}, {and} \bibinfo{person}{Gregory
  Wilson}.} \bibinfo{year}{2008}\natexlab{}.
\newblock \showarticletitle{On the difficulty of replicating human subjects
  studies in software engineering}. In \bibinfo{booktitle}{\emph{2008 ACM/IEEE
  30th International Conference on Software Engineering}}. IEEE,
  \bibinfo{pages}{191--200}.
\newblock


\bibitem[\protect\citeauthoryear{Ma, Xing, Chen, Chen, Qu, and Li}{Ma
  et~al\mbox{.}}{2019}]%
        {ma2019easy}
\bibfield{author}{\bibinfo{person}{Suyu Ma}, \bibinfo{person}{Zhenchang Xing},
  \bibinfo{person}{Chunyang Chen}, \bibinfo{person}{Cheng Chen},
  \bibinfo{person}{Lizhen Qu}, {and} \bibinfo{person}{Guoqiang Li}.}
  \bibinfo{year}{2019}\natexlab{}.
\newblock \showarticletitle{Easy-to-Deploy API Extraction by Multi-Level
  Feature Embedding and Transfer Learning}.
\newblock \bibinfo{journal}{\emph{IEEE Transactions on Software Engineering}}
  (\bibinfo{year}{2019}).
\newblock


\bibitem[\protect\citeauthoryear{Mahmood, Bowes, Hall, Lane, and
  Petri{\'c}}{Mahmood et~al\mbox{.}}{2018}]%
        {mahmood2018reproducibility}
\bibfield{author}{\bibinfo{person}{Zaheed Mahmood}, \bibinfo{person}{David
  Bowes}, \bibinfo{person}{Tracy Hall}, \bibinfo{person}{Peter~CR Lane}, {and}
  \bibinfo{person}{Jean Petri{\'c}}.} \bibinfo{year}{2018}\natexlab{}.
\newblock \showarticletitle{Reproducibility and replicability of software
  defect prediction studies}.
\newblock \bibinfo{journal}{\emph{Information and Software Technology}}
  \bibinfo{volume}{99} (\bibinfo{year}{2018}), \bibinfo{pages}{148--163}.
\newblock


\bibitem[\protect\citeauthoryear{Malhotra, Kaur, and Singh}{Malhotra
  et~al\mbox{.}}{2010}]%
        {malhotra2010application}
\bibfield{author}{\bibinfo{person}{Ruchika Malhotra}, \bibinfo{person}{Arvinder
  Kaur}, {and} \bibinfo{person}{Yogesh Singh}.}
  \bibinfo{year}{2010}\natexlab{}.
\newblock \showarticletitle{Application of machine learning methods for
  software effort prediction}.
\newblock \bibinfo{journal}{\emph{ACM SIGSOFT Software Engineering Notes}}
  \bibinfo{volume}{35}, \bibinfo{number}{3} (\bibinfo{year}{2010}),
  \bibinfo{pages}{1--6}.
\newblock


\bibitem[\protect\citeauthoryear{Malik, Patra, and Pradel}{Malik
  et~al\mbox{.}}{2019}]%
        {malik2019nl2type}
\bibfield{author}{\bibinfo{person}{Rabee~Sohail Malik}, \bibinfo{person}{Jibesh
  Patra}, {and} \bibinfo{person}{Michael Pradel}.}
  \bibinfo{year}{2019}\natexlab{}.
\newblock \showarticletitle{NL2Type: inferring JavaScript function types from
  natural language information}. In \bibinfo{booktitle}{\emph{2019 IEEE/ACM
  41st International Conference on Software Engineering (ICSE)}}. IEEE,
  \bibinfo{pages}{304--315}.
\newblock


\bibitem[\protect\citeauthoryear{Manning, Surdeanu, Bauer, Finkel, Bethard, and
  McClosky}{Manning et~al\mbox{.}}{2014}]%
        {manning2014stanford}
\bibfield{author}{\bibinfo{person}{Christopher~D Manning},
  \bibinfo{person}{Mihai Surdeanu}, \bibinfo{person}{John Bauer},
  \bibinfo{person}{Jenny~Rose Finkel}, \bibinfo{person}{Steven Bethard}, {and}
  \bibinfo{person}{David McClosky}.} \bibinfo{year}{2014}\natexlab{}.
\newblock \showarticletitle{The Stanford CoreNLP natural language processing
  toolkit}. In \bibinfo{booktitle}{\emph{Proceedings of 52nd annual meeting of
  the association for computational linguistics: system demonstrations}}.
  \bibinfo{pages}{55--60}.
\newblock


\bibitem[\protect\citeauthoryear{Mi, Keung, Xiao, Mensah, and Gao}{Mi
  et~al\mbox{.}}{2018}]%
        {mi2018improving}
\bibfield{author}{\bibinfo{person}{Qing Mi}, \bibinfo{person}{Jacky Keung},
  \bibinfo{person}{Yan Xiao}, \bibinfo{person}{Solomon Mensah}, {and}
  \bibinfo{person}{Yujin Gao}.} \bibinfo{year}{2018}\natexlab{}.
\newblock \showarticletitle{Improving code readability classification using
  convolutional neural networks}.
\newblock \bibinfo{journal}{\emph{Information and Software Technology}}
  \bibinfo{volume}{104} (\bibinfo{year}{2018}), \bibinfo{pages}{60--71}.
\newblock


\bibitem[\protect\citeauthoryear{Mikolov, Karafi{\'a}t, Burget,
  {\v{C}}ernock{\`y}, and Khudanpur}{Mikolov et~al\mbox{.}}{2010}]%
        {mikolov2010recurrent}
\bibfield{author}{\bibinfo{person}{Tom{\'a}{\v{s}} Mikolov},
  \bibinfo{person}{Martin Karafi{\'a}t}, \bibinfo{person}{Luk{\'a}{\v{s}}
  Burget}, \bibinfo{person}{Jan {\v{C}}ernock{\`y}}, {and}
  \bibinfo{person}{Sanjeev Khudanpur}.} \bibinfo{year}{2010}\natexlab{}.
\newblock \showarticletitle{Recurrent neural network based language model}. In
  \bibinfo{booktitle}{\emph{Eleventh annual conference of the international
  speech communication association}}.
\newblock


\bibitem[\protect\citeauthoryear{Mikolov, Kombrink, Burget, {\v{C}}ernock{\`y},
  and Khudanpur}{Mikolov et~al\mbox{.}}{2011}]%
        {mikolov2011extensions}
\bibfield{author}{\bibinfo{person}{Tom{\'a}{\v{s}} Mikolov},
  \bibinfo{person}{Stefan Kombrink}, \bibinfo{person}{Luk{\'a}{\v{s}} Burget},
  \bibinfo{person}{Jan {\v{C}}ernock{\`y}}, {and} \bibinfo{person}{Sanjeev
  Khudanpur}.} \bibinfo{year}{2011}\natexlab{}.
\newblock \showarticletitle{Extensions of recurrent neural network language
  model}. In \bibinfo{booktitle}{\emph{2011 IEEE international conference on
  acoustics, speech and signal processing (ICASSP)}}. IEEE,
  \bibinfo{pages}{5528--5531}.
\newblock


\bibitem[\protect\citeauthoryear{Molina, Degiovanni, Ponzio, Regis, Aguirre,
  and Frias}{Molina et~al\mbox{.}}{2019}]%
        {molina2019training}
\bibfield{author}{\bibinfo{person}{Facundo Molina}, \bibinfo{person}{Renzo
  Degiovanni}, \bibinfo{person}{Pablo Ponzio}, \bibinfo{person}{Germ{\'a}n
  Regis}, \bibinfo{person}{Nazareno Aguirre}, {and} \bibinfo{person}{Marcelo
  Frias}.} \bibinfo{year}{2019}\natexlab{}.
\newblock \showarticletitle{Training binary classifiers as data structure
  invariants}. In \bibinfo{booktitle}{\emph{2019 IEEE/ACM 41st International
  Conference on Software Engineering (ICSE)}}. IEEE, \bibinfo{pages}{759--770}.
\newblock


\bibitem[\protect\citeauthoryear{Moran, Bernal-C{\'a}rdenas, Curcio, Bonett,
  and Poshyvanyk}{Moran et~al\mbox{.}}{2018}]%
        {moran2018machine}
\bibfield{author}{\bibinfo{person}{Kevin Moran}, \bibinfo{person}{Carlos
  Bernal-C{\'a}rdenas}, \bibinfo{person}{Michael Curcio},
  \bibinfo{person}{Richard Bonett}, {and} \bibinfo{person}{Denys Poshyvanyk}.}
  \bibinfo{year}{2018}\natexlab{}.
\newblock \showarticletitle{Machine learning-based prototyping of graphical
  user interfaces for mobile apps}.
\newblock \bibinfo{journal}{\emph{arXiv preprint arXiv:1802.02312}}
  (\bibinfo{year}{2018}).
\newblock


\bibitem[\protect\citeauthoryear{Mou, Li, Zhang, Wang, and Jin}{Mou
  et~al\mbox{.}}{2016}]%
        {mou2016convolutional}
\bibfield{author}{\bibinfo{person}{Lili Mou}, \bibinfo{person}{Ge Li},
  \bibinfo{person}{Lu Zhang}, \bibinfo{person}{Tao Wang}, {and}
  \bibinfo{person}{Zhi Jin}.} \bibinfo{year}{2016}\natexlab{}.
\newblock \showarticletitle{Convolutional neural networks over tree structures
  for programming language processing}. In \bibinfo{booktitle}{\emph{Thirtieth
  AAAI Conference on Artificial Intelligence}}.
\newblock


\bibitem[\protect\citeauthoryear{Mu, Guo, Cuevas, Chen, Gai, Xing, Mao, and
  Song}{Mu et~al\mbox{.}}{2019}]%
        {mu2019renn}
\bibfield{author}{\bibinfo{person}{Dongliang Mu}, \bibinfo{person}{Wenbo Guo},
  \bibinfo{person}{Alejandro Cuevas}, \bibinfo{person}{Yueqi Chen},
  \bibinfo{person}{Jinxuan Gai}, \bibinfo{person}{Xinyu Xing},
  \bibinfo{person}{Bing Mao}, {and} \bibinfo{person}{Chengyu Song}.}
  \bibinfo{year}{2019}\natexlab{}.
\newblock \showarticletitle{RENN: Efficient Reverse Execution with
  Neural-network-assisted Alias Analysis}. In \bibinfo{booktitle}{\emph{2019
  34th IEEE/ACM International Conference on Automated Software Engineering
  (ASE)}}. IEEE, \bibinfo{pages}{924--935}.
\newblock


\bibitem[\protect\citeauthoryear{Nafi, Kar, Roy, Roy, and Schneider}{Nafi
  et~al\mbox{.}}{2019}]%
        {nafi2019clcdsa}
\bibfield{author}{\bibinfo{person}{Kawser~Wazed Nafi},
  \bibinfo{person}{Tonny~Shekha Kar}, \bibinfo{person}{Banani Roy},
  \bibinfo{person}{Chanchal~K Roy}, {and} \bibinfo{person}{Kevin~A Schneider}.}
  \bibinfo{year}{2019}\natexlab{}.
\newblock \showarticletitle{CLCDSA: Cross Language Code Clone Detection using
  Syntactical Features and API Documentation}. In
  \bibinfo{booktitle}{\emph{2019 34th IEEE/ACM International Conference on
  Automated Software Engineering (ASE)}}. IEEE, \bibinfo{pages}{1026--1037}.
\newblock


\bibitem[\protect\citeauthoryear{Nakkiran, Kaplun, Bansal, Yang, Barak, and
  Sutskever}{Nakkiran et~al\mbox{.}}{2019}]%
        {nakkiran2019deep}
\bibfield{author}{\bibinfo{person}{Preetum Nakkiran}, \bibinfo{person}{Gal
  Kaplun}, \bibinfo{person}{Yamini Bansal}, \bibinfo{person}{Tristan Yang},
  \bibinfo{person}{Boaz Barak}, {and} \bibinfo{person}{Ilya Sutskever}.}
  \bibinfo{year}{2019}\natexlab{}.
\newblock \showarticletitle{Deep double descent: Where bigger models and more
  data hurt}.
\newblock \bibinfo{journal}{\emph{arXiv preprint arXiv:1912.02292}}
  (\bibinfo{year}{2019}).
\newblock


\bibitem[\protect\citeauthoryear{Neto}{Neto}{2019}]%
        {neto2019strategy}
\bibfield{author}{\bibinfo{person}{Amadeu~Anderlin Neto}.}
  \bibinfo{year}{2019}\natexlab{}.
\newblock \showarticletitle{A Strategy to Support Replications of Controlled
  Experiments in Software Engineering}.
\newblock \bibinfo{journal}{\emph{ACM SIGSOFT Software Engineering Notes}}
  \bibinfo{volume}{44}, \bibinfo{number}{3} (\bibinfo{year}{2019}),
  \bibinfo{pages}{23--23}.
\newblock


\bibitem[\protect\citeauthoryear{Nguyen, Nguyen, Phan, and Nguyen}{Nguyen
  et~al\mbox{.}}{2018}]%
        {nguyen2018deep}
\bibfield{author}{\bibinfo{person}{Anh~Tuan Nguyen}, \bibinfo{person}{Trong~Duc
  Nguyen}, \bibinfo{person}{Hung~Dang Phan}, {and} \bibinfo{person}{Tien~N
  Nguyen}.} \bibinfo{year}{2018}\natexlab{}.
\newblock \showarticletitle{A deep neural network language model with contexts
  for source code}. In \bibinfo{booktitle}{\emph{2018 IEEE 25th International
  Conference on Software Analysis, Evolution and Reengineering (SANER)}}. IEEE,
  \bibinfo{pages}{323--334}.
\newblock


\bibitem[\protect\citeauthoryear{Poggio, Kawaguchi, Liao, Miranda, Rosasco,
  Boix, Hidary, and Mhaskar}{Poggio et~al\mbox{.}}{2017}]%
        {poggio2017theory}
\bibfield{author}{\bibinfo{person}{Tomaso Poggio}, \bibinfo{person}{Kenji
  Kawaguchi}, \bibinfo{person}{Qianli Liao}, \bibinfo{person}{Brando Miranda},
  \bibinfo{person}{Lorenzo Rosasco}, \bibinfo{person}{Xavier Boix},
  \bibinfo{person}{Jack Hidary}, {and} \bibinfo{person}{Hrushikesh Mhaskar}.}
  \bibinfo{year}{2017}\natexlab{}.
\newblock \showarticletitle{Theory of deep learning III: explaining the
  non-overfitting puzzle}.
\newblock \bibinfo{journal}{\emph{arXiv preprint arXiv:1801.00173}}
  (\bibinfo{year}{2017}).
\newblock


\bibitem[\protect\citeauthoryear{Rani and Mahapatra}{Rani and
  Mahapatra}{2018}]%
        {rani2018neural}
\bibfield{author}{\bibinfo{person}{Pooja Rani} {and} \bibinfo{person}{GS
  Mahapatra}.} \bibinfo{year}{2018}\natexlab{}.
\newblock \showarticletitle{Neural network for software reliability analysis of
  dynamically weighted NHPP growth models with imperfect debugging}.
\newblock \bibinfo{journal}{\emph{Software Testing, Verification and
  Reliability}} \bibinfo{volume}{28}, \bibinfo{number}{5}
  (\bibinfo{year}{2018}), \bibinfo{pages}{e1663}.
\newblock


\bibitem[\protect\citeauthoryear{Ren, Xing, Xia, Lo, Wang, and Grundy}{Ren
  et~al\mbox{.}}{2019}]%
        {ren2019neural}
\bibfield{author}{\bibinfo{person}{Xiaoxue Ren}, \bibinfo{person}{Zhenchang
  Xing}, \bibinfo{person}{Xin Xia}, \bibinfo{person}{David Lo},
  \bibinfo{person}{Xinyu Wang}, {and} \bibinfo{person}{John Grundy}.}
  \bibinfo{year}{2019}\natexlab{}.
\newblock \showarticletitle{Neural Network-based Detection of Self-Admitted
  Technical Debt: From Performance to Explainability}.
\newblock \bibinfo{journal}{\emph{ACM Transactions on Software Engineering and
  Methodology (TOSEM)}} \bibinfo{volume}{28}, \bibinfo{number}{3}
  (\bibinfo{year}{2019}), \bibinfo{pages}{1--45}.
\newblock


\bibitem[\protect\citeauthoryear{Romansky, Borle, Chowdhury, Hindle, and
  Greiner}{Romansky et~al\mbox{.}}{2017}]%
        {romansky2017deep}
\bibfield{author}{\bibinfo{person}{Stephen Romansky}, \bibinfo{person}{Neil~C
  Borle}, \bibinfo{person}{Shaiful Chowdhury}, \bibinfo{person}{Abram Hindle},
  {and} \bibinfo{person}{Russ Greiner}.} \bibinfo{year}{2017}\natexlab{}.
\newblock \showarticletitle{Deep green: Modelling time-series of software
  energy consumption}. In \bibinfo{booktitle}{\emph{2017 IEEE International
  Conference on Software Maintenance and Evolution (ICSME)}}. IEEE,
  \bibinfo{pages}{273--283}.
\newblock


\bibitem[\protect\citeauthoryear{Ruan, Chen, Peng, and Zhao}{Ruan
  et~al\mbox{.}}{2019}]%
        {ruan2019deeplink}
\bibfield{author}{\bibinfo{person}{Hang Ruan}, \bibinfo{person}{Bihuan Chen},
  \bibinfo{person}{Xin Peng}, {and} \bibinfo{person}{Wenyun Zhao}.}
  \bibinfo{year}{2019}\natexlab{}.
\newblock \showarticletitle{DeepLink: Recovering issue-commit links based on
  deep learning}.
\newblock \bibinfo{journal}{\emph{Journal of Systems and Software}}
  \bibinfo{volume}{158} (\bibinfo{year}{2019}), \bibinfo{pages}{110406}.
\newblock


\bibitem[\protect\citeauthoryear{Sak, Senior, and Beaufays}{Sak
  et~al\mbox{.}}{2014}]%
        {sak2014long}
\bibfield{author}{\bibinfo{person}{Hasim Sak}, \bibinfo{person}{Andrew~W
  Senior}, {and} \bibinfo{person}{Fran{\c{c}}oise Beaufays}.}
  \bibinfo{year}{2014}\natexlab{}.
\newblock \showarticletitle{Long short-term memory recurrent neural network
  architectures for large scale acoustic modeling}.
\newblock  (\bibinfo{year}{2014}).
\newblock


\bibitem[\protect\citeauthoryear{Schmidhuber}{Schmidhuber}{2015}]%
        {schmidhuber2015deep}
\bibfield{author}{\bibinfo{person}{J{\"u}rgen Schmidhuber}.}
  \bibinfo{year}{2015}\natexlab{}.
\newblock \showarticletitle{Deep learning in neural networks: An overview}.
\newblock \bibinfo{journal}{\emph{Neural networks}}  \bibinfo{volume}{61}
  (\bibinfo{year}{2015}), \bibinfo{pages}{85--117}.
\newblock


\bibitem[\protect\citeauthoryear{See, Liu, and Manning}{See
  et~al\mbox{.}}{2017}]%
        {see2017get}
\bibfield{author}{\bibinfo{person}{Abigail See}, \bibinfo{person}{Peter~J Liu},
  {and} \bibinfo{person}{Christopher~D Manning}.}
  \bibinfo{year}{2017}\natexlab{}.
\newblock \showarticletitle{Get to the point: Summarization with
  pointer-generator networks}.
\newblock \bibinfo{journal}{\emph{arXiv preprint arXiv:1704.04368}}
  (\bibinfo{year}{2017}).
\newblock


\bibitem[\protect\citeauthoryear{Shuai, Xu, Liu, Yan, Xia, and Lei}{Shuai
  et~al\mbox{.}}{[n.d.]}]%
        {shuaiimproving}
\bibfield{author}{\bibinfo{person}{Jianhang Shuai}, \bibinfo{person}{Ling Xu},
  \bibinfo{person}{Chao Liu}, \bibinfo{person}{Meng Yan}, \bibinfo{person}{Xin
  Xia}, {and} \bibinfo{person}{Yan Lei}.} \bibinfo{year}{[n.d.]}\natexlab{}.
\newblock \showarticletitle{Improving Code Search with Co-Attentive
  Representation Learning}.
\newblock  (\bibinfo{year}{[n.\,d.]}).
\newblock


\bibitem[\protect\citeauthoryear{Sj{\o}berg, Hannay, Hansen, Kampenes,
  Karahasanovic, Liborg, and Rekdal}{Sj{\o}berg et~al\mbox{.}}{2005}]%
        {sjoberg2005survey}
\bibfield{author}{\bibinfo{person}{Dag~IK Sj{\o}berg},
  \bibinfo{person}{Jo~Erskine Hannay}, \bibinfo{person}{Ove Hansen},
  \bibinfo{person}{Vigdis~By Kampenes}, \bibinfo{person}{Amela Karahasanovic},
  \bibinfo{person}{N-K Liborg}, {and} \bibinfo{person}{Anette~C Rekdal}.}
  \bibinfo{year}{2005}\natexlab{}.
\newblock \showarticletitle{A survey of controlled experiments in software
  engineering}.
\newblock \bibinfo{journal}{\emph{IEEE transactions on software engineering}}
  \bibinfo{volume}{31}, \bibinfo{number}{9} (\bibinfo{year}{2005}),
  \bibinfo{pages}{733--753}.
\newblock


\bibitem[\protect\citeauthoryear{Szeliski}{Szeliski}{2010}]%
        {szeliski2010computer}
\bibfield{author}{\bibinfo{person}{Richard Szeliski}.}
  \bibinfo{year}{2010}\natexlab{}.
\newblock \bibinfo{booktitle}{\emph{Computer vision: algorithms and
  applications}}.
\newblock \bibinfo{publisher}{Springer Science \& Business Media}.
\newblock


\bibitem[\protect\citeauthoryear{Thaller, Linsbauer, and Egyed}{Thaller
  et~al\mbox{.}}{2019}]%
        {thaller2019feature}
\bibfield{author}{\bibinfo{person}{Hannes Thaller}, \bibinfo{person}{Lukas
  Linsbauer}, {and} \bibinfo{person}{Alexander Egyed}.}
  \bibinfo{year}{2019}\natexlab{}.
\newblock \showarticletitle{Feature maps: A comprehensible software
  representation for design pattern detection}. In
  \bibinfo{booktitle}{\emph{2019 IEEE 26th International Conference on Software
  Analysis, Evolution and Reengineering (SANER)}}. IEEE,
  \bibinfo{pages}{207--217}.
\newblock


\bibitem[\protect\citeauthoryear{Tian, Pei, Jana, and Ray}{Tian
  et~al\mbox{.}}{2018}]%
        {tian2018deeptest}
\bibfield{author}{\bibinfo{person}{Yuchi Tian}, \bibinfo{person}{Kexin Pei},
  \bibinfo{person}{Suman Jana}, {and} \bibinfo{person}{Baishakhi Ray}.}
  \bibinfo{year}{2018}\natexlab{}.
\newblock \showarticletitle{Deeptest: Automated testing of
  deep-neural-network-driven autonomous cars}. In
  \bibinfo{booktitle}{\emph{Proceedings of the 40th international conference on
  software engineering}}. \bibinfo{pages}{303--314}.
\newblock


\bibitem[\protect\citeauthoryear{Tong, Liu, and Wang}{Tong
  et~al\mbox{.}}{2018}]%
        {tong2018software}
\bibfield{author}{\bibinfo{person}{Haonan Tong}, \bibinfo{person}{Bin Liu},
  {and} \bibinfo{person}{Shihai Wang}.} \bibinfo{year}{2018}\natexlab{}.
\newblock \showarticletitle{Software defect prediction using stacked denoising
  autoencoders and two-stage ensemble learning}.
\newblock \bibinfo{journal}{\emph{Information and Software Technology}}
  \bibinfo{volume}{96} (\bibinfo{year}{2018}), \bibinfo{pages}{94--111}.
\newblock


\bibitem[\protect\citeauthoryear{Tufano, Pantiuchina, Watson, Bavota, and
  Poshyvanyk}{Tufano et~al\mbox{.}}{2019a}]%
        {tufano2019learning}
\bibfield{author}{\bibinfo{person}{Michele Tufano}, \bibinfo{person}{Jevgenija
  Pantiuchina}, \bibinfo{person}{Cody Watson}, \bibinfo{person}{Gabriele
  Bavota}, {and} \bibinfo{person}{Denys Poshyvanyk}.}
  \bibinfo{year}{2019}\natexlab{a}.
\newblock \showarticletitle{On learning meaningful code changes via neural
  machine translation}. In \bibinfo{booktitle}{\emph{2019 IEEE/ACM 41st
  International Conference on Software Engineering (ICSE)}}. IEEE,
  \bibinfo{pages}{25--36}.
\newblock


\bibitem[\protect\citeauthoryear{Tufano, Watson, Bavota, Di~Penta, White, and
  Poshyvanyk}{Tufano et~al\mbox{.}}{2018}]%
        {tufano2018learning}
\bibfield{author}{\bibinfo{person}{Michele Tufano}, \bibinfo{person}{Cody
  Watson}, \bibinfo{person}{Gabriele Bavota}, \bibinfo{person}{Massimiliano
  Di~Penta}, \bibinfo{person}{Martin White}, {and} \bibinfo{person}{Denys
  Poshyvanyk}.} \bibinfo{year}{2018}\natexlab{}.
\newblock \showarticletitle{Learning how to mutate source code from bug-fixes}.
  In \bibinfo{booktitle}{\emph{2019 IEEE International Conference on Software
  Maintenance and Evolution (ICSME)}}. IEEE, \bibinfo{pages}{301--312}.
\newblock


\bibitem[\protect\citeauthoryear{Tufano, Watson, Bavota, Penta, White, and
  Poshyvanyk}{Tufano et~al\mbox{.}}{2019b}]%
        {tufano2019empirical}
\bibfield{author}{\bibinfo{person}{Michele Tufano}, \bibinfo{person}{Cody
  Watson}, \bibinfo{person}{Gabriele Bavota}, \bibinfo{person}{Massimiliano~Di
  Penta}, \bibinfo{person}{Martin White}, {and} \bibinfo{person}{Denys
  Poshyvanyk}.} \bibinfo{year}{2019}\natexlab{b}.
\newblock \showarticletitle{An empirical study on learning bug-fixing patches
  in the wild via neural machine translation}.
\newblock \bibinfo{journal}{\emph{ACM Transactions on Software Engineering and
  Methodology (TOSEM)}} \bibinfo{volume}{28}, \bibinfo{number}{4}
  (\bibinfo{year}{2019}), \bibinfo{pages}{1--29}.
\newblock


\bibitem[\protect\citeauthoryear{Vaswani, Bengio, Brevdo, Chollet, Gomez,
  Gouws, Jones, Kaiser, Kalchbrenner, Parmar, et~al\mbox{.}}{Vaswani
  et~al\mbox{.}}{2018}]%
        {vaswani2018tensor2tensor}
\bibfield{author}{\bibinfo{person}{Ashish Vaswani}, \bibinfo{person}{Samy
  Bengio}, \bibinfo{person}{Eugene Brevdo}, \bibinfo{person}{Francois Chollet},
  \bibinfo{person}{Aidan~N Gomez}, \bibinfo{person}{Stephan Gouws},
  \bibinfo{person}{Llion Jones}, \bibinfo{person}{{\L}ukasz Kaiser},
  \bibinfo{person}{Nal Kalchbrenner}, \bibinfo{person}{Niki Parmar},
  {et~al\mbox{.}}} \bibinfo{year}{2018}\natexlab{}.
\newblock \showarticletitle{Tensor2tensor for neural machine translation}.
\newblock \bibinfo{journal}{\emph{arXiv preprint arXiv:1803.07416}}
  (\bibinfo{year}{2018}).
\newblock


\bibitem[\protect\citeauthoryear{Wan, Shu, Sui, Xu, Zhao, Wu, and Yu}{Wan
  et~al\mbox{.}}{2019}]%
        {wan2019multi}
\bibfield{author}{\bibinfo{person}{Yao Wan}, \bibinfo{person}{Jingdong Shu},
  \bibinfo{person}{Yulei Sui}, \bibinfo{person}{Guandong Xu},
  \bibinfo{person}{Zhou Zhao}, \bibinfo{person}{Jian Wu}, {and}
  \bibinfo{person}{Philip Yu}.} \bibinfo{year}{2019}\natexlab{}.
\newblock \showarticletitle{Multi-modal attention network learning for semantic
  source code retrieval}. In \bibinfo{booktitle}{\emph{2019 34th IEEE/ACM
  International Conference on Automated Software Engineering (ASE)}}. IEEE,
  \bibinfo{pages}{13--25}.
\newblock


\bibitem[\protect\citeauthoryear{Wan, Zhao, Yang, Xu, Ying, Wu, and Yu}{Wan
  et~al\mbox{.}}{2018}]%
        {wan2018improving}
\bibfield{author}{\bibinfo{person}{Yao Wan}, \bibinfo{person}{Zhou Zhao},
  \bibinfo{person}{Min Yang}, \bibinfo{person}{Guandong Xu},
  \bibinfo{person}{Haochao Ying}, \bibinfo{person}{Jian Wu}, {and}
  \bibinfo{person}{Philip~S Yu}.} \bibinfo{year}{2018}\natexlab{}.
\newblock \showarticletitle{Improving automatic source code summarization via
  deep reinforcement learning}. In \bibinfo{booktitle}{\emph{Proceedings of the
  33rd ACM/IEEE International Conference on Automated Software Engineering}}.
  \bibinfo{pages}{397--407}.
\newblock


\bibitem[\protect\citeauthoryear{Wang, Liu, Nam, and Tan}{Wang
  et~al\mbox{.}}{2018}]%
        {wang2018deep}
\bibfield{author}{\bibinfo{person}{Song Wang}, \bibinfo{person}{Taiyue Liu},
  \bibinfo{person}{Jaechang Nam}, {and} \bibinfo{person}{Lin Tan}.}
  \bibinfo{year}{2018}\natexlab{}.
\newblock \showarticletitle{Deep semantic feature learning for software defect
  prediction}.
\newblock \bibinfo{journal}{\emph{IEEE Transactions on Software Engineering}}
  (\bibinfo{year}{2018}).
\newblock


\bibitem[\protect\citeauthoryear{Wang, Liu, and Tan}{Wang
  et~al\mbox{.}}{2016}]%
        {wang2016automatically}
\bibfield{author}{\bibinfo{person}{Song Wang}, \bibinfo{person}{Taiyue Liu},
  {and} \bibinfo{person}{Lin Tan}.} \bibinfo{year}{2016}\natexlab{}.
\newblock \showarticletitle{Automatically learning semantic features for defect
  prediction}. In \bibinfo{booktitle}{\emph{2016 IEEE/ACM 38th International
  Conference on Software Engineering (ICSE)}}. IEEE, \bibinfo{pages}{297--308}.
\newblock


\bibitem[\protect\citeauthoryear{Wang, Chen, and Xing}{Wang
  et~al\mbox{.}}{2019a}]%
        {wang2019domain}
\bibfield{author}{\bibinfo{person}{Xu Wang}, \bibinfo{person}{Chunyang Chen},
  {and} \bibinfo{person}{Zhenchang Xing}.} \bibinfo{year}{2019}\natexlab{a}.
\newblock \showarticletitle{Domain-specific machine translation with recurrent
  neural network for software localization}.
\newblock \bibinfo{journal}{\emph{Empirical Software Engineering}}
  \bibinfo{volume}{24}, \bibinfo{number}{6} (\bibinfo{year}{2019}),
  \bibinfo{pages}{3514--3545}.
\newblock


\bibitem[\protect\citeauthoryear{Wang, Xu, Zhou, Lyu, and Wang}{Wang
  et~al\mbox{.}}{2019b}]%
        {wang2019textout}
\bibfield{author}{\bibinfo{person}{Yaohui Wang}, \bibinfo{person}{Hui Xu},
  \bibinfo{person}{Yangfan Zhou}, \bibinfo{person}{Michael~R Lyu}, {and}
  \bibinfo{person}{Xin Wang}.} \bibinfo{year}{2019}\natexlab{b}.
\newblock \showarticletitle{Textout: Detecting Text-Layout Bugs in Mobile Apps
  via Visualization-Oriented Learning}. In \bibinfo{booktitle}{\emph{2019 IEEE
  30th International Symposium on Software Reliability Engineering (ISSRE)}}.
  IEEE, \bibinfo{pages}{239--249}.
\newblock


\bibitem[\protect\citeauthoryear{Wen, Wu, and Cheung}{Wen
  et~al\mbox{.}}{2018}]%
        {wen2018well}
\bibfield{author}{\bibinfo{person}{Ming Wen}, \bibinfo{person}{Rongxin Wu},
  {and} \bibinfo{person}{Shing-Chi Cheung}.} \bibinfo{year}{2018}\natexlab{}.
\newblock \showarticletitle{How well do change sequences predict defects?
  sequence learning from software changes}.
\newblock \bibinfo{journal}{\emph{IEEE Transactions on Software Engineering}}
  (\bibinfo{year}{2018}).
\newblock


\bibitem[\protect\citeauthoryear{White, Tufano, Martinez, Monperrus, and
  Poshyvanyk}{White et~al\mbox{.}}{2019}]%
        {white2019sorting}
\bibfield{author}{\bibinfo{person}{Martin White}, \bibinfo{person}{Michele
  Tufano}, \bibinfo{person}{Matias Martinez}, \bibinfo{person}{Martin
  Monperrus}, {and} \bibinfo{person}{Denys Poshyvanyk}.}
  \bibinfo{year}{2019}\natexlab{}.
\newblock \showarticletitle{Sorting and transforming program repair ingredients
  via deep learning code similarities}. In \bibinfo{booktitle}{\emph{2019 IEEE
  26th International Conference on Software Analysis, Evolution and
  Reengineering (SANER)}}. IEEE, \bibinfo{pages}{479--490}.
\newblock


\bibitem[\protect\citeauthoryear{White, Tufano, Vendome, and Poshyvanyk}{White
  et~al\mbox{.}}{2016}]%
        {white2016deep}
\bibfield{author}{\bibinfo{person}{Martin White}, \bibinfo{person}{Michele
  Tufano}, \bibinfo{person}{Christopher Vendome}, {and} \bibinfo{person}{Denys
  Poshyvanyk}.} \bibinfo{year}{2016}\natexlab{}.
\newblock \showarticletitle{Deep learning code fragments for code clone
  detection}. In \bibinfo{booktitle}{\emph{2016 31st IEEE/ACM International
  Conference on Automated Software Engineering (ASE)}}. IEEE,
  \bibinfo{pages}{87--98}.
\newblock


\bibitem[\protect\citeauthoryear{Wilcoxon}{Wilcoxon}{1945}]%
        {wilcoxon1945individual}
\bibfield{author}{\bibinfo{person}{Frank Wilcoxon}.}
  \bibinfo{year}{1945}\natexlab{}.
\newblock \showarticletitle{Individual comparisons by ranking methods}.
\newblock \bibinfo{journal}{\emph{Biometrics bulletin}} \bibinfo{volume}{1},
  \bibinfo{number}{6} (\bibinfo{year}{1945}), \bibinfo{pages}{80--83}.
\newblock


\bibitem[\protect\citeauthoryear{Xiao, Keung, Bennin, and Mi}{Xiao
  et~al\mbox{.}}{2019}]%
        {xiao2019improving}
\bibfield{author}{\bibinfo{person}{Yan Xiao}, \bibinfo{person}{Jacky Keung},
  \bibinfo{person}{Kwabena~E Bennin}, {and} \bibinfo{person}{Qing Mi}.}
  \bibinfo{year}{2019}\natexlab{}.
\newblock \showarticletitle{Improving bug localization with word embedding and
  enhanced convolutional neural networks}.
\newblock \bibinfo{journal}{\emph{Information and Software Technology}}
  \bibinfo{volume}{105} (\bibinfo{year}{2019}), \bibinfo{pages}{17--29}.
\newblock


\bibitem[\protect\citeauthoryear{Xie, Chen, Ye, Li, Hu, Du, and Zhang}{Xie
  et~al\mbox{.}}{2019}]%
        {xie2019deeplink}
\bibfield{author}{\bibinfo{person}{Rui Xie}, \bibinfo{person}{Long Chen},
  \bibinfo{person}{Wei Ye}, \bibinfo{person}{Zhiyu Li},
  \bibinfo{person}{Tianxiang Hu}, \bibinfo{person}{Dongdong Du}, {and}
  \bibinfo{person}{Shikun Zhang}.} \bibinfo{year}{2019}\natexlab{}.
\newblock \showarticletitle{DeepLink: A code knowledge graph based deep
  learning approach for issue-commit link recovery}. In
  \bibinfo{booktitle}{\emph{2019 IEEE 26th International Conference on Software
  Analysis, Evolution and Reengineering (SANER)}}. IEEE,
  \bibinfo{pages}{434--444}.
\newblock


\bibitem[\protect\citeauthoryear{Xu, Ye, Xing, Xia, Chen, and Li}{Xu
  et~al\mbox{.}}{2016}]%
        {xu2016predicting}
\bibfield{author}{\bibinfo{person}{Bowen Xu}, \bibinfo{person}{Deheng Ye},
  \bibinfo{person}{Zhenchang Xing}, \bibinfo{person}{Xin Xia},
  \bibinfo{person}{Guibin Chen}, {and} \bibinfo{person}{Shanping Li}.}
  \bibinfo{year}{2016}\natexlab{}.
\newblock \showarticletitle{Predicting semantically linkable knowledge in
  developer online forums via convolutional neural network}. In
  \bibinfo{booktitle}{\emph{2016 31st IEEE/ACM International Conference on
  Automated Software Engineering (ASE)}}. IEEE, \bibinfo{pages}{51--62}.
\newblock


\bibitem[\protect\citeauthoryear{Xu, Li, Xu, Liu, Luo, Zhang, Zhang, Keung, and
  Tang}{Xu et~al\mbox{.}}{2019}]%
        {xu2019ldfr}
\bibfield{author}{\bibinfo{person}{Zhou Xu}, \bibinfo{person}{Shuai Li},
  \bibinfo{person}{Jun Xu}, \bibinfo{person}{Jin Liu}, \bibinfo{person}{Xiapu
  Luo}, \bibinfo{person}{Yifeng Zhang}, \bibinfo{person}{Tao Zhang},
  \bibinfo{person}{Jacky Keung}, {and} \bibinfo{person}{Yutian Tang}.}
  \bibinfo{year}{2019}\natexlab{}.
\newblock \showarticletitle{LDFR: Learning deep feature representation for
  software defect prediction}.
\newblock \bibinfo{journal}{\emph{Journal of Systems and Software}}
  \bibinfo{volume}{158} (\bibinfo{year}{2019}), \bibinfo{pages}{110402}.
\newblock


\bibitem[\protect\citeauthoryear{Yan, Yang, Liu, and Zhang}{Yan
  et~al\mbox{.}}{2016}]%
        {yan2016self}
\bibfield{author}{\bibinfo{person}{Meng Yan}, \bibinfo{person}{Mengning Yang},
  \bibinfo{person}{Chao Liu}, {and} \bibinfo{person}{Xiaohong Zhang}.}
  \bibinfo{year}{2016}\natexlab{}.
\newblock \showarticletitle{Self-learning Change-prone Class Prediction}. In
  \bibinfo{booktitle}{\emph{SEKE}}. \bibinfo{pages}{134--140}.
\newblock


\bibitem[\protect\citeauthoryear{Yan, Zhang, Liu, Xu, Yang, and Yang}{Yan
  et~al\mbox{.}}{2017a}]%
        {yan2017automated}
\bibfield{author}{\bibinfo{person}{Meng Yan}, \bibinfo{person}{Xiaohong Zhang},
  \bibinfo{person}{Chao Liu}, \bibinfo{person}{Ling Xu},
  \bibinfo{person}{Mengning Yang}, {and} \bibinfo{person}{Dan Yang}.}
  \bibinfo{year}{2017}\natexlab{a}.
\newblock \showarticletitle{Automated change-prone class prediction on
  unlabeled dataset using unsupervised method}.
\newblock \bibinfo{journal}{\emph{Information and Software Technology}}
  \bibinfo{volume}{92} (\bibinfo{year}{2017}), \bibinfo{pages}{1--16}.
\newblock


\bibitem[\protect\citeauthoryear{Yan, Zhang, Liu, Zou, Xu, and Xia}{Yan
  et~al\mbox{.}}{2017b}]%
        {yan2017learning}
\bibfield{author}{\bibinfo{person}{Meng Yan}, \bibinfo{person}{Xiaohong Zhang},
  \bibinfo{person}{Chao Liu}, \bibinfo{person}{Jie Zou}, \bibinfo{person}{Ling
  Xu}, {and} \bibinfo{person}{Xin Xia}.} \bibinfo{year}{2017}\natexlab{b}.
\newblock \showarticletitle{Learning to aggregate: an automated aggregation
  method for software quality model}. In \bibinfo{booktitle}{\emph{2017
  IEEE/ACM 39th International Conference on Software Engineering Companion
  (ICSE-C)}}. IEEE, \bibinfo{pages}{268--270}.
\newblock


\bibitem[\protect\citeauthoryear{Yan, Xiao, Hu, Peng, and Jiang}{Yan
  et~al\mbox{.}}{2018}]%
        {yan2018new}
\bibfield{author}{\bibinfo{person}{Ruibo Yan}, \bibinfo{person}{Xi Xiao},
  \bibinfo{person}{Guangwu Hu}, \bibinfo{person}{Sancheng Peng}, {and}
  \bibinfo{person}{Yong Jiang}.} \bibinfo{year}{2018}\natexlab{}.
\newblock \showarticletitle{New deep learning method to detect code injection
  attacks on hybrid applications}.
\newblock \bibinfo{journal}{\emph{Journal of Systems and Software}}
  \bibinfo{volume}{137} (\bibinfo{year}{2018}), \bibinfo{pages}{67--77}.
\newblock


\bibitem[\protect\citeauthoryear{Yu, Lam, Chen, Li, Xie, and Wang}{Yu
  et~al\mbox{.}}{2019}]%
        {yu2019neural}
\bibfield{author}{\bibinfo{person}{Hao Yu}, \bibinfo{person}{Wing Lam},
  \bibinfo{person}{Long Chen}, \bibinfo{person}{Ge Li}, \bibinfo{person}{Tao
  Xie}, {and} \bibinfo{person}{Qianxiang Wang}.}
  \bibinfo{year}{2019}\natexlab{}.
\newblock \showarticletitle{Neural detection of semantic code clones via
  tree-based convolution}. In \bibinfo{booktitle}{\emph{2019 IEEE/ACM 27th
  International Conference on Program Comprehension (ICPC)}}. IEEE,
  \bibinfo{pages}{70--80}.
\newblock


\bibitem[\protect\citeauthoryear{Zannier, Melnik, and Maurer}{Zannier
  et~al\mbox{.}}{2006}]%
        {zannier2006success}
\bibfield{author}{\bibinfo{person}{Carmen Zannier}, \bibinfo{person}{Grigori
  Melnik}, {and} \bibinfo{person}{Frank Maurer}.}
  \bibinfo{year}{2006}\natexlab{}.
\newblock \showarticletitle{On the success of empirical studies in the
  international conference on software engineering}. In
  \bibinfo{booktitle}{\emph{Proceedings of the 28th international conference on
  Software engineering}}. \bibinfo{pages}{341--350}.
\newblock


\bibitem[\protect\citeauthoryear{Zhang, Wang, Zhang, Sun, Wang, and Liu}{Zhang
  et~al\mbox{.}}{2019b}]%
        {zhang2019novel}
\bibfield{author}{\bibinfo{person}{Jian Zhang}, \bibinfo{person}{Xu Wang},
  \bibinfo{person}{Hongyu Zhang}, \bibinfo{person}{Hailong Sun},
  \bibinfo{person}{Kaixuan Wang}, {and} \bibinfo{person}{Xudong Liu}.}
  \bibinfo{year}{2019}\natexlab{b}.
\newblock \showarticletitle{A novel neural source code representation based on
  abstract syntax tree}. In \bibinfo{booktitle}{\emph{2019 IEEE/ACM 41st
  International Conference on Software Engineering (ICSE)}}. IEEE,
  \bibinfo{pages}{783--794}.
\newblock


\bibitem[\protect\citeauthoryear{Zhang, Lei, Mao, and Li}{Zhang
  et~al\mbox{.}}{2019a}]%
        {zhang2019cnn}
\bibfield{author}{\bibinfo{person}{Zhuo Zhang}, \bibinfo{person}{Yan Lei},
  \bibinfo{person}{Xiaoguang Mao}, {and} \bibinfo{person}{Panpan Li}.}
  \bibinfo{year}{2019}\natexlab{a}.
\newblock \showarticletitle{CNN-FL: An effective approach for localizing faults
  using convolutional neural networks}. In \bibinfo{booktitle}{\emph{2019 IEEE
  26th International Conference on Software Analysis, Evolution and
  Reengineering (SANER)}}. IEEE, \bibinfo{pages}{445--455}.
\newblock


\bibitem[\protect\citeauthoryear{Zhao, Xing, Chen, Xia, and Li}{Zhao
  et~al\mbox{.}}{2019}]%
        {zhao2019actionnet}
\bibfield{author}{\bibinfo{person}{Dehai Zhao}, \bibinfo{person}{Zhenchang
  Xing}, \bibinfo{person}{Chunyang Chen}, \bibinfo{person}{Xin Xia}, {and}
  \bibinfo{person}{Guoqiang Li}.} \bibinfo{year}{2019}\natexlab{}.
\newblock \showarticletitle{ActionNet: vision-based workflow action recognition
  from programming screencasts}. In \bibinfo{booktitle}{\emph{2019 IEEE/ACM
  41st International Conference on Software Engineering (ICSE)}}. IEEE,
  \bibinfo{pages}{350--361}.
\newblock


\bibitem[\protect\citeauthoryear{Zhao and Huang}{Zhao and Huang}{2018}]%
        {zhao2018deepsim}
\bibfield{author}{\bibinfo{person}{Gang Zhao} {and} \bibinfo{person}{Jeff
  Huang}.} \bibinfo{year}{2018}\natexlab{}.
\newblock \showarticletitle{Deepsim: deep learning code functional similarity}.
  In \bibinfo{booktitle}{\emph{Proceedings of the 2018 26th ACM Joint Meeting
  on European Software Engineering Conference and Symposium on the Foundations
  of Software Engineering}}. \bibinfo{pages}{141--151}.
\newblock


\bibitem[\protect\citeauthoryear{Zhao, Albarghouthi, Rastogi, Jha, and
  Octeau}{Zhao et~al\mbox{.}}{2018}]%
        {zhao2018neural}
\bibfield{author}{\bibinfo{person}{Jinman Zhao}, \bibinfo{person}{Aws
  Albarghouthi}, \bibinfo{person}{Vaibhav Rastogi}, \bibinfo{person}{Somesh
  Jha}, {and} \bibinfo{person}{Damien Octeau}.}
  \bibinfo{year}{2018}\natexlab{}.
\newblock \showarticletitle{Neural-augmented static analysis of Android
  communication}. In \bibinfo{booktitle}{\emph{Proceedings of the 2018 26th ACM
  Joint Meeting on European Software Engineering Conference and Symposium on
  the Foundations of Software Engineering}}. \bibinfo{pages}{342--353}.
\newblock


\bibitem[\protect\citeauthoryear{Zheng, Xie, Su, Ma, Hao, Meng, Liu, Shen,
  Chen, and Fan}{Zheng et~al\mbox{.}}{2019}]%
        {zheng2019wuji}
\bibfield{author}{\bibinfo{person}{Yan Zheng}, \bibinfo{person}{Xiaofei Xie},
  \bibinfo{person}{Ting Su}, \bibinfo{person}{Lei Ma}, \bibinfo{person}{Jianye
  Hao}, \bibinfo{person}{Zhaopeng Meng}, \bibinfo{person}{Yang Liu},
  \bibinfo{person}{Ruimin Shen}, \bibinfo{person}{Yingfeng Chen}, {and}
  \bibinfo{person}{Changjie Fan}.} \bibinfo{year}{2019}\natexlab{}.
\newblock \showarticletitle{Wuji: Automatic online combat game testing using
  evolutionary deep reinforcement learning}. In \bibinfo{booktitle}{\emph{2019
  34th IEEE/ACM International Conference on Automated Software Engineering
  (ASE)}}. IEEE, \bibinfo{pages}{772--784}.
\newblock


\bibitem[\protect\citeauthoryear{Zhongxin, Xin, David, Zhenchang, Ahmed, and
  Shanping}{Zhongxin et~al\mbox{.}}{2019}]%
        {liu2019log}
\bibfield{author}{\bibinfo{person}{Liu Zhongxin}, \bibinfo{person}{Xia Xin},
  \bibinfo{person}{Lo David}, \bibinfo{person}{Xing Zhenchang},
  \bibinfo{person}{E.~Hassan Ahmed}, {and} \bibinfo{person}{Li Shanping}.}
  \bibinfo{year}{2019}\natexlab{}.
\newblock \showarticletitle{Which Variables Should I Log?}
\newblock \bibinfo{journal}{\emph{IEEE Transactions on Software Engineering}}
  (\bibinfo{year}{2019}).
\newblock


\bibitem[\protect\citeauthoryear{Zhou, Liu, Liu, Yang, and Grundy}{Zhou
  et~al\mbox{.}}{2019a}]%
        {zhou2019deep}
\bibfield{author}{\bibinfo{person}{Pingyi Zhou}, \bibinfo{person}{Jin Liu},
  \bibinfo{person}{Xiao Liu}, \bibinfo{person}{Zijiang Yang}, {and}
  \bibinfo{person}{John Grundy}.} \bibinfo{year}{2019}\natexlab{a}.
\newblock \showarticletitle{Is deep learning better than traditional approaches
  in tag recommendation for software information sites?}
\newblock \bibinfo{journal}{\emph{Information and software technology}}
  \bibinfo{volume}{109} (\bibinfo{year}{2019}), \bibinfo{pages}{1--13}.
\newblock


\bibitem[\protect\citeauthoryear{Zhou, Yan, Yang, Chen, and Huang}{Zhou
  et~al\mbox{.}}{2019b}]%
        {zhou2019augmenting}
\bibfield{author}{\bibinfo{person}{Yu Zhou}, \bibinfo{person}{Xin Yan},
  \bibinfo{person}{Wenhua Yang}, \bibinfo{person}{Taolue Chen}, {and}
  \bibinfo{person}{Zhiqiu Huang}.} \bibinfo{year}{2019}\natexlab{b}.
\newblock \showarticletitle{Augmenting Java method comments generation with
  context information based on neural networks}.
\newblock \bibinfo{journal}{\emph{Journal of Systems and Software}}
  \bibinfo{volume}{156} (\bibinfo{year}{2019}), \bibinfo{pages}{328--340}.
\newblock


\end{thebibliography}

\end{document}